\documentclass[11pt,english]{article}
\usepackage[T1]{fontenc}
\usepackage{amsmath}
\usepackage{amssymb}
\usepackage{graphicx}

\makeatletter
\usepackage{jheppub}
\usepackage{multicol}
\usepackage{bbm}\usepackage{enumerate}
\usepackage{bibentry}
\usepackage[toc,page]{appendix}

\author[a]{Zhi-Zhong Li}
\author[a]{Hung-Hwa Lin}
\author[a]{Shun-Qing Zhang}
\affiliation{$^a$ Department of Physics and Astronomy, National Taiwan University, Taipei 10617, Taiwan}
\emailAdd{b02202003@ntu.edu.tw}
\emailAdd{hunghwalin@gmail.com}
\emailAdd{e24019025@gmail.com}
\abstract{Double-soft theorems, like its single-soft counterparts, arises from the underlying symmetry principles that constrain the interactions of massless particles. While single soft theorems can be derived in a non-perturbative fashion by employing current algebras, recent attempts of extending such an approach to known double soft theorems has been met with difficulties. In this work, we have traced the difficulty to two inequivalent expansion schemes, depending on whether the soft limit is taken asymmetrically or symmetrically, which we denote as type A and B respectively. The soft-behaviour for type A scheme can simply be derived from single soft theorems, and are thus non-perturbatively protected. For type B, the information of the four-point vertex is required to determine the corresponding soft theorems, and thus are in general not protected. 
This argument can be readily extended to general multi-soft theorems. 
We also ask whether unitarity can be emergent from locality together with the two kinds of soft theorems, which has not been fully investigated before.  }

\@ifundefined{showcaptionsetup}{}{%
 \PassOptionsToPackage{caption=false}{subfig}}
\usepackage{subfig}
\makeatother

\usepackage{babel}
\begin{document}

\title{On the Symmetry Foundation of Double Soft Theorems}
\maketitle

\section{Introduction}

Symmetry principles of massless theories tend to manifest themselves
as soft behavior of scattering amplitudes, where one considers an
expansion on small momenta for some of the external legs. Some classic
examples are gauge symmetry manifested as Weinberg's soft theorem
\cite{Weinberg}, and spontaneously broken symmetry reflected as Adler's
zero \cite{Adler}. Many other theorems involving photons and gravitons have been derived \cite{Low, photon_Kroll, photon_Duca, grav_Gross, grav_Jackiw, grav_White}, including some later works \cite{recent_Cachazo, recent_Casali}.

Recently, some new double soft theorems of higher soft order has been
derived for theories with spontaneously broken symmetry, using CHY
representation \cite{soft2He}. Involved theories include nonlinear
sigma model (NLSM), Dirac-Born-Infield (DBI) theory, and special Galilean
(sGal), to name a few. Since the derivation is valid only for tree
amplitudes, while underlying symmetry principle should persist into
loop order, it is worth investigating whether these theorems withstand
loop correction. Using current algebra of the spontaneously broken
symmetry and explicit loop calculation of specific theories, some
of these theorems have been reproduced and their behavior under loop
correction discussed \cite{DiVecchia, soft2Huang}.

One notable shortfall of \cite{soft2Huang} is that not all double soft theorems of  \cite{soft2He} were derived. For example,
DBI amplitudes has double soft theorem up to $\tau^{3}$ \cite{soft2He},
where $\tau$ is the soft expansion parameter. Explicit loop calculation
shows that theorems up to $\tau^{2}$ are not modified, indicating
that they may be symmetry protected \cite{soft2Huang}. However, current
algebra has only reproduced theorems up to $\tau^{1}$, leaving $\tau^{2}$
unaccounted for. It is also unclear the origin of $\tau^{3}$. Second, the current algebra structure of conformal symmetry breaking is similar to DBI, one reflecting the symmetry breaking due to a brane in AdS space while the other, flat space. However, as opposed to DBI, up
to now double soft theorems for dilaton only exists up to $\tau^{1}$
order. It is not clear whether further double soft theorems exist.
If it does not, the discrepancy seems difficult to be explained from the view point of 
current algebras. Furthermore, since some theorems (in particular $\tau^{2}$
of DBI) are not derived, it is still not clear whether (or why and
why not) these theorems will be modified by loop corrections.

We try to address these issues in this work. 
First, the difficulty in using current algebra to derive existing double
soft theorems arises from ambiguity of double soft expansion. 
There are actually two inequivalent ways to perform
double soft expansion: either parametrizing the soft-limit by two independent, or with a single parameter. 
We will refer the former as type A scheme, and later type B scheme. 
Such distinction has already been explored for gluon and graviton amplitudes (e.g. \cite{gluon_AB, grav_AB, current_is_A}), 
and is also familiar to phenomenologists working on soft physics (e.g. in soft-collinear effective theory \cite{SCET}).  
While for well-behaved functions these two schemes are equivalent, scattering
amplitudes involve pieces that become singular under double soft limit, such as three-particle factorization poles, which will lead to discrepancies. 
Therefore, current algebra, which naturally gives type A theorems \cite{current_is_A}, 
cannot directly reproduce the new double soft theorems in \cite{soft2He}, which are of type B.

To circumvent the problem, we establish a procedure to derive double soft theorems, using two relations:
First, type A double soft theorems for type A scheme can be directly derived from single soft theorems \cite{gluon_AB, grav_AB}.  
Second, type B scheme can be obtained from type A by adding information of four point vertex, known in SCET theory \cite{SCET}. 
Though these relations are separately known in literature, they have not been used together to address the derivation and loop correction of double soft theorems. 
We combine the two into a general framework and fill out some explicit computation steps, borrowing some techniques in \cite{nonlinear}. 
The result is a procedure that brings us from single soft theorems to type A theorems, and eventually to type B double soft theorems. 
Our approach proves to be an efficient way to derive double soft theorems from symmetry principle, 
and facilitates definite prediction on whether double soft theorems will be modified by loop corrections. 
Since single soft theorems can be derived clearly from current algebra, soft-theorems for the A type should be protected from loop corrections. 
For soft-theorems of the B type, the modifications of which can be unambiguously identified from the four-point vertex. 
We are able to derive the theorems
left out in \cite{soft2Huang} as well a new double soft theorem for dilaton, 
and clarify the behavior of those theorems under loop correction. 
Though only single massless scalar theories are considered here, this
can be generalized to theories with color or flavor group structure.

Our analysis also clarifies the distinction between spontaneous symmetry breaking, and the corresponding non-linear realization on fields. In the usual discussion of effective field theories (EFT) we only retain a subset of operators valid under certain approximations such as leading derivative or constant field. The non-linear realization of the broken symmetries is then only valid under such approximations, with the understanding that higher order corrections will demand further modification of the transformation rules. However as we've discussed previously, there are soft-theorems that are universal and not subject to quantum corrections, and hence apply to the effective field theory irrespective to which order in the approximation one is considering. Thus this implies that part of the non-linear transformation has to be universal. Indeed, the nonlinear transformation of
Goldstone bosons under spontaneously broken symmetries can often be
split into field independent and field dependent parts. It has been
shown that vanishing single soft theorems, which is derivable from current algebra, only comes from the field
independent part \cite{periodic}, which indeed is expected to be universal. The field dependent part determines (or determined by) the structure of the leading four-point operator, and thus cannot be universal.\footnote{We are considering the single scalar part of the EFT, for example the dilaton part of the effective action for broken CFTs.}    
 Therefore, in the above sense, type B double soft theorems contains additional information,
while type A theorems are equivalent to single soft theorems. For theories with non-vanishing single soft theorems, however, some
technicality prevents double soft theorems from fully incorporating
information from single soft theorems. Therefore, double soft theorems
contain less information in such cases. 

The procedure we propose can be readily extended to multiple soft theorems. Similar to double soft theorems, type A multiple soft theorems can be derived from single soft theorems directly, hence symmetry protected. The derivation of type B theorems is more cumbersome, since there are more diagrams behaving differently when expansion scheme is changed. However, once up to $(m-1)$-soft theorems are known, $m$-soft theorem can still be derived by adding information from some higher-point vertices. We can thus recursively construct type B theorems. They would be modified by loop corrections, since the explicit vertices are used. 

We test our relation numerically by using both kinds of theorems as
constraints on explicit amplitudes. We build up an ansatz for the $n$-point amplitude with a set of Feynman propagators whose numerators satisfy power counting and the symmetry properties dictated by the propagator, as well as polynomial terms. We analyze the strength of the soft-constraints by considering how much the space of the ansatz is reduced. Note that since we don't assume factorization for the numerators, this allows us to ask to what extent does unitarity (factorization) emerges from locality (the presence of Feynman propagators) and soft constraints. The latter has been proved correct on DBI amplitudes using
vanishing single soft theorems \cite{DBI_locality,DBI_locality2}, but not yet investigated
for other cases. We found that for both DBI and dilaton amplitudes,
double soft and single soft can enforce unitarity from locality, but
the coefficient of factorization channels can only be fixed by type
B double soft theorems. The remaining coefficient can be completely
fixed by both kinds of double soft theorems for DBI and single soft
theorems for both theories, but not by double soft theorems for dilaton.
This indicates that type B theorems indeed contains additional information,
while the information from nonvanishing single soft theorems may not
be fully incorporated into double soft theorems of either type. This
is consistent with our analytical results. 

This work is organized as follows: In Section \ref{sec:single},
we discuss how nonlinear symmetries induced by spontaneously broken
symmetry give single soft theorems and constrain four-point vertex.
Then in Section \ref{sec:double_from_single}, we clarify the two
different schemes of soft expansion, show their relation with single
soft theorems, and discuss the implication on their loop correction.
Unanswered questions in previous works would be addressed here. In
Section \ref{sec:fix_ampitude} we discuss how to use soft theorems
as constraints on amplitudes, and compare the results given by single
and double soft theorems. While fixing the amplitudes, we dicuss the
relation among locality, soft theorem and unitarity. Finally, in Section
\ref{sec:conclusion} we present our conclusions. Some technical issues
of calculation are dicussed in the appendices.

\section{Nonlinear Symmetry, Single Soft Theorems, and Four-point vertex\label{sec:single}}

The soft theorems we will discuss arise from spontaneously broken
symmetry, which induce nonlinear symmetry transformation on Goldstone
bosons. Here, without referring to the underlying spontaneously broken
symmetry, we discuss how nonlinear symmetry leads to single soft theorems
and constrains the four-point vertex, since these two ingredients
will be used to derive double soft theorems. We will see that part
of the information from symmetries is not incorporated in single soft
theorems, but shows up as constraints on four-point vertices. 

\subsection{Single Soft Theorems from Nonlinear Symmetry}

The nonlinear symmetry transformation induced by spontaneously broken
symmetries often takes the form 
\begin{equation}
\phi\rightarrow\phi+\theta^{(0)}+\theta_{\mu}^{(1)}\left[x^{\mu}+\Delta_{(1)}^{\mu}\left(\phi,x\right)\right]+\cdots+\theta_{\mu_{1}\mu_{2}\cdots\mu_{m}}^{(m)}\left[x^{\mu_{1}}x^{\mu_{2}}\cdots x^{\mu_{m}}+\Delta_{(m)}^{\mu_{1}\mu_{2}\cdots\mu_{m}}\left(\phi,x\right)\right]
\end{equation}
where $\Delta_{(i)}^{\mu_{1}\mu_{2}\cdots\mu_{i}}\left(\phi,x\right)$
is linear combination of local operators comprised of $\phi$. It
has been shown that the constant shift symmetry $\delta\phi=\theta^{(0)}$
gives rise to Adler's zero \cite{nonlinear}, and the other part of
the symmetry leads to higher order vanishing soft theorems, irrespective
of the exact form of the various $\Delta_{(i)}^{\mu_{1}\mu_{2}\cdots\mu_{i}}$
\cite{periodic}. Here we review the entire statement and provide
an alternative derivation for the higher order soft theorems. 

The above symmetry can be rewritten into a combination of independent
symmetries, 
\begin{equation}
\delta\phi=\delta_{0}\phi+\delta_{1}\phi+\cdots+\delta_{m}\phi
\end{equation}
where 
\begin{equation}
\delta_{0}\phi=\theta^{(0)},\hspace{0.3in}\delta_{i}\phi=\theta_{\mu_{1}\mu_{2}\cdots\mu_{i}}^{(n)}\left[x^{\mu_{1}}x^{\mu_{2}}\cdots x^{\mu_{i}}+\Delta_{(i)}^{\mu_{1}\mu_{2}\cdots\mu_{i}}\left(\phi,x\right)\right] .
\end{equation}
The corresponding Noether currents are therefore
\begin{align}
j_{0}^{\mu} & =\frac{\partial L}{\partial(\partial_{\mu}\phi)}, & j_{i}^{\mu\mu_{1}\mu_{2}\cdots\mu_{i}}= & j_{0}^{\mu}\left[x^{\mu_{1}}x^{\mu_{2}}\cdots x^{\mu_{i}}+\Delta_{(i)}^{\mu_{1}\mu_{2}\cdots\mu_{i}}\left(\phi,x\right)\right],
\end{align}
and current coservation implies 
\begin{equation}
\partial_{\mu}j_{i}^{\mu\mu_{1}\mu_{2}\cdots\mu_{i}}=0
\end{equation}
for $0\leq i\leq m$. For example, for DBI model, 
\begin{align}
 & \begin{aligned}\delta\phi_{\text{DBI}}= & \theta^{(0)}+\theta_{\mu}^{(1)}\left(x^{\mu}-F^{-d}\phi\partial^{\mu}\phi\right)\end{aligned} \nonumber
\\
 & \begin{aligned}j_{0}^{\mu} & =\frac{\partial L}{\partial(\partial_{\mu}\phi)}, & j_{1}^{\mu\nu}= & j_{0}^{\mu}\left[x^{\nu}-F^{-d}\phi\partial^{\nu}\phi\right]\end{aligned}
\label{eq:sym_DBI}
\end{align}
for Galileon theory, 
\begin{align}
 & \begin{aligned}\delta\phi_{\text{Gal}}= & \theta^{(0)}+\theta_{\mu}^{(1)}x^{\mu}\end{aligned} \nonumber
\\
 & \begin{aligned}j_{0}^{\mu} & =\frac{\partial L}{\partial(\partial_{\mu}\phi)}, & j_{1}^{\mu\nu}= & j_{0}^{\mu}x^{\nu}\end{aligned}
 \label{eq:sym_Gal}
\end{align}
and for special Galileon theory, 
\begin{align}
 & \begin{aligned}\delta\phi_{\text{sGal}} & =\theta^{(0)}+\theta_{\mu}^{(1)}x^{\mu}+\theta_{\mu\nu}^{(2)}\left(\alpha^{2}x^{\mu}x^{\nu}-\partial^{\mu}\phi\partial^{\nu}\phi\right)\end{aligned} \nonumber
\\
 & \begin{aligned}j_{0}^{\mu} & =\frac{\partial L}{\partial(\partial_{\mu}\phi)}, & j_{1}^{\mu\nu}= & j_{0}^{\mu}\left[x^{\nu}-F^{-d}\phi\partial^{\nu}\phi\right], & j_{2}^{\mu\nu\sigma} & =j_{0}^{\mu}(x^{\nu}x^{\sigma}-G\partial^{\nu}\phi\partial^{\sigma}\phi)\end{aligned}.
 \label{eq:sym_sGal}
\end{align}
Following \cite{nonlinear}, we first consider the Ward identity of
$\delta_{0} \phi$ and $j_{0}^{\mu}$, 
\begin{equation}
\partial_{\nu}\langle j_{0}^{\nu}(x)\phi(x_{1})\cdots\phi(x_{n})\rangle=\sum_{r}\delta(x-x_{r})\langle\phi(x_{1})\cdots\delta_{0}\phi(x_{r})\cdots\phi(x_{n})\rangle\label{eq:Ward_0}
\end{equation}
for the case $n=1$ and performing Fourier transform with respect
to $x_{1}$, the Lehmann-Symanzik-Zimmermann (LSZ) reduction formula
then implies 
\begin{equation}
\partial_{\mu} \langle 0 |j_{0}^{\mu} (x) |\phi(p_{1})\rangle \propto ie^{-ip_{1}\cdot x}
\end{equation}
leading to 
\begin{equation}
\left\langle 0\left|j_{0}^{\mu}\left(x\right)\right|\phi\left(p_{1}\right)\right\rangle \propto ip^{\mu}e^{-ip_{1}\cdot x}\label{eq:j_create}
\end{equation}
Therefore, the Noether current for the constant shift symmetry creates
a one-particle state $\phi$. Note that in this derivation we do not
need to know that the current $j_{0}^{\mu}$ is associated with some
spontaneously broken symmetry. For general $n$, performing Fourier
transform on $x$ and LSZ reduction on $x_{1}\sim x_{n}$ on \eqref{eq:Ward_0},
the RHS is zero since $\delta_{0}\phi=\theta^{(0)}$ is not a physical
state, and the LHS can be expressed as
\begin{align}
 & [\text{LSZ}]\int dx\,e^{ipx}\langle j_{1}^{\nu}(x)\phi(x_{1})\cdots\phi(x_{n})\rangle \nonumber \\
= & [\text{LSZ}]\int dx\,e^{ipx}\langle0|j_{1}^{\nu}|\phi(p)\rangle\frac{1}{p^{2}}\langle\phi(p)|T^{*}\{\phi(x_{1})\cdots\phi(x_{n})\}|0\rangle+R^{\nu}\nonumber \\
= & \frac{p^{\nu}}{p^{2}}M_{n+1}(p;k_{1},\cdots,k_{n})+R^{\nu}.
\label{eq:M+R}
\end{align}
Equating with RHS of \eqref{eq:Ward_0} gives 
\begin{equation}
M_{n+1}=-p\cdot R\label{eq:MtoR}
\end{equation}
where we used 
\begin{equation}
\int dx\,e^{ipx}\langle0|j_{0}^{\nu}|\phi(p)\rangle=p^{\nu}
\end{equation}
derived from \eqref{eq:j_create}, and 
\begin{equation}
[\text{LSZ}]=\prod_{i}\int dx_{i}e^{ip_{i}x_{i}}p_{i}^{2}\,.
\end{equation}
Since $R^{\nu}$ should be finite at $p=0$, we have 
\begin{equation}
M_{n+1}=-p\cdot R=\mathcal{O}(p)
\end{equation}
giving the famous Adler's zero. The above derivation has been done
in \cite{nonlinear}. 

Higher order double soft theorems come from conservation of the currents
$j_{i}^{\mu\mu_{1}\mu_{2}\cdots\mu_{i}}$, which impose further constraints
on $j_{0}^{\mu}$. First, if the theory further satisfies $\delta_{1}\phi=0$,
current conservation of $j_{1}^{\mu\nu}$ gives 
\begin{equation}
\partial_{\mu}j_{1}^{\mu\nu}=j_{0}^{\nu}-\partial_{\mu}\left[\Delta_{(1)}^{\nu}\left(\phi,x\right)j_{0}^{\mu}\right]=0,
\end{equation}
or 
\begin{equation}
j_{0}^{\nu}=\partial_{\mu}\left[\Delta_{(1)}^{\nu}\left(\phi,x\right)j_{0}^{\mu}\right].
\end{equation}
Plugging this into the first line of \eqref{eq:M+R} gives 
\begin{align}
 & [\text{LSZ}]\int dx\,e^{ipx}\langle j_{1}^{\nu}(x)\phi(x_{1})\cdots\phi(x_{n})\rangle\nonumber \\
= & [\text{LSZ}]\int dx\,e^{ipx}\langle\partial_{\mu}\left[\Delta_{(1)}^{\nu}\left(\phi,x\right)j_{0}^{\mu}(x)\right]\phi(x_{1})\cdots\phi(x_{n})\rangle\nonumber \\
= & [\text{LSZ}]\int dx\,e^{ipx}\bigg[\partial_{\mu}\langle\Delta_{(1)}^{\nu}\left(\phi,x\right)j_{0}^{\mu}(x)\phi(x_{1})\cdots\phi(x_{n})\rangle\nonumber \\
 & -\sum_{i}\langle\phi(x_{1})\cdots[\Delta_{(1)}^{\nu}\left(\phi,x\right)j_{0}^{0}(x),\phi(x_{i})]\cdots\phi(x_{n})\rangle\delta(t-t_{i})\bigg]\nonumber \\
= & [\text{LSZ}]\int dx\,e^{ipx}\bigg[\partial_{\mu}\langle\Delta_{(1)}^{\nu}\left(\phi,x\right)j_{0}^{\mu}(x)\phi(x_{1})\cdots\phi(x_{n})\rangle\nonumber \\
 & -\sum_{i}\langle\phi(x_{1})\cdots[\Delta_{(1)}^{\nu}\left(\phi,x\right),\phi(x_{i})]j_{0}^{0}(x)\cdots\phi(x_{n})\rangle\delta(t-t_{i})\bigg].
\end{align}
As long as $[\Delta_{(1)}^{\nu}\left(\phi,x\right),\phi(x_{i})]j_0^0(x)$
does not generate a physical state, which is the case for DBI, Galileon
and special Galileon theories, the last term would not survive LSZ
reduction, and we have 
\begin{align}
 & [\text{LSZ}]\int dx\,e^{ipx}\langle j_{1}^{\nu}(x)\phi(x_{1})\cdots\phi(x_{n})\rangle\nonumber \\
= & [\text{LSZ}]\int dx\,e^{ipx}\partial_{\mu}\langle\Delta_{(1)}^{\nu}\left(\phi,x\right)j_{0}^{\mu}(x)\phi(x_{1})\cdots\phi(x_{n})\rangle\nonumber \\
= & p_{\mu}N^{\mu\nu}\nonumber \\
= & -\frac{p^{\nu}p_{\mu}}{p^{2}}R^{\mu}+R^{\nu}\label{eq:Ward_1}
\end{align}
where \eqref{eq:M+R} and \eqref{eq:MtoR} are used. Since there is
no 3-point vertex and $\Delta_{(1)}^{\nu}\left(\phi,x\right)j_{0}^{\mu}(x)$
does not create physical states, $N^{\mu\nu}$ cannot have $k\cdot p$
or $p^{2}$ in the denominator, so $N^{\mu\nu}$ can only be a polynomial.
Thus, $R^{\mu}=\mathcal{O}\left(p^{1}\right)$ by \eqref{eq:Ward_1}.
Then we arrive at a conclusion that 
\begin{equation}
M_{n+1}=-p\cdot R=\mathcal{O}(p^{2}),
\end{equation}
giving the vanishing subleading single soft theorem. 

If the theory also satisfies $\delta_{2}\phi=0$, then current conservation
$\partial_{\mu}j_{2}^{\mu\nu\sigma}=0$ provides further constraint
on $j_{0}^{\mu}$, 
\begin{equation}
j_{0}^{\sigma}=\partial_{\nu}\left[\partial_{\mu}(\Delta_{(2)}^{\nu\sigma}j_{0}^{\mu})-j_{0}^{\sigma}x^{\nu}\right].
\end{equation}
Plugging back to \eqref{eq:Ward_0} gives
\begin{align}
 & [\text{LSZ}]\int dx\,e^{ipx}\langle j_{0}^{\sigma}(x)\phi(x_{1})\cdots\phi(x_{n})\rangle\nonumber \\
= & [\text{LSZ}]\int dx\,e^{ipx}\langle\partial_{\nu}\left[\partial_{\mu}(\Delta_{(2)}^{\nu\sigma}j_{0}^{\mu})-j_{0}^{\sigma}x^{\nu}\right]\phi(x_{1})\cdots\phi(x_{n})\rangle\nonumber \\
= & [\text{LSZ}]\int dx\,e^{ipx}\partial_{\nu}\langle\partial_{\mu}(\Delta_{(2)}^{\nu\sigma}j_{0}^{\mu})\phi(x_{1})\cdots\phi(x_{n})\rangle\nonumber \\
 & -[\text{LSZ}]\int dx\,e^{ipx}\sum_{i}\langle\phi(x_{1})\cdots\left[\partial_{\mu}(\Delta_{(2)}^{0\sigma}j_{0}^{\mu})(x),\phi(x_{i})\right]\cdots\phi(x_{n})\rangle\delta(t-t_{i})\nonumber \\
 & +p\cdot\partial_{p}\,[\text{LSZ}]\int dx\,e^{ipx}\langle j_{0}^{\sigma}(x)\phi(x_{1})\cdots\phi(x_{n})\rangle.
\end{align}
As in the previous case, as long as $\left[\partial_{\mu}(\Delta_{(2)}^{0\sigma}j_{1}^{\mu})(x),\phi(x_{i})\right]$
does not create physical states, which is the case for special Galileon
theory, for example, we have 
\begin{align}
 & \left(1-p\cdot\partial_{p}\right)[\text{LSZ}]\int dx\,e^{ipx}\langle j_{0}^{\sigma}(x)\phi(x_{1})\cdots\phi(x_{n})\rangle\nonumber \\
= & [\text{LSZ}]\int dx\,e^{ipx}\partial_{\nu}\langle\partial_{\mu}(\Delta_{(2)}^{\nu\sigma}j_{0}^{\mu})\phi(x_{1})\cdots\phi(x_{n})\rangle\nonumber \\
= & [\text{LSZ}]\int dx\,e^{ipx}\partial_{\nu}\partial_{\mu}\langle(\Delta_{(2)}^{\nu\sigma}j_{0}^{\mu})\phi(x_{1})\cdots\phi(x_{n})\rangle\nonumber \\
 & -[\text{LSZ}]\int dx\,e^{ipx}\partial_{\nu}\left[\sum_{i}\langle\phi(x_{1})\cdots\left[(\Delta_{(2)}^{\nu\sigma}j_{0}^{0})(x),\phi(x_{i})\right]\cdots\phi(x_{n})\rangle\delta(t-t_{i})\right]\nonumber \\
= & [\text{LSZ}]\int dx\,e^{ipx}\partial_{\nu}\partial_{\mu}\langle(\Delta_{(2)}^{\nu\sigma}j_{0}^{\mu})\phi(x_{1})\cdots\phi(x_{n})\rangle,
\end{align}
or, from \eqref{eq:M+R} and \eqref{eq:MtoR}, 
\begin{equation}
\left(1-p\cdot\partial_{p}\right)\left(-\frac{p^{\sigma}p_{\mu}}{p^{2}}R^{\mu}+R^{\sigma}\right)=p_{\nu}p_{\mu}N^{\mu\nu\sigma},
\end{equation}
with $N^{\mu\nu\sigma}$ a polynomial by the same argument. This pushes
vanishing single soft theorem to the next order, 
\begin{equation}
M_{n+1}=-p\cdot R=\mathcal{O}(p^{3}).
\end{equation}
The derivation can be readily extended to arbitrary order of $\delta_{i}\phi=0$,
which would imply $M_{n+1}=\mathcal{O}(p^{i+1})$

From the derivation, we see that single soft theorems can be derived
from the induced nonlinear symmetries without the knowledge of the
underlying spontaneously broken symmetry. Also, the derivation does
not depend on the field-dependent part of the transformation, $\Delta_{(i)}^{\mu_{1}\mu_{2}\cdots\mu_{i}}\left(\phi,x\right)$.
This piece of information is not included in single soft theorems. 

\subsection{Path Integrals}
Actually, there exists another derivation of the vanishing single soft theorems above, using path integrals. We will do them for DBI model as an example. The derivation for other models respecting shift symmetry together with any higher order of nonlinear symmetries should be a trivial generalization of our example.

Starting with the generating functional
\begin{equation}
Z[J]=\int D\phi \, e^{iS+\int dx \, J(x)\phi(x)},
\end{equation}
we can easily get 
\begin{equation}\label{gen}
\langle \phi(x_1) \cdots \phi(x_n) \rangle = {1\over Z[0]}\left\lbrace {\partial \over \partial J(x_1)} \cdots {\partial \over \partial J(x_n)} Z[J] \right\rbrace_{J=0}.
\end{equation}
However, that our theorem has following symmetry
\begin{equation}
\phi \rightarrow \phi ' = \phi + \theta^{(0)} + \theta_\mu^{(1)} \left( x^\mu + \Delta_{(1)}^\mu \right)
\end{equation}
suggests two other forms of $Z[J]$:
\begin{equation}\label{z1}
Z'[J]= \int D\phi \, e^{iS+\int dx \, J(x) \left[ \phi(x) + \theta^{(0)} \right]},
\end{equation}
\begin{equation}\label{z2}
Z''[J]= \int D\phi \left(1+ \theta_\mu^{(1)} {\partial \Delta_{(1)}^\mu \over \partial \phi }\right) e^{iS+\int dx \, J(x) \left[ \phi(x) + \theta_\mu^{(1)} \left( x^\mu + \Delta_{(1)}^\mu \right) \right]} ,
\end{equation}
which give us leading and subleading soft theorem at the end of day.

Performing the same operation on $Z'[J]$ instead of $Z[J]$, identifying them as the same, we have
\begin{equation}
\langle \phi(x_1) \cdots \phi(x_n) \rangle = \langle \phi(x_1) \cdots \phi(x_n) \rangle + \sum_i \langle \phi(x_1) \cdots \phi(x_{i-1}) \theta^{(0)} \phi(x_{i+1}) \cdots \phi(x_n) \rangle.
\end{equation}
That is, 
\begin{equation}
\sum_i \langle \phi(x_1) \cdots \phi(x_{i-1}) \theta^{(0)} \phi(x_{i+1}) \cdots \phi(x_n) \rangle = 0
\end{equation}
Doing LSZ reductions on the legs 2 to n eliminate all except the first term in the summation:
\begin{equation}
\prod_{i=2}^n \left[ \int dx_i \, e^{ik_i x_i} k_i^2 \right] \langle \theta^{(0)} \phi(x_2) \cdots \phi(x_n) \rangle = 0
\end{equation}
Note that we cannot interpret $\theta^{(0)}$ as a constant number here, or it gives us a physically impossible answer
\begin{equation}
M_{n-1}(k_2,\cdots,k_n) = 0,
\end{equation}
for all $k_i$'s. On the other hand, we should identify $\theta^{(0)}$ as a one particle state which has zero momentum because at LSZ region ($t=\pm \infty$), the field is effectively free so that a field with zero momentum is a constant state.\footnote{For free field, $ \phi(x) = \int dp \, a_p \, e^{-ipx} + a_p^\dagger \, e^{ipx}$, where $a_p$ and $a_p^\dagger$ are constants.} Therefore, we have the vanishing leading soft theorem
\begin{equation}
\lim\limits_{k_1 \rightarrow 0} M_n(k_1,\cdots k_n)=0,
\end{equation}
or by power expansion,
\begin{equation}
M_n(k_1, \cdots, k_n)=\mathcal{O}(k_1),
\end{equation}
for $k_1 \rightarrow 0$.
Likewise, doing the same trick for $Z''[J]$ gives us
\begin{equation}
\sum_i \langle \phi(x_1) \cdots \phi(x_{i-1}) \theta_\mu^{(1)} \left( x_i^\mu + \phi {\partial \Delta_{(1)}^\mu \over \partial \phi} (x_i) + \Delta_{(1)}^\mu (x_i) \right) \, \phi(x_{i+1}) \cdots \phi(x_n) \rangle = 0.
\end{equation}
Given that nonlinear transformation factor $\Delta_{(1)}^\mu$ consists of more than one field operators, after LSZ reduction on legs 2 to $n$, we are left with
\begin{equation}
\prod_{i=2}^n \left[ \int dx_i \, e^{ik_i x_i} k_i^2 \right] \langle \theta_\mu^{(1)} \left( x_1^\mu + \phi {\partial \Delta_{(1)}^\mu \over \partial \phi} (x_1) + \Delta_{(1)}^\mu (x_1) \right) \, \phi(x_2) \cdots \phi(x_n) \rangle = 0
\end{equation}
For DBI model, $\Delta_{(1)}^\mu = - F^{-d} \phi \partial^\mu \phi$ gives $k_1$ after Fourier transformation,\footnote{For theorems respecting traditional shift symmetry $\phi \rightarrow \phi + \theta^{(0)}$, its nonlinear transformation factor $\Delta$ can have no more than one field operator without $\partial$, so there is at least one $\partial$ in $\Delta$.} so we have
\begin{equation}
\lim\limits_{k_1\rightarrow 0} \xi^\mu {\partial \over \partial k_1^\mu} M_n(k_1, \cdots k_n) = 0,
\end{equation}
with $\xi^\mu$ an arbitrary constant. In expansion, it means
\begin{equation}
M_n(k_1,\cdots,k_n) = \mathcal{O}(k_1^2),
\end{equation}
for $k_1 \rightarrow 0$. Like the previous subsection, there is no need for the nonlinear shift symmetry to be a spontaneously broken one. 

\subsection{Nonlinear Symmetry Constrains Four-Point Vertex}

Although the field dependent part of nonlinear transformation does
not affect single soft theorem, it does constrain the form of the
four-point vertex. To illustrate this, consider DBI and Galileon theory,
which share the same single soft theorem, $\left.M_{n+1}\left(p,\cdots\right)\right|_{p\to0}=\mathcal{O}\left(p^{2}\right)$.
Suppose we only consider single soft theorems, the $s^{2}$ vertex
\begin{equation}
M_{4}^{(2)}=s^{2}+t^{2}+u^{2}
\end{equation}
seems permissible in both theories. However, if we examine the respective
Lagrangians (e.g. \cite{periodic}), the four-point interaction terms
give different fundamental vertices, 
\begin{align}
M_{4,\text{DBI}} & \propto M_{4}^{(2)}=s^{2}+t^{2}+u^{2} \nonumber \\
M_{4,\text{Gal}} & \propto M_{4}^{(3)}=s^{3}+t^{3}+u^{3}\,.
\end{align}
Indeed, the Lagrangians follow the symmetries \eqref{eq:sym_DBI}
and \eqref{eq:sym_Gal} respectively, with different $\Delta^{\mu_{1}\mu_{2}\cdots\mu_{n}}\left(x\right)$.
In particular, consider the DBI Lagrangian, 
\begin{equation}
\mathcal{L}=\frac{1}{2}\partial\phi\cdot\partial\phi+\frac{1}{4F^{d}}\left(\partial\phi\cdot\partial\phi\right)^{2}+\cdots
\end{equation}
where we present the kinetic term and the four-point interaction term.
Under the transformation \eqref{eq:sym_DBI}, the $x^{\mu}$ transformation
on $\left(\partial\phi\cdot\partial\phi\right)^{2}$ gives a three-point
operator. The kinetic term under the transformation $\Delta^{\mu}\left(x\right)=F^{-d}\phi\partial^{\mu}\phi$
will also generate an identical three-point operator. The two terms
cancel, keeping the Lagrangian invariant. Were it not for the term
$\Delta^{\mu}\left(x\right)=F^{-d}\phi\partial^{\mu}\phi$, this three-point
operator cannot be canceled. Therefore, the field dependent term
is crucial to allow the four-derivative interaction term at four-point,
giving the vertex $s^{2}+t^{2}+u^{2}$. For Galileon theory, $\Delta^{\mu}\left(x\right)$
is zero, so $s^{2}$ vertex is not permissible. This example illustrates
that the four-point vertex indeed contains information of the field
dependent part of the transformation. 

\section{Double Soft Theorems from Single Soft Theorems\label{sec:double_from_single}}

In this section, we show how double soft theorems are completely determined
by single soft theorems and fundamental four point vertex. First,
we introduce the two different schemes of double soft expansion present in literature (e.g. \cite{gluon_AB, grav_AB, current_is_A, SCET}), calling
them type A and type B. 
We then illustrate how type A double soft theorems
can directly implied by single soft theorems, and how type B can be
readily obtained from type A using information of the four point vertex.
Though these two facts have been known separately in literature (e.g. \cite{gluon_AB, grav_AB, current_is_A, SCET}), we connect them here into a general derivation procedure, from single soft theorem all the way to type B double soft theorem, and specify the necessary computation steps explicitly. 
It is very general, independent of theories, although those with vanishing single soft limit possess
some additional properties.
Apart from showing that type A theorems contain no further information than single soft theorems,
our procedure can also determine whether
the double soft theorem withstands loop correction. We explicitly
compute the double soft operators of dilaton, DBI, and special Galileon
amplitudes to illustrate our results.

\subsection{Different Expansion Schemes: Type A and Type B}

A general double soft theorem can be expressed as an expansion over
soft momenta, 
\begin{equation}
\left.M_{n+2}\left(\tau p,\tau q,\cdots\right)\right|_{\tau=0}=\left(S^{(0)}+\tau S^{(1)}+\cdots+\tau^{\lambda_{d}}S^{(\lambda_{d})}\right)M_{n}\left(\cdots\right)+\mathcal{O}\left(\tau^{\lambda_{d}+1}\right)\,,
\end{equation}
indicating that the amplitude $M_{n+2}$ reduces to lower point amplitude
$M_{n}$ up to order $\lambda_{d}$. Though this seems unambiguous,
there are two schemes to expand an amplitude $M_{n+2}\left(p,q,\cdots\right)$
in terms of soft legs $p$ and $q$, as have been recognized in the literature \cite{gluon_AB, grav_AB, SCET}. The first one, denoted by type
A here, is to give the soft legs distinct perturbation parameters,
$\tau_{p}$ and $\tau_{q}$, and perform bivariate Taylor expansion.
The second one, type B, is to assign the same parameter, $\tau$,
to both soft legs, and perform single-variate Taylor expansion. More
explicitly, 
\begin{itemize}
\item Type A: 
\end{itemize}
\begin{align}
M_{n+2} & \left.\left(\tau p,\tau q,\cdots\right)\right|_{\text{type A}}=\left.M_{n+2}\left(\tau_{p}p,\tau_{q}q,\cdots\right)\right|_{\tau_{p}=\tau_{q}=\tau=0}\nonumber \\
= & M_{n+2}\left(0,0,\cdots\right) \nonumber \\
 & +\left[\left(\tau_{p}\frac{\partial}{\partial\tau_{p}}\right)+\left(\tau_{q}\frac{\partial}{\partial\tau_{q}}\right)\right]\left.M_{n+2}\left(\tau_{p}p,\tau_{q}q,\cdots\right)\right|_{\tau_{p}=\tau_{q}=0} \nonumber \\
 & +\left[\frac{1}{2}\left(\tau_{p}^{2}\frac{\partial}{\partial\tau_{p}^{2}}\right)+\left(\tau_{p}\frac{\partial}{\partial\tau_{p}}\right)\left(\tau_{q}\frac{\partial}{\partial\tau_{q}}\right)+\frac{1}{2}\left(\tau_{q}^{2}\frac{\partial}{\partial\tau_{q}^{2}}\right)\right]\left.M_{n+2}\left(\tau_{p}p,\tau_{q}q,\cdots\right)\right|_{\tau_{p}=\tau_{q}=0} \nonumber \\
 & +\cdots \nonumber \\
= & M_{n+2}\left(0,0,\cdots\right) \nonumber \\
 & +\tau\left[\left(p\cdot\frac{\partial}{\partial p}\right)+\left(q\cdot\frac{\partial}{\partial q}\right)\right]M_{n+2}\left(0,0,\cdots\right)\nonumber \\
 & +\tau^{2}\left[\frac{1}{2}\left(p^{2}\cdot\frac{\partial}{\partial p^{2}}\right)+\left(p\cdot\frac{\partial}{\partial p}\right)\left(q\cdot\frac{\partial}{\partial q}\right)+\frac{1}{2}\left(q^{2}\cdot\frac{\partial}{\partial q^{2}}\right)\right]M_{n+2}\left(0,0,\cdots\right)\nonumber \\
 & +\cdots\label{eq:type_A}
\end{align}
\begin{itemize}
\item Type B: 
\end{itemize}
\begin{align}
\left.M_{n+2}\left(\tau p,\tau q,\cdots\right)\right|_{\text{type B}} & =M_{n+2}\left(0,0,\cdots\right)\nonumber \\
 & +\left.\tau\frac{\partial}{\partial\tau}M_{n+2}\left(\tau p,\tau q,\cdots\right)\right|_{\tau=0}\nonumber \\
 & +\left.\frac{1}{2}\tau^{2}\frac{\partial^{2}}{\partial\tau^{2}}M_{n+2}\left(\tau p,\tau q,\cdots\right)\right|_{\tau=0}\nonumber \\
 & +\cdots
\,.\label{eq:type_B}
\end{align}
Therefore, type A and B start to differ from order $\tau^{1}$. For
example, if $S^{(1)}$ exists in both expansion schemes, its form
will likely be different. Sometimes $\tau$ is suppressed by absorbing
it into soft momenta $p$ and $q$.

In the original proof of the double soft theorems of theories arising from spontaneous
broken symmetry, including NLSM, DBI and special Galileon \cite{soft2He}, 
type B scheme was used, where a single parameter $\tau$ was assigned to
both soft legs. 
On the other hand, type A scheme naturally arises from derivation
using current algebra, the approach adopted in \cite{soft2Huang}. 
That it is of type A has been recognized in the literature \cite{current_is_A}, but we illustrate it here using DBI theory, an example more relevant to us. Following the approach in \cite{soft2Huang}
to derive double soft theorem up to order $\tau^{1}$, consider the
following correlation functions 
\begin{align}
&\left\langle J_{P_{\nu}}\left(x\right)J_{P_{\nu}}\left(y\right)\phi\left(x_{1}\right)\cdots\phi\left(x_{n}\right)\right\rangle \nonumber \\
&\left\langle J_{L_{0\mu}}\left(x\right)J_{P_{\nu}}\left(y\right)\phi\left(x_{1}\right)\cdots\phi\left(x_{n}\right)\right\rangle \nonumber \\
&\left\langle J_{P_{\nu}}\left(x\right)J_{L_{0\mu}}\left(y\right)\phi\left(x_{1}\right)\cdots\phi\left(x_{n}\right)\right\rangle \,.
\end{align}
Currents of broken Poincare symmetries are related by 
\begin{equation}
\left[L_{0\mu},P_{\nu}\right]=\eta_{\mu\nu}P_{0}
\end{equation}
so that, roughly, 
\begin{equation}
\partial_{\mu}j_{L_{0\nu}}^{\mu}\sim x^{\nu}\partial_{\mu}j_{P_{0}}^{\mu}\,.
\end{equation}
After LSZ reduction and Fourier transform, the factor $x^{\nu}$ becomes
$\partial/\partial p$ in momentum space, giving subleading factor
in soft momentum, 
\begin{align}
\left\langle J_{P_{\nu}}\left(x\right)J_{P_{\nu}}\left(y\right)\phi\left(x_{1}\right)\cdots\phi\left(x_{n}\right)\right\rangle  & \rightarrow M_{n+2}\left(0,0,\cdots\right) \nonumber \\
\left\langle J_{L_{0\mu}}\left(x\right)J_{P_{\nu}}\left(y\right)\phi\left(x_{1}\right)\cdots\phi\left(x_{n}\right)\right\rangle  & \rightarrow\frac{\partial}{\partial p}\left.M_{n+2}\left(p,0,\cdots\right)\right|_{p=0} \nonumber \\
\left\langle J_{P_{\nu}}\left(x\right)J_{L_{0\mu}}\left(y\right)\phi\left(x_{1}\right)\cdots\phi\left(x_{n}\right)\right\rangle  & \rightarrow\frac{\partial}{\partial q}\left.M_{n+2}\left(0,q,\cdots\right)\right|_{q=0}\,,
\end{align}
where we associate $p$ with $x$ and $q$ with $y$. See \cite{soft2Huang}
for details. From these formulas it is clear that this approach is
of type A expansion scheme, as was the case for dilaton soft theorems
derived there. The motivation for this alternative derivation was to
investigate if double soft theorems can be modified by loop correction.

We have performed numerical tests on explicit amplitudes of DBI and
dilaton, showing that the two schemes are indeed inequivalent. It
is thus important to specify which scheme is used. For example, using
current algebra to reproduce theorems derived from CHY representation
is bound to failure, as was the case for $\tau^{2}$ and $\tau^{3}$
double soft theorems for DBI amplitudes in \cite{soft2Huang}, since
the expansion schemes involved are different. Although it was claimed
that the $\tau^{1}$ theorem is reproduced, the information of four-point
vertex is used in the proof, as straightforward current algebra manipulation
shouldn't have produced type B expansion. We shall see, however, that
the two schemes can be related. That is, for every type A double soft
theorem, a type B theorem of the same soft order can always be derived.

\subsection{Type A Double Soft Theorems}

\begin{figure}
\begin{centering}
\includegraphics[height=4cm]{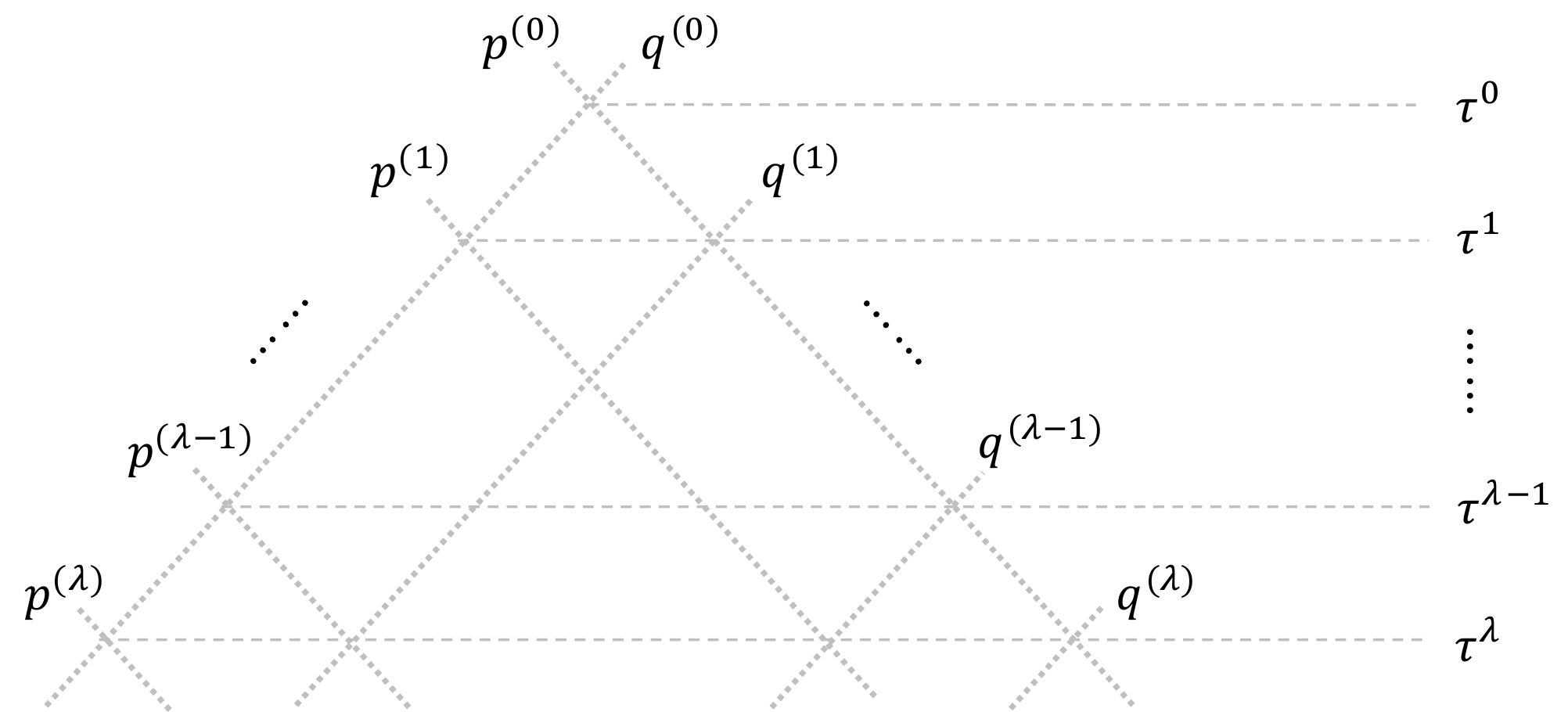}
\par\end{centering}
\caption{The Pascal triangle of single and double soft expansion. $\tau$ denotes
the double soft order\label{fig:pascal}}
\end{figure}

Type A double soft theorems can be derived simply by applying the
single soft theorem twice, a fact recognized for gluon and graviton amplitudes \cite{gluon_AB, grav_AB}. Here we show how this can be carried out for a general theory. 

\begin{figure}
\subfloat[First single soft expansion. \label{fig:typeA_non0_p}]{\begin{centering}
\includegraphics[height=7cm]{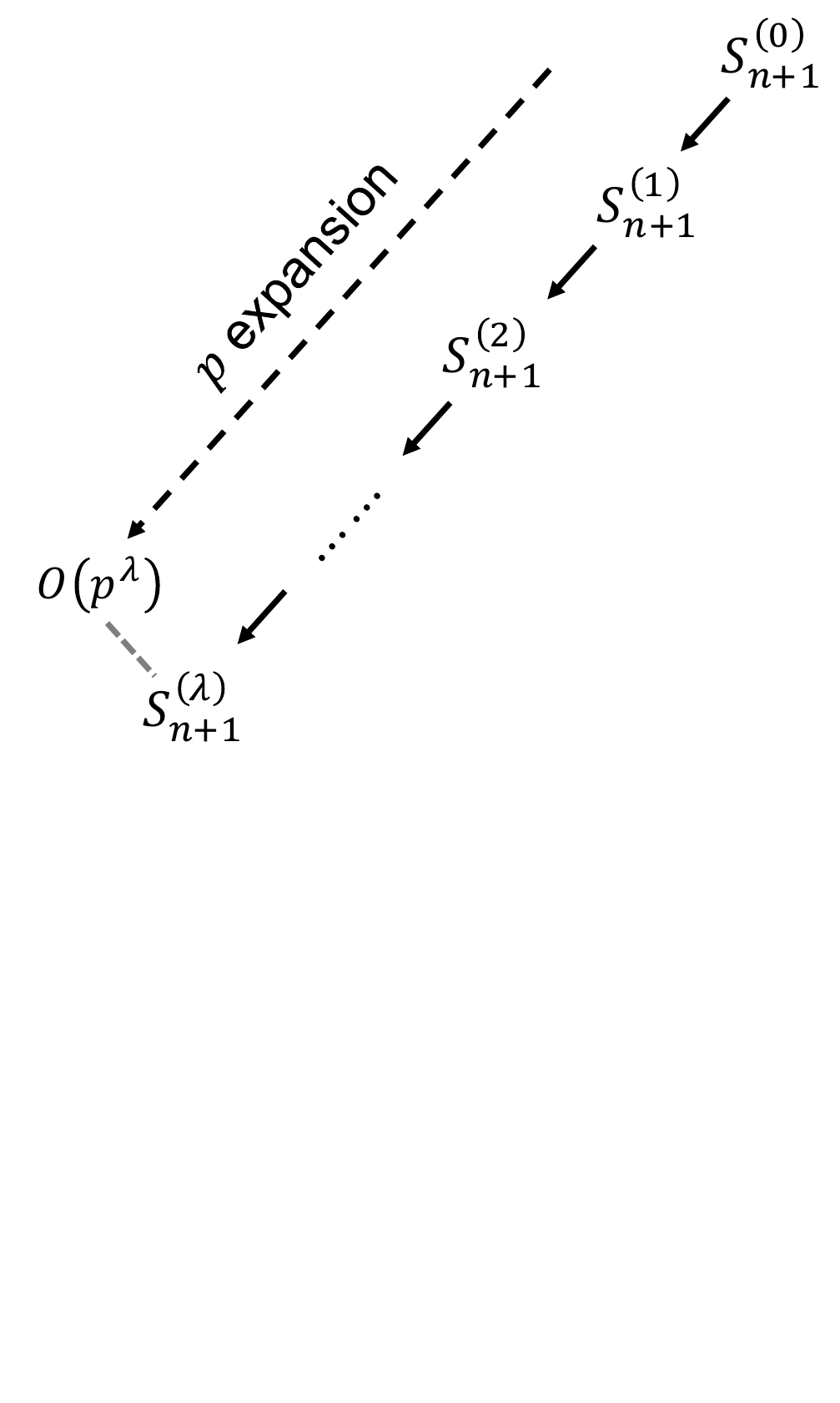}
\par\end{centering}
}~~\subfloat[Second single soft expansion, in $q$, also up to order $\mathcal{O}\left(q^{\lambda}\right)$.\label{fig:typeA_non0_pq}]{\begin{centering}
\includegraphics[height=7cm]{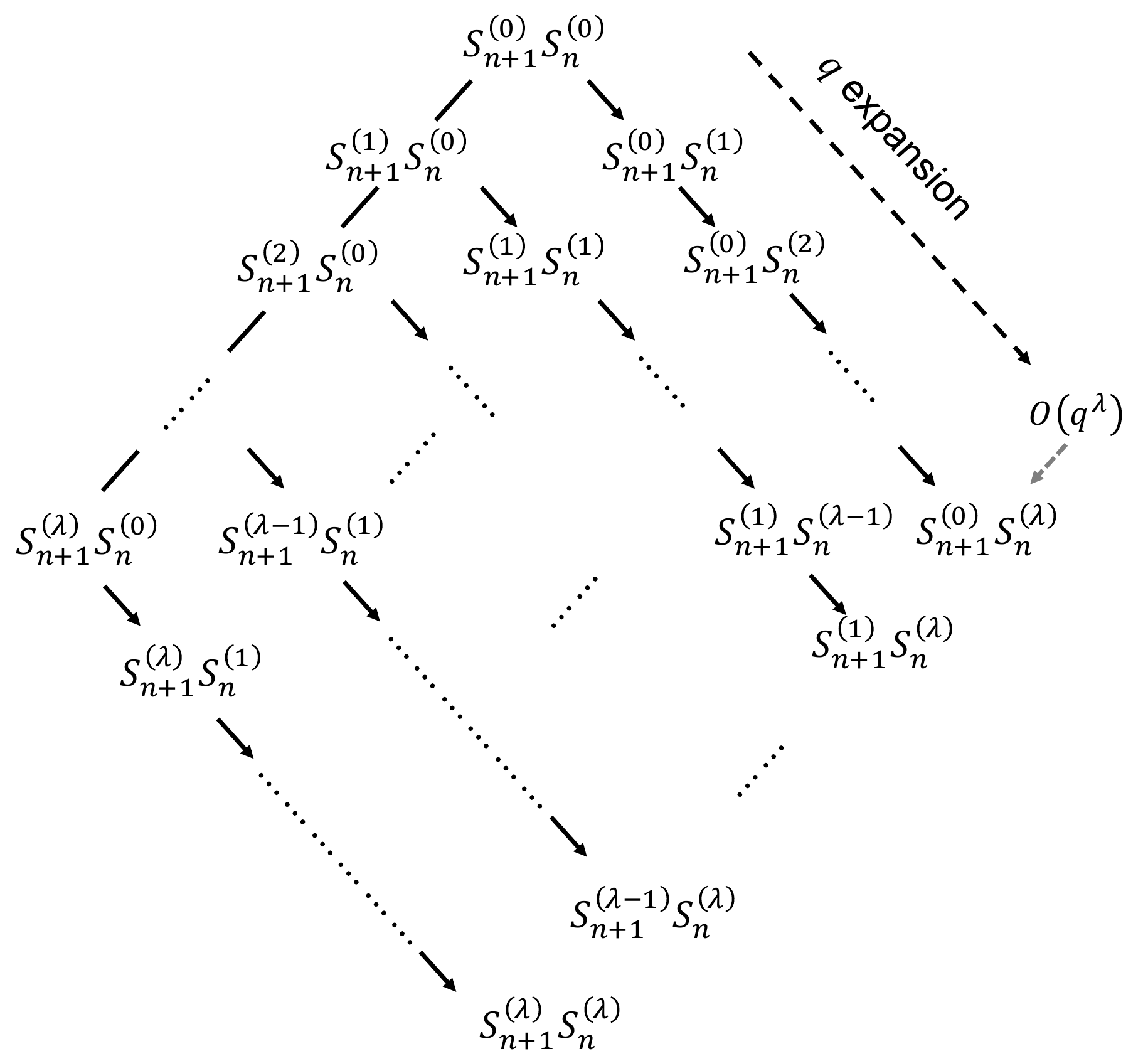}
\par\end{centering}
}
\centering{}\newline\subfloat[Collecting double soft factors. \label{fig:typeA_non0_double}]{\begin{centering}
\includegraphics[height=7cm]{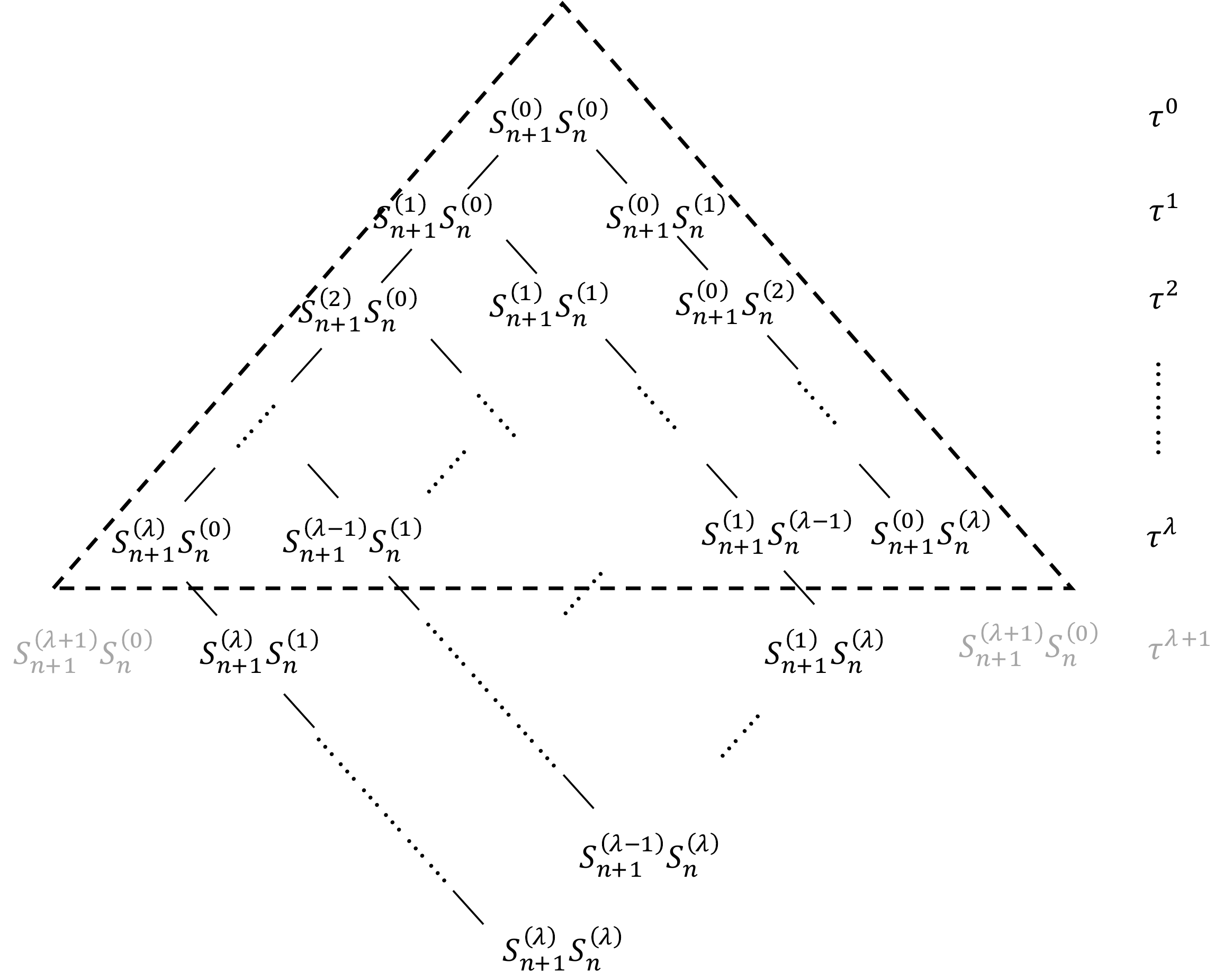}
\par\end{centering}
}\caption{Deriving type A double soft from single soft theorems. The coordinate
follows the Pascal triangle in Fig.\ref{fig:pascal}. (a) First single
soft expansion, in $p$, up to order $\mathcal{O}\left(p^{\lambda}\right)$.
(b) Second single soft expansion, in $q$, also up to order $\mathcal{O}\left(q^{\lambda}\right)$.
(c) For double soft order higher than $\tau^{\lambda+1}$, there are
terms that cannot be derived from single soft theorems (e.g. the ones
in gray). Therefore, only the terms in the dashed triangle can be
incorporated into double soft theorems. The remaining terms would
be discarded.\label{fig:typeA_non0}}
\end{figure}

The procedure resembles the Pascal triangle
produced by binomial expansion, Fig.\ref{fig:pascal}. Suppose the
amplitude of a theory has single soft theorems up to order $\lambda$,
as in Fig.\ref{fig:typeA_non0_p}, 
\begin{equation}
M_{n+1}\left(p,\cdots\right)=\left(S^{(0)}+p\cdot S^{(1)}+p_{\mu}p_{\nu}S^{(2)\mu\nu}+\cdots+p^{\lambda}\cdot S^{(\lambda)}\right)M_{n}\left(\cdots\right)+O\left(p^{\lambda+1}\right)
\end{equation}
where $p^{i}\cdot S^{(i)}$ is a short hand of $p_{\mu_{1}}\cdots p_{\mu_{i}}\left(S^{(i)}\right)^{\mu_{1}\cdots\mu_{i}}$,
and we absorb $\tau$ into the soft momentum $p$. The operators $S^{(k)}$
generally do not vanish. Performing an additional single soft expansion
as in Fig.\ref{fig:typeA_non0_pq} gives the double soft as 
\begin{align}
M_{n+2} & \left.\left(p,q,\cdots\right)\right|_{\text{type A}}=\left(S_{n+1}^{(0)}+p\cdot S_{n+1}^{(1)}+\cdots+p^{\lambda}\cdot S_{n+1}^{(\lambda)}\right)M_{n+1}\left(q,\cdots\right)+O\left(p^{\lambda+1}\right)\nonumber \\
= & \left(S_{n+1}^{(0)}+p\cdot S_{n+1}^{(1)}+\cdots p^{\lambda}\cdot S_{n+1}^{(\lambda)}\right)\left(S_{n}^{(0)}+q\cdot S_{n}^{(1)}+\cdots q^{\lambda}\cdot S_{n}^{(\lambda)}\right)M_{n}\left(\cdots\right)\nonumber \\
 & +\left(S_{n+1}^{(0)}+p\cdot S_{n+1}^{(1)}+\cdots p^{\lambda}\cdot S^{(\lambda)}\right)O\left(q^{\lambda+1}\right)+O\left(p^{\lambda+1}\right)\nonumber \\
= & \left[S_{n+1}^{(0)}S_{n}^{(0)}+q^{\mu}S_{n+1}^{(0)}S_{n,\mu}^{(1)}+p^{\mu}S_{n+1,\mu}^{(1)}S_{n}^{(0)}+\cdots+\left(p^{\lambda}\cdot S_{n+1}^{(\lambda)}\right)\left(q^{\lambda}\cdot S_{n}^{(\lambda)}\right)\right]M_{n}\left(\cdots\right)\nonumber \\
 & +\cdots\nonumber \\
= & \left[S_{\text{d}}^{(0)}+S_{\text{d}}^{(1)}+\cdots+S_{\text{d}}^{(\lambda_{d})}\right]M_{n}\left(\cdots\right)+\cdots\label{eq:type_A_expand}
\end{align}
where $S_{m}^{(i)}$ is an order $p^{i}$ or $q^{i}$ operator acting
on an $m$ point amplitude, and $S_{\text{d}}^{(m)}$ are double soft
operators. Collecting terms into $S_{\text{d}}^{(m)}$ according to
(\ref{eq:type_A}) gives type A double soft theorems, as illustrated
in Fig.\ref{fig:typeA_non0_double}. Naively, (\ref{eq:type_A_expand})
implies that $M_{n+2}$ reduces to $M_{n}$ for orders $\left(p^{i},q^{j}\right)$,
$0\leq i,j\leq\lambda$, with soft factor $\left(p^{i}\cdot S_{n+1}^{(i)}\right)\left(q^{j}\cdot S_{n}^{(j)}\right)$.
It is thus tempting to identify $S_{\text{d}}^{(m)}=\sum_{i=0}^{m}\left(p^{i}\cdot S_{n+1}^{(i)}\right)\left(q^{(m-i)}\cdot S_{n}^{(m-i)}\right)$,
which will give double soft theorem up to order $\lambda_{d}=\lambda$.
However, the factor $\left(p^{i}\cdot S_{n+1}^{(i)}\right)\left(q^{j}\cdot S_{n}^{(j)}\right)$
may contain additional pieces with order different from $\left(p^{i},q^{j}\right)$,
\begin{equation}
\left(p^{i}\cdot S_{n+1}^{(i)}\right)\left(q^{j}\cdot S_{n}^{(j)}\right)\neq\mathcal{O}\left(p^{i}q^{j}\right)
\end{equation}
since $S_{n+1}^{(i)}$, which operates on $M_{n+1}\left(q,\cdots\right)$,
may contain factors or derivatives of $q$. This can modify the soft
order (in $q$) of pieces following it. For example, it is possible
that $\left(p^{i}\cdot S_{n+1}^{(i)}\right)\left(q^{\lambda+1}\cdot\frac{\partial}{\partial q^{\lambda+1}}M_{n+1}\right)$
not only contributes to $\left(p^{i},q^{\lambda+1-i}\right)$ but
also to $\left(p^{i},q^{\lambda-i}\right)$, hence to double soft
order $\lambda$. However, $\left(q^{\lambda+1}\cdot\frac{\partial}{\partial q^{\lambda+1}}M_{n+1}\right)$
cannot be expressed in terms of $M_{n}$ since single soft theorem
in $q$ only exists up to order $q^{\lambda}$, so double soft theorem
at order $\lambda$ cannot be obtained. Therefore, double soft theorems
may only exist up to a smaller order $\lambda_{d}\leq\lambda$. It
would be more accurate to explicitly work out all the soft factors
and group them into appropriate double soft orders, as will be done
for dilaton amplitudes below \footnote{For some cases such as gluon amplitude, the expansion order of $p$ and $q$ matters. One should thus further specify the choice of order, or incorporate both kind ofpossibilities, as in \cite{gluon_AB, current_is_A}. In any case, ambiguities can be resolved by specifying one's choice, and the procedure here can still be used.}.

Since the derivation of type A double soft theorems only requires
single soft theorems, it obviously does not contain further information
than single soft theorems. Moreover, since we cutoff at double soft
order $\lambda_{d}\leq\lambda$ for technical reasons shown above,
the terms $\left(p^{i}\cdot S_{n+1}^{(i)}\right)\left(q^{j}\cdot S_{n}^{(j)}\right)$
given by single soft theorems with $i,j\leq\lambda$ but $i+j>\lambda$
are discarded, as illustrated in Fig.\ref{fig:typeA_non0_double}.
Therefore, type A double soft theorem may actually contain less information.
We illustrate these points by explicit examples.

\subsubsection{Type A double soft from non-vanishing single soft \label{subsec:type_A_cDBI}}

An example with non-vanishing single soft theorems is dilaton amplitudes.
Its single soft theorems exist up to $p^{\lambda}$ with $\lambda=1$
\cite{cDBI_single}, 
\begin{align}
M_{n+1}\left(p,\cdots\right) & =\left(S^{(0)}+p\cdot S^{(1)}\right)M_{n}\left(\cdots\right)+O\left(p^{2}\right)\nonumber \\
S^{(0)} & =-\sum_{i=1}^{n}\left(k_{i}\cdot\frac{\partial}{\partial k_{i}}+\frac{D-2}{2}\right)+D\nonumber \\
S_{\mu}^{(1)} & =-\sum_{i=1}^{n}\left[k_{i}^{\nu}\frac{\partial^{2}}{\partial k_{i}^{\nu}\partial k_{i}^{\mu}}-\frac{k_{i\mu}}{2}\frac{\partial^{2}}{\partial k_{i\nu}\partial k_{i}^{\nu}}+\frac{D-2}{2}\frac{\partial}{\partial k_{i}^{\mu}}\right]\label{eq:cDBI_single}\,.
\end{align}
Naively, this directly gives us expansion terms of $\left(p^{0},q^{0}\right)$,
$\left(p^{0},q^{1}\right)$, $\left(q^{0},p^{1}\right)$, and $\left(p^{1},q^{1}\right)$,
with soft operators $\left(p^{i}\cdot S_{n+1}^{(i)}\right)\left(q^{j}\cdot S_{n}^{(j)}\right)$
of corresponding order. However, the operator $S_{n+1}^{(1)}$ contains
derivatives with respect to $q$, 
\begin{align*}
S_{n+1,\mu}^{(1)}= & -\sum_{i=1}^{n}\left[k_{i}^{\nu}\frac{\partial^{2}}{\partial k_{i}^{\nu}\partial k_{i}^{\mu}}-\frac{k_{i\mu}}{2}\frac{\partial^{2}}{\partial k_{i\nu}\partial k_{i}^{\nu}}+\frac{D-2}{2}\frac{\partial}{\partial k_{i}^{\mu}}\right]-\left[q^{\nu}\frac{\partial^{2}}{\partial q^{\nu}\partial q^{\mu}}-\frac{q_{\mu}}{2}\frac{\partial^{2}}{\partial q_{\nu}\partial q^{\nu}}+\frac{D-2}{2}\frac{\partial}{\partial q^{\mu}}\right].
\end{align*}
Therefore, a term following it will have its order of $q$ reduced
by 1. For example, 
\begin{equation}
S_{n+1,\mu}^{(1)}q^{\nu}=-\frac{D-2}{2}\delta_{\mu}^{\nu}
\end{equation}
and its operation on higher order of $q$ would be more complicated.
Thus the term $\left(p\cdot S_{n+1}^{(1)}\right)\left(q\cdot S_{n}^{(1)}\right)$
actually contains a term of order $\left(p^{1},q^{0}\right)$, in
addition to order $\left(p^{1},q^{1}\right)$. This problem does not
exist for $S_{n+1}^{(0)}$ since its $q$ dependence is $q\cdot\frac{\partial}{\partial q}$,
which does not change the order of $q$. After taking care of this,
and noting that $\frac{\partial}{\partial q}M_{n}=0$ and $S^{(0)}M_{n}$
is simply a number, the leading and subleading double soft theorem
in \cite{soft2Huang} can be reproduced from (\ref{eq:type_A_expand}),
\begin{align}
M_{n+2} & \left.\left(p,q,\cdots\right)\right|_{\text{type A}}=\left[n\frac{\left(D-2\right)}{2}-D+\sum_{i}k_{i}\cdot\partial_{i}\right]\left[\left(n+1\right)\frac{\left(D-2\right)}{2}-D+\sum_{i}k_{i}\cdot\partial_{i}\right]M_{n}\nonumber \\
 & +\left(p+q\right)^{\mu}\sum_{i}\left[\left(\frac{D-2}{2}+k_{i}\cdot\partial_{i}\right)\partial_{i,\mu}-\frac{1}{2}k_{i,\mu}\partial_{i}^{2}\right]\left[\left(n+1\right)\frac{D-2}{2}-D+\sum_{i}k_{i}\cdot\partial_{i}\right]M_{n}\nonumber \\
 & +\text{higher order}\label{eq:dilaton_A}
\end{align}
where it is obvious that the zeroth order corresponds to $S_{n+1}^{(0)}S_{n}^{(0)}$
and the first order to $S_{n+1}^{(0)}S_{n}^{(1)}$. The result exactly
matches that in \cite{soft2Huang}, indicating that the expansion
scheme used there is of type A. These soft operators are dimensionless,
relating the $s^{m}$ terms of $M_{n+2}$ to $s^{m}$ terms of $M_{n}$.

We can also comment on the existence of subsubleading double soft
theorem. Since the single soft theorems for dilaton only exist up
to first order, the terms $\left(p^{0},q^{2}\right)$ and $\left(q^{0},q^{2}\right)$
cannot be obtained. This obstructs a fully second order type A double
soft theorem, i.e. the $\tau^{2}$ terms of (\ref{eq:type_A}), as
have been discussed for general $\lambda$. While it is tempting to
consider a restricted theorem concerning only the term $\left(p^{1},q^{1}\right)$
since $\left(p\cdot S_{n+1}^{(1)}\right)\left(q\cdot S_{n}^{(1)}\right)$
does contain a piece of this order, it is incomplete since $\left(p\cdot S_{n+1}^{(1)}\right)$
can modify the order of $q$, as discussed above. Thus, the term $\left(p\cdot S_{n+1}^{(1)}\right)\left.\frac{1}{2}q^{2}\frac{\partial^{2}}{\partial q^{2}}M_{n+1}\left(q,\cdots\right)\right|_{q=0}$
also contributes, whereas $\left.\frac{1}{2}q^{2}\frac{\partial^{2}}{\partial q^{2}}M_{n+1}\left(q,\cdots\right)\right|_{q=0}$
cannot be expressed in terms of $M_{n}$ since subsubleading single
soft theorem does not exist. Therefore, there is no subsubleading
double soft theorem, even if only the piece of order $\left(p^{1},q^{1}\right)$
is considered. This piece of information is thus discarded, so that
double soft theorems of dilaton contain less information than single
soft theorems. 

\begin{figure}
\begin{centering}
\subfloat[First single soft expansion.\label{fig:typeA_dilaton_p}]{\begin{centering}
\includegraphics[height=2.4cm]{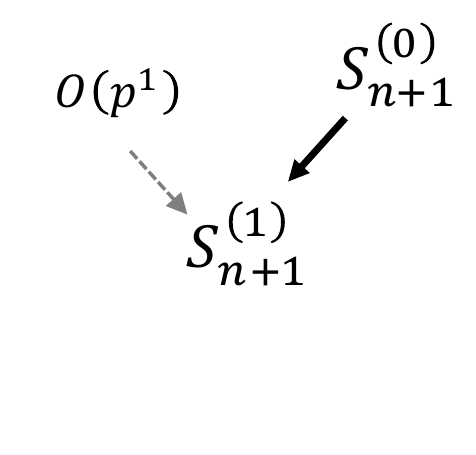}
\par\end{centering}
}\subfloat[Second single soft expansion.\label{fig:typeA_dilaton_pq}]{\begin{centering}
\includegraphics[height=2.4cm]{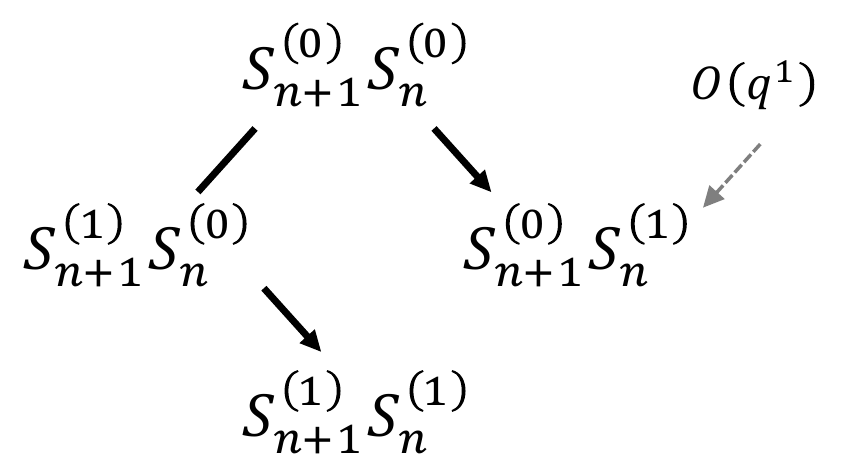}
\par\end{centering}
}\subfloat[Collecting double soft orders.\label{fig:typeA_dilaton_double}]{\begin{centering}
\includegraphics[height=3.2cm]{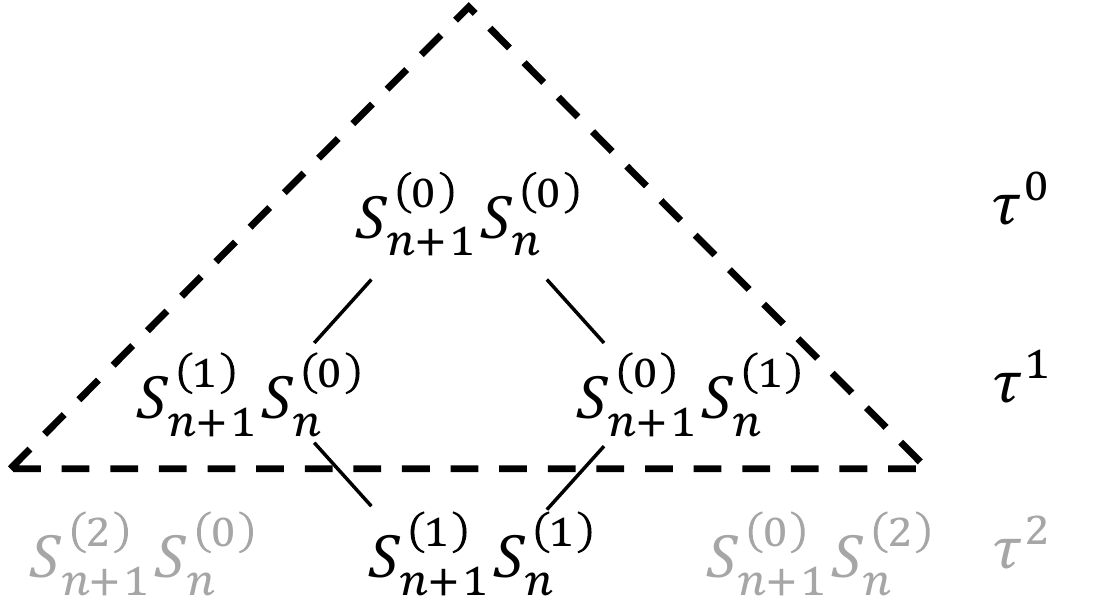}
\par\end{centering}
}
\par\end{centering}
\caption{A special case of Fig.\ref{fig:typeA_non0} for dilaton amplitudes,
with nonvanishing single soft theorem and $\lambda=1$. For double
soft order $\lambda+1=2$, there are terms unattainable from single
soft expansion. Therefore, double soft theorem only exists up to order
$\lambda=1$, and the information of $\left(p\cdot S_{n+1}^{(1)}\right)\left(q\cdot S_{n}^{(1)}\right)$
is discarded. \label{fig:typeA_non0_dilaton}}
\end{figure}

\subsubsection{Type A double soft from vanishing single soft}

\begin{figure}
\subfloat[First single soft expansion.\label{fig:typeA_0_p}]{\begin{centering}
\includegraphics[height=6cm]{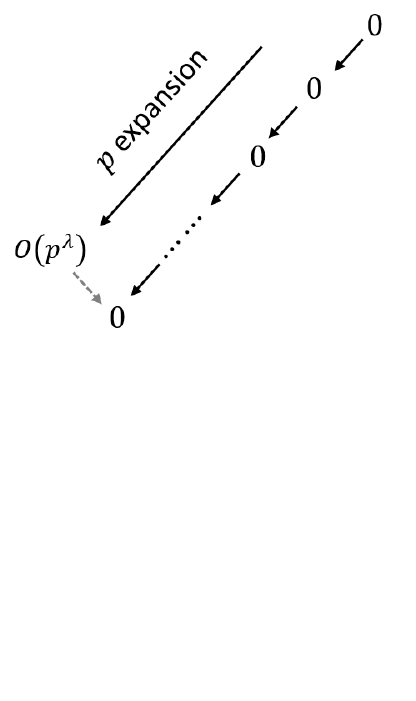}
\par\end{centering}
}\subfloat[Second single soft expansion.\label{fig:typeA_0_pq}]{\begin{centering}
\includegraphics[height=6cm]{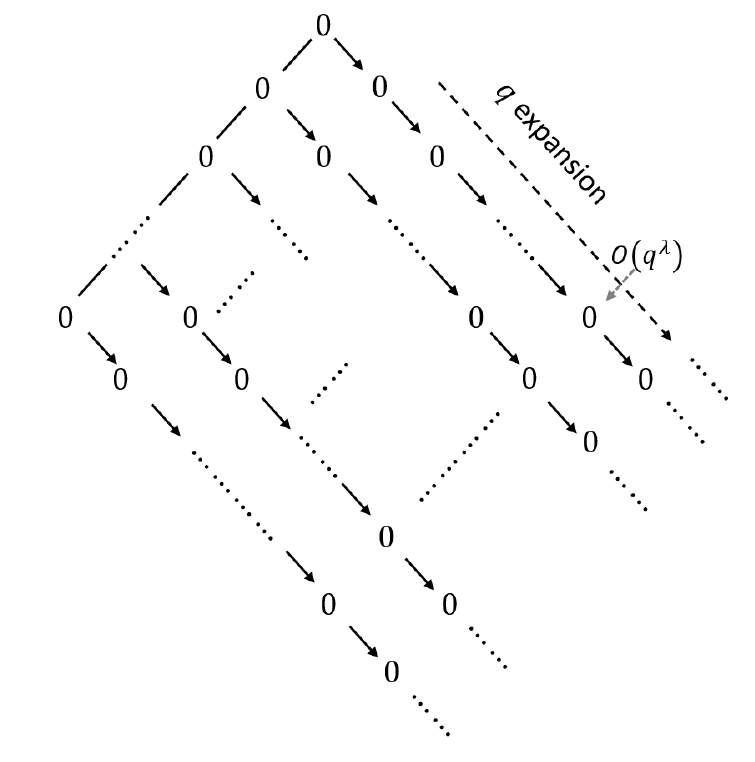}
\par\end{centering}
}\subfloat[Reversed ordering.\label{fig:typeA_0_qp}]{\begin{centering}
\includegraphics[height=6cm]{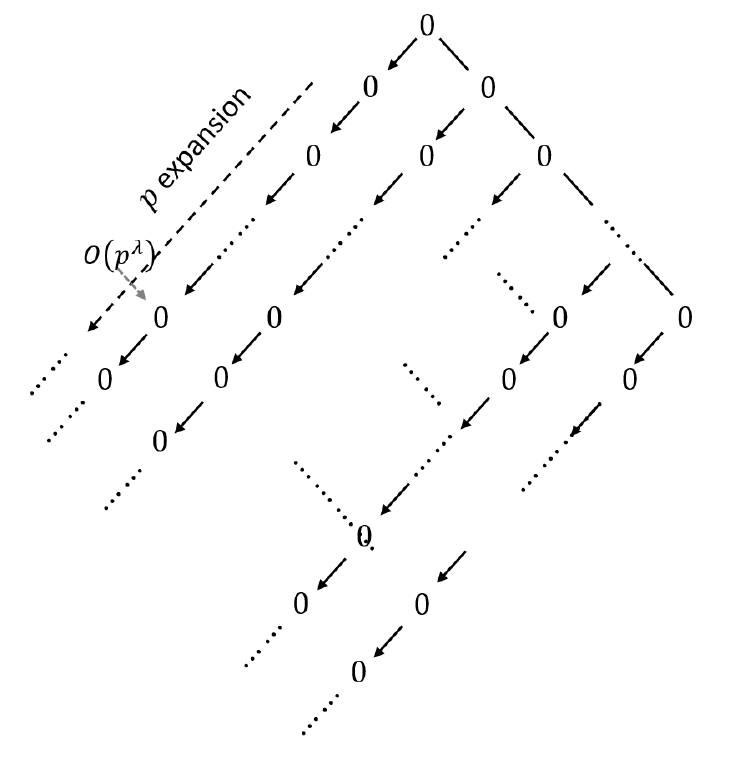}
\par\end{centering}
}

\subfloat[Collecting double soft factors.\label{fig:typeA_0_double} ]{\begin{centering}
\includegraphics[height=6cm]{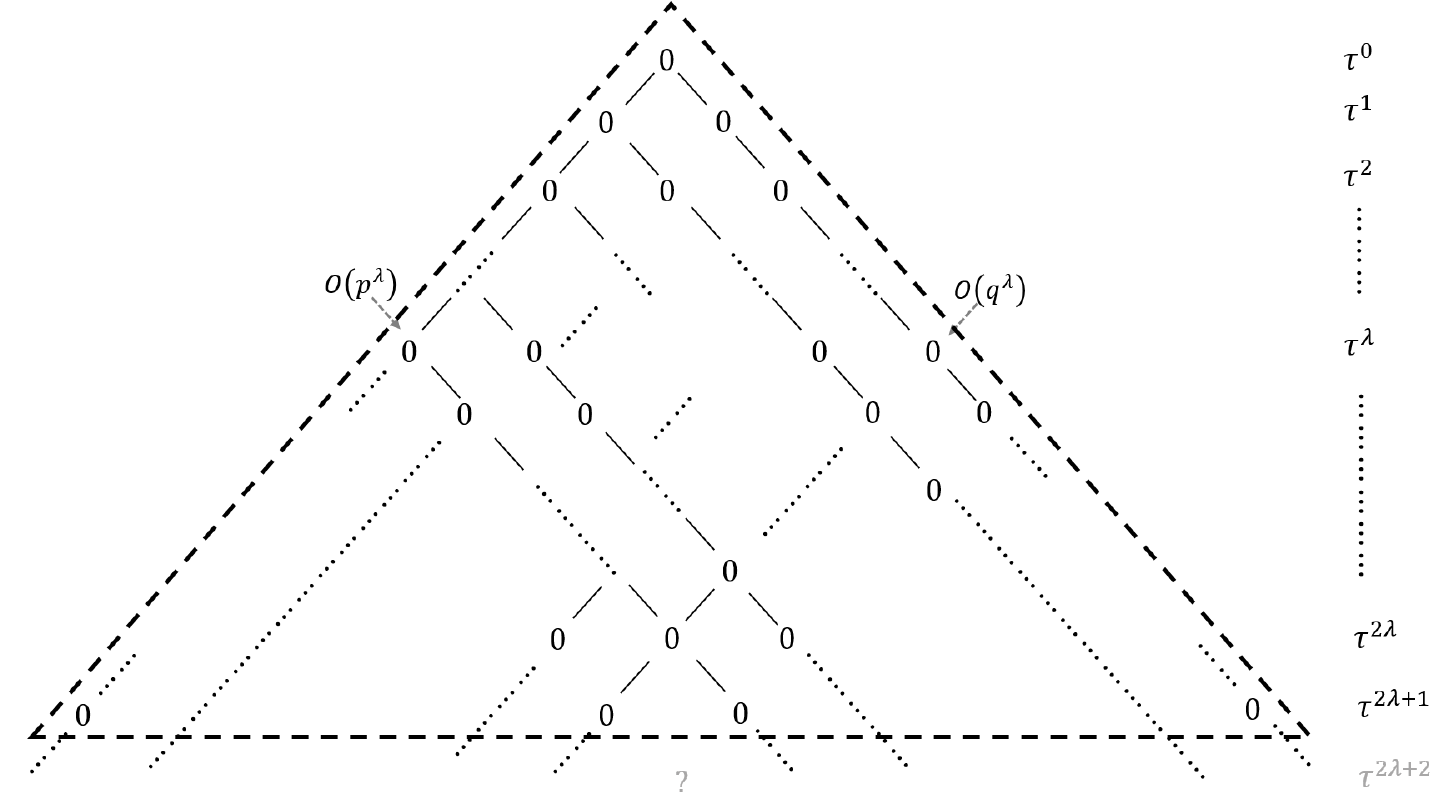}
\par\end{centering}
}
\centering{}\caption{Deriving double soft theorems from vanishing single soft theorems.
(a) First single soft expansion, in $p$, vanishes up to order $\mathcal{O}\left(p^{\lambda}\right)$.
(b) Second single soft expansion, in $q$, does not stop at order
$\mathcal{O}\left(q^{\lambda}\right)$ but continues to arbitrary
order. (c) Expanding first in $q$ and then in $p$ gives similar
result. (d) Vanishing single soft gives more higher order terms, so
double soft theorems can be obtained up to order $\tau^{2\lambda+1}$.
Note the comparison with non-vanishing single soft limit, given in
Fig.\ref{fig:typeA_non0_double}, which stops at order $\tau^{\lambda}$.
\label{fig:typeA_0}}
\end{figure}

For amplitudes with vanishing single soft theorems, 
\begin{equation}
M_{n+1}\left(p,\cdots\right)=\sum_{i=1}^{\lambda}\left(p^{i}\cdot S^{(i)} \right)M_{n}\left(\cdots\right)+\mathcal{O}\left(p^{\lambda+1}\right)\label{eq:0_single}
\end{equation}
with all $S^{(i)}=0$. DBI and special Galileon amplitudes are examples
with $\lambda=$1, 2, respectively \cite{periodic}. Since the single
soft limit vanishes, 
\begin{equation}
M_{n+2}\left(p,q,\cdots\right)=0+\mathcal{O}\left(p^{\lambda+1}\right)
\end{equation}
the subsequent expansion (\ref{eq:type_A_expand}) of the second soft
momentum $q$ is trivial, as Fig.\ref{fig:typeA_0_pq} illustrates.
It is valid up to arbitrary order, not restricted by the order $\lambda$
of single soft theorems. This gives us vanishing double soft limits
of order up to $\left(p^{\lambda},q^{\infty}\right)$. Of course,
the potential problem of shifted order of $q$, present in dilaton,
does not arise here. Similarly, double soft limits vanish up to order
$\left(q^{\lambda},p^{\infty}\right)$, as in Fig.\ref{fig:typeA_0_qp}.
This gives us vanishing type A double soft limit up to order $\lambda_{d}=2\lambda+1$,
since all the terms needed, $\left(p^{i},q^{\lambda_{d}-i}\right)$,
can be evaluated, as in Fig.\ref{fig:typeA_0_double}. That is, 
\begin{equation}
\left.M_{n+2}\left(p,q,\cdots\right)\right|_{\text{type A}}=0+\mathcal{O}\left(p^{i},q^{j};i+j=2\lambda+2\right).
\label{eq:type_A_0}
\end{equation}
For example, $\lambda_{d}=3$, 5 for DBI and special Galileon, respectively.
The order of double soft theorem is higher than cases with non-vanishing
single soft limits, where $\lambda_{d}\leq\lambda$. This explains,
for example, why DBI double soft theorem exists up to $\tau^{3}$
but dilaton only to $\tau^{1}$, while both are associated with two
kinds of spontaneously broken symmetries that are derivatively related.
It is simply because the vanishing single soft limit of DBI facilitates
further expansion. Also, since no term needs to be discarded as in
the case of non-vanishing single soft theorem, double soft theorems
here should contain equivalent information as do  vanishing single
soft theorems.

\begin{figure}
\begin{centering}
\subfloat[First single soft expansion.]{\begin{centering}
\includegraphics[height=4cm]{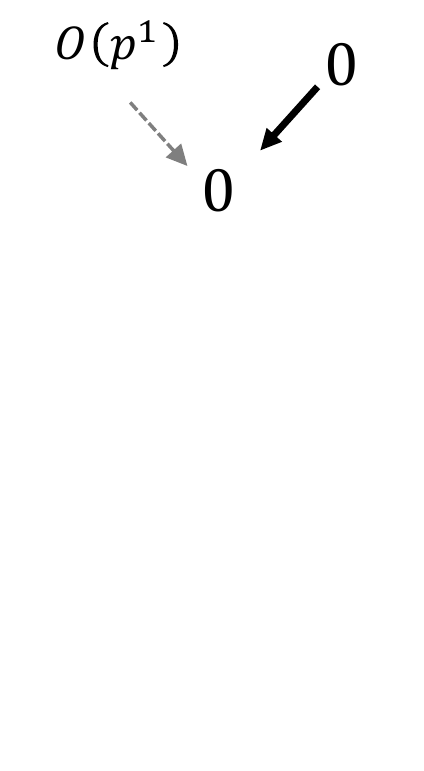}
\par\end{centering}
}\subfloat[Second single soft expansion.]{\begin{centering}
\includegraphics[height=4cm]{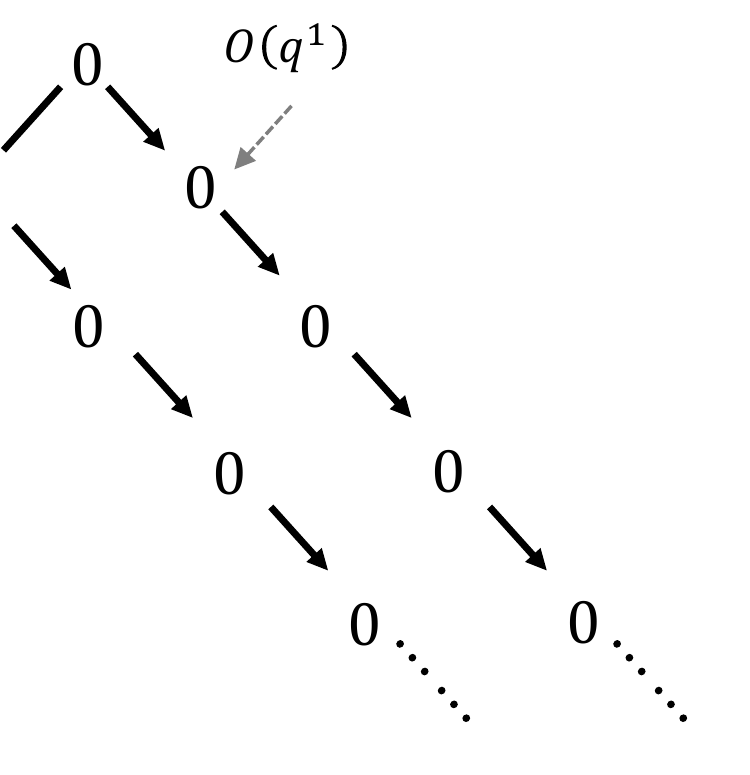}
\par\end{centering}
}\subfloat[Reversed expansion order.]{\begin{centering}
\includegraphics[height=4cm]{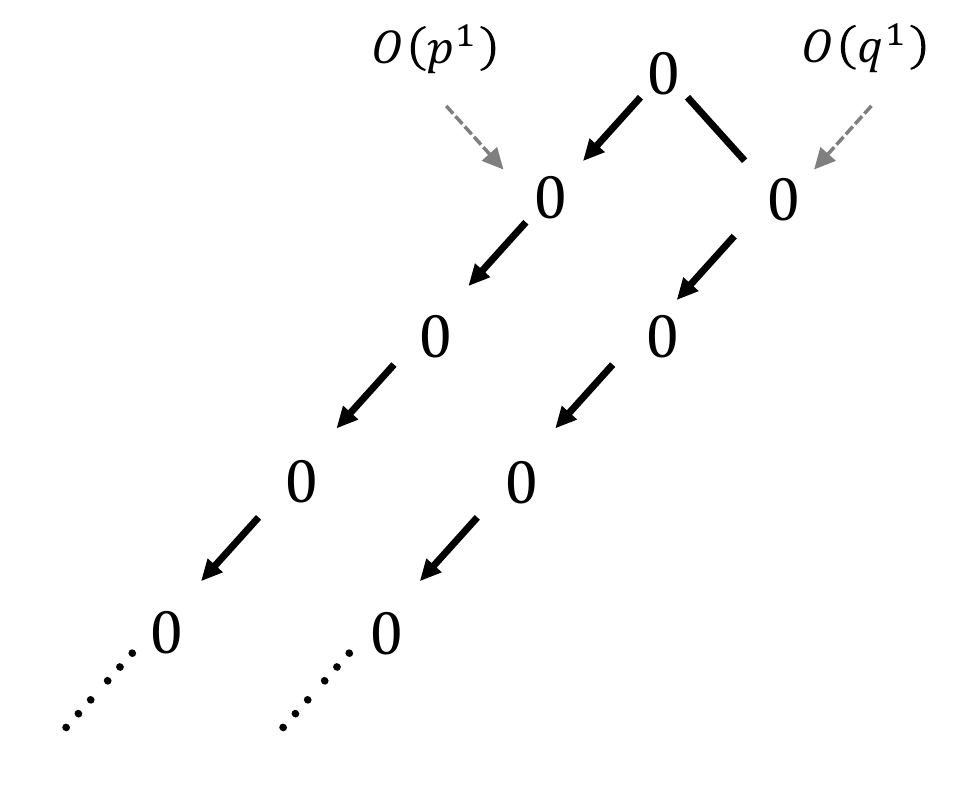}
\par\end{centering}
}
\par\end{centering}
\begin{centering}
\subfloat[Collecting double soft order.]{\begin{centering}
\includegraphics[height=4cm]{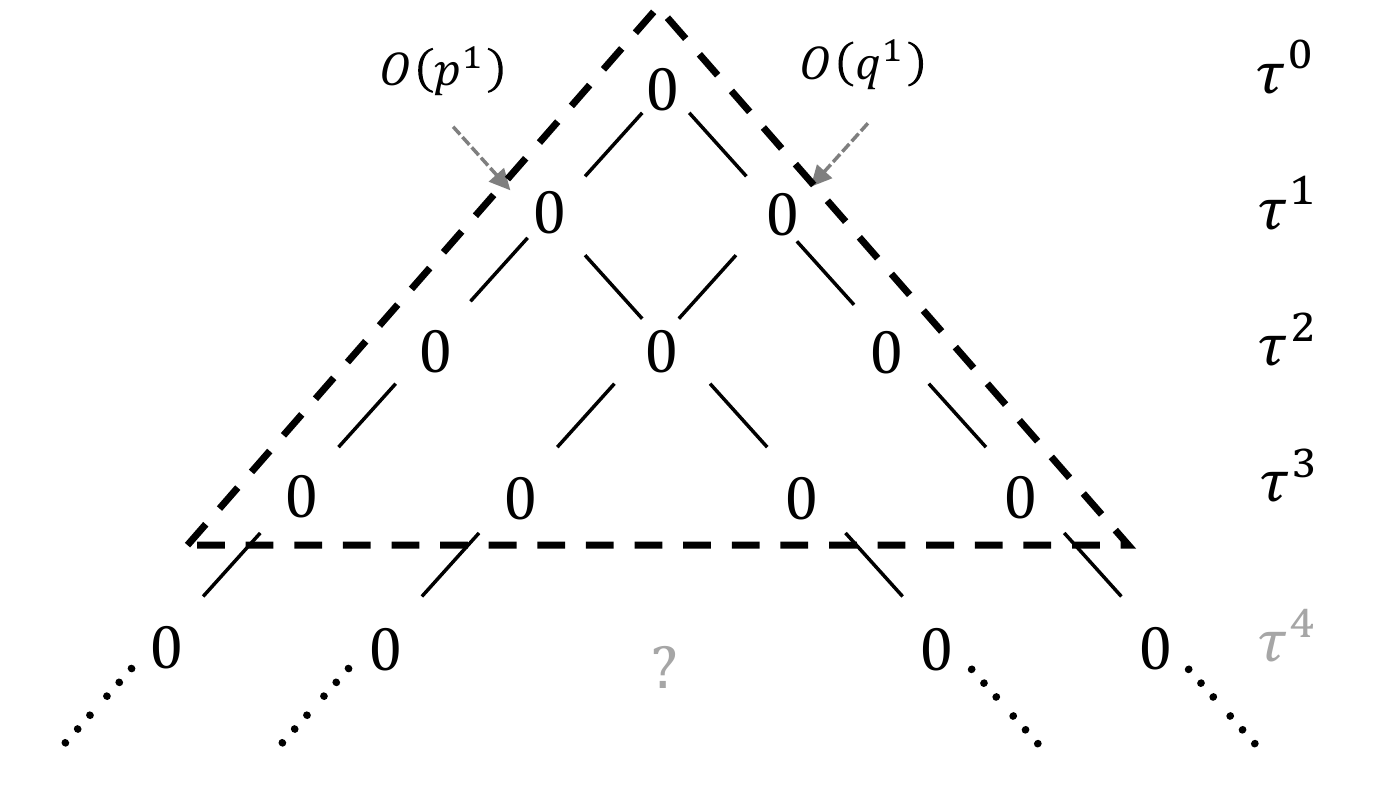}
\par\end{centering}
}
\par\end{centering}
\caption{A special case of Fig.\ref{fig:typeA_0} for DBI amplitudes, with
vanishing single soft theorem and $\lambda=1$. Compared with Fig.\ref{fig:typeA_non0_dilaton},
more expansion terms can be derived, hence higher order double soft
theorems. }
\end{figure}

\subsection{Type B Double Soft Theorems}

Since type B double soft limit requires expansion in terms of identical
perturbation parameter for both soft legs, it cannot be obtained from
single soft limits directly. However, it can be derived from type
A double soft theorems, as the distinction of the two expansion schemes
only affects certain kinds of diagrams. Taking care of the changes
on such diagrams, we can go from type A to type B soft theorems. 
Our approach here is a modification of derivations in \cite{nonlinear},
where double soft theorems of NLSM are derived from single soft theorems.
Our version illustrates more clearly the distinction between the two
schemes. 

\subsubsection{Deriving Type B from Type A Theorems}

For an amplitude $M_{n+2}\left(p,q,\cdots\right)$, we single out
the diagrams where both soft legs are attached to the same four point
vertex, denoting them as pole diagrams and the rest as gut diagrams,
\begin{equation}
M_{n+2}\left(\cdots,p,q\right)=M_{\text{pole}}\left(\cdots,p,q\right)+N\left(\cdots,p,q\right).
\label{eq:pole+gut}
\end{equation}
\begin{figure}
\centering
\subfloat[A pole diagram.]{\includegraphics[height=3.8cm]{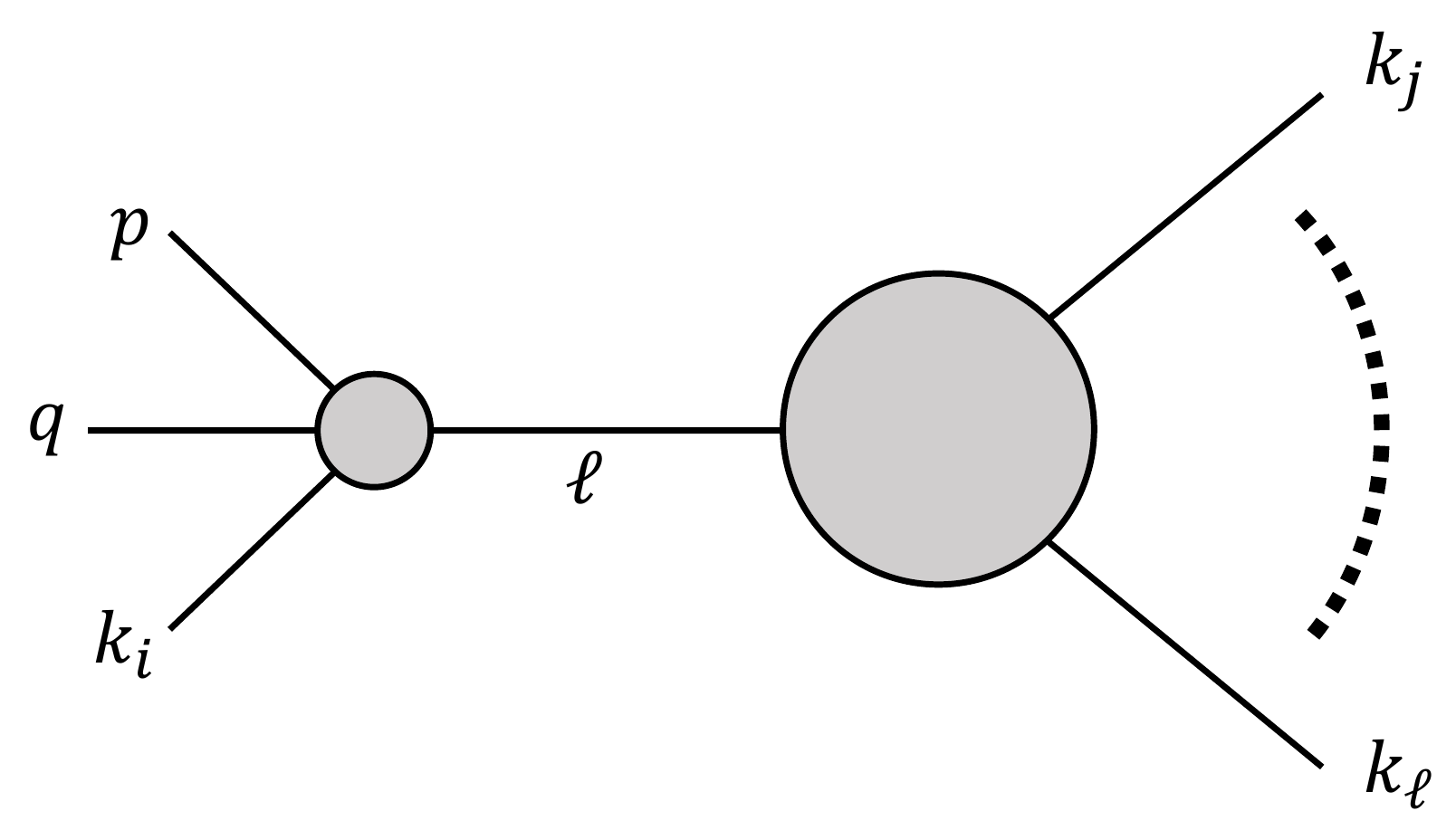}

}\subfloat[A gut diagram with $p$ and $q$ on different side of a propagator. ]{\includegraphics[height=3.8cm]{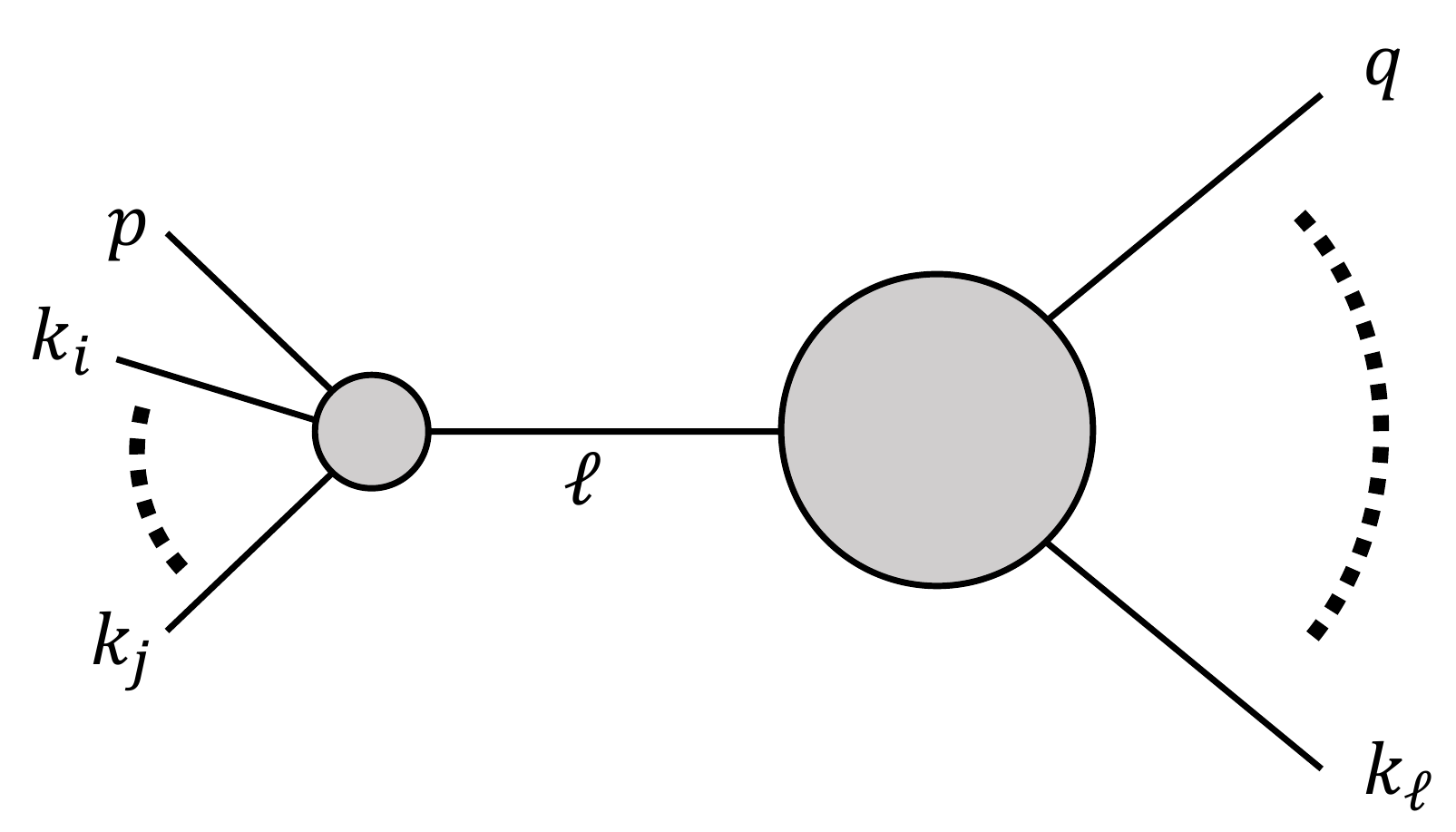}

} 

\subfloat[A gut diagram with $p$ and $q$ attached to a higher point vertex.]{\includegraphics[height=3.8cm]{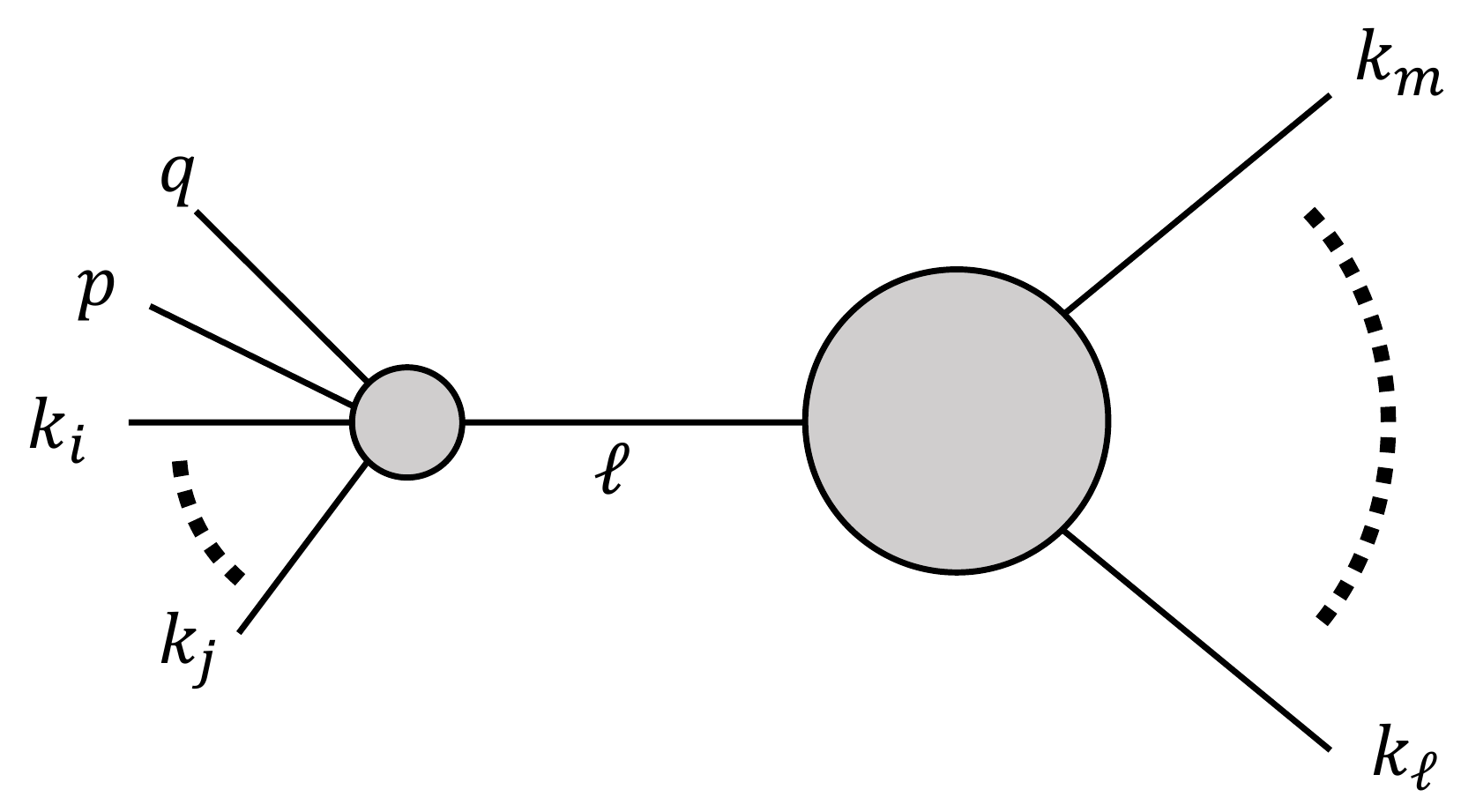}

} \subfloat[A gut diagram involving only a contact term.]{\includegraphics[height=3.8cm]{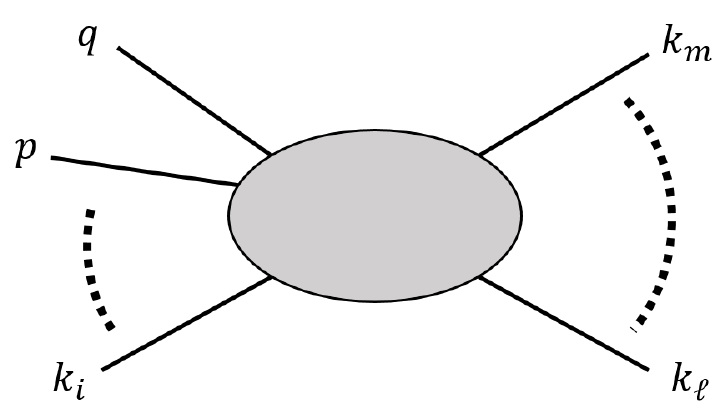}

}

\caption{Pole diagrams and gut diagrams. }
\label{fig:diag} 
\end{figure}
Fig.\ref{fig:diag} shows the form of pole diagrams and some possible
gut diagrams\footnote{In the original usage, a gut diagram has soft legs are attached to
inner propagators. Here it simply denotes anything that is not a pole
diagram. }. This distinction is the classic approach in the discussion of Weinberg
soft theorems, also used in \cite{nonlinear} on which we based our
proof. However, our derivation is different in that we do not need
the full information of pole and gut diagrams. Instead, we only focus
on pole diagrams, which can be expressed explicitly as 
\begin{equation}
M_{\text{pole}}\left(\cdots,p,q\right)=\sum_{i}\tilde{M}_{4}\left(k_{i},p,q\right)\frac{1}{\ell_{i}^{2}}\tilde{M}_{n}\left(\cdots,\ell_{i},\cdots\right)
\end{equation}
where the internal propagator is $\ell_{i}=k_{i}+p_{n+1}+p_{n+2}$
with

\begin{equation}
\ell_{i}^{2}=2\left[k_{i}\cdot\left(p+q\right)+p\cdot q\right]
\end{equation}
and $\tilde{M}_{m}$ denotes off-shell amplitudes which goes on-shell
after taking double soft limit. Since $l_{i}^{2}\rightarrow0$ as
$p$ and $q$ vanish, pole diagrams develop singularity under double
soft limit. Thus, they may behave differently under type A and type
B expansions. Gut diagrams, on the other hand, have the two soft legs
attached either to different vertices, to the same higher point vertex
($n>4$), to an inner propagator, or simply to a single contact term
without propagators. In all possible cases, they are well-behaved
rational functions under double soft limit, since no propagator would
be put on-shell, as can be verified by computing the propagators $\ell$
in Fig.\ref{fig:diag}. The expansion on them is thus the same under
both schemes. More explicitly, 
\begin{align}
\left.M_{\text{pole}}\right|_{\text{type B}}\neq & \left.M_{\text{pole}}\right|_{\text{type A}}\nonumber \\
\left.N\right|_{\text{type B}}= & \left.N\right|_{\text{type A}}\,.
\label{eq:pole_gut_AB}
\end{align}
As a result, the distinction between type B and type A double soft
theorems only comes from pole diagrams, so that, using (\ref{eq:pole+gut})
and (\ref{eq:pole_gut_AB}), 
\begin{align}
 & \left.M_{n+2}\right|_{\text{type B}}-\left.M_{n+2}\right|_{\text{type A}}\nonumber \\
= & \left(\left.M_{\text{pole}}\right|_{\text{type B}}-\left.M_{\text{pole}}\right|_{\text{type A}}\right)+\left(\left.N\right|_{\text{type B}}-\left.N\right|_{\text{type A}}\right)\nonumber \\
= & \left(\left.M_{\text{pole}}\right|_{\text{type B}}-\left.M_{\text{pole}}\right|_{\text{type A}}\right)
\end{align}
or 
\begin{equation}
\left.M_{n+2}\right|_{\text{type B}}=\left.M_{n+2}\right|_{\text{type A}}+\left(\left.M_{\text{pole}}\right|_{\text{type B}}-\left.M_{\text{pole}}\right|_{\text{type A}}\right)\label{eq:AtoB_raw}
\end{equation}
Fortunately, pole diagrams are already reducible to the lower point
amplitude $M_{n}$. Their expansion in both schemes can be evaluated
and compared up to arbitrary order. Therefore, we can derive a corresponding
type B theorem from each type A theorem using (\ref{eq:AtoB_raw}).

For practical computation, we illustrate the steps for a generic single scalar theory\footnote{Other theories (with spin, color etc) will introduce more complicated terms into the four point vertex $M_4$. After incorporating those terms, the computation is similar.}. Its four point vertex has the form 
\begin{equation}
M_{4}(k_{1},k_{2},k_{3},k_{4})=c_{4}^{(2)}\left(s^{2}+t^{2}+u^{2}\right)+c_{4}^{(3)}\left(s^{3}+t^{3}+u^{3}\right)+\cdots
\end{equation}
where $c_{n}^{(m)}$ indicates the coefficient of $s^{m}$ term in
$n$ point amplitude, which may or may not vanish. Alternative representations
such as $st+tu+us$ may be used, which only adds a normalization factor to $c_{4}^{(m)}$. We split pole diagrams into pieces involving different
terms of four point vertex, 
\begin{equation}
M_{\text{pole}}=M_{\text{pole},s^{2}}+M_{\text{pole},s^{3}}+\cdots
\end{equation}
where $M_{\text{pole},s^{2}}\left(\cdots,p,q\right)=\sum_{i}c_{4}^{(2)}\left(s^{2}+t^{2}+u^{2}\right)\frac{1}{\ell_{i}^{2}}\tilde{M}_{n}\left(\cdots,\ell_{i},\cdots\right)$
etc. We then evaluate each term with different order of $s$ separately.
Type B expansion can be performed straightforwardly, while type A
involves some issues of mixed derivatives. We list the results here,
with the calculation details relegated to Appendix \ref{sec:type_a_calc}.
For order $s^{2}$, denoting the perturbation parameter as $\tau$,
we evaluate up to double soft order $\tau^{3}$, with 
\begin{align}
 & \left.M_{\text{pole},s^{2}}\right|_{\text{type B}}-\left.M_{\text{pole},s^{2}}\right|_{\text{type A}}\nonumber \\
= & 0 &  & \tau^{0}\nonumber \\
+ & c_{4}^{(2)}\sum_{i}\left[\frac{\left[k_{i}\cdot\left(p-q\right)\right]^{2}}{k_{i}\cdot\left(p+q\right)}+k_{i}\cdot\left(p+q\right)\right]M_{n}\left(\cdots,p_{i},\cdots\right) &  & \tau^{1}\nonumber \\
+ & 2c_{4}^{(2)}\sum_{i}\left\{ \frac{2\left(k_{i}\cdot p\right)\left(k_{i}\cdot q\right)}{\left[k_{i}\cdot\left(p+q\right)\right]^{2}}\left(p\cdot q\right)\right.\nonumber \\
 & +\left.\frac{k_{i}\cdot\left(p-q\right)}{\left[k_{i}\cdot\left(p+q\right)\right]}\left(p_{\mu}^{\text{ }}q_{\nu}J_{i}^{\mu\nu}\right)+k_{i}\cdot\left(p+q\right)\right\} M_{n}\left(\cdots k_{i},\cdots\right) &  & \tau^{2}\nonumber \\
+ & 2c_{4}^{(2)}\sum_{i}\left\{ \frac{\left(p_{\mu}q_{\nu}J^{\mu\nu}\right)^{2}}{k_{i}\cdot\left(p+q\right)}\right.\nonumber \\
 & \left.+\frac{(p\cdot q)^{2}}{\left[k_{i}\cdot\left(p+q\right)\right]}\left(\frac{3}{2}+\frac{\left[k_{i}\cdot\left(p-q\right)\right]{}^{2}}{\left[k_{i}\cdot\left(p+q\right)\right]^{2}}\right)\right\} M_{n}\left(\cdots,k_{i},\cdots\right). &  & \tau^{3}\label{eq:AtoB_s2}
\end{align}
These soft operators are of order $s^{1}$, which is obvious from
counting the dimensions of $\tilde{M}_{4,s^{2}}$ and the propagator
$1/l_{i}^{2}$. Also, the effect of $M_{\text{pole},s^{2}}$ starts
to contribute at order $\tau^{1}$. For order $s^{3}$, we evaluate
up to order $\tau^{5}$, 
\begin{align}
 & \left.M_{\text{pole},s^{3}}\right|_{\text{type B}}-\left.M_{\text{pole},s^{3}}\right|_{\text{type A}}\nonumber \\
= & 0 &  & \tau^{0},\tau^{1},\tau^{2}\nonumber \\
+ & 3c_{4}^{(3)}(p\cdot q)\sum_{i}\left(\frac{[k_{i}\cdot(p-q)]^{2}}{\left[k_{i}\cdot\left(p+q\right)\right]}\right)M_{n}(\cdots k_{i},\cdots) &  & \tau^{3}\nonumber \\
+ & 6c_{4}^{(3)}(p\cdot q)\sum_{i}\left\{ \frac{k_{i}\cdot\left(p-q\right)}{k_{i}\cdot\left(p+q\right)}\left(p_{\mu}^{\text{ }}q_{\nu}J_{i}^{\mu\nu}\right)\right.\nonumber \\
 & \left.+2\frac{(k_{i}\cdot p)(k_{i}\cdot q)(p\cdot q)}{\left[k_{i}\cdot\left(p+q\right)\right]^{2}}+k_{i}\cdot(p+q)\right\} M_{n}(\cdots k_{i},\cdots) &  & \tau^{4}\nonumber \\
+ & 6c_{4}^{(3)}\left(p\cdot q\right)\left\{ \sum_{i}\frac{\left(p_{\mu}q_{\nu}J^{\mu\nu}\right)^{2}}{k_{i}\cdot\left(p+q\right)}+2\frac{(k_{i}\cdot p)(k_{i}\cdot q)(p\cdot q)}{\left[k_{i}\cdot\left(p+q\right)\right]^{2}}\left[\left(p+q\right)\cdot\frac{\partial}{\partial k_{i}}\right]\right.\nonumber \\
 & \left.+\frac{\left(p\cdot q\right)^{2}}{k_{i}\cdot\left(p+q\right)}\left(-\frac{1}{2}+\frac{\left[k_{i}\cdot\left(p-q\right)\right]^{2}}{2\left[k_{i}\cdot\left(p+q\right)\right]^{2}}\right)\right\} M_{n}\left(\cdots k_{i},\cdots\right). &  & \tau^{5}\label{eq:AtoB_s3}
\end{align}
These operators are of order $s^{2}$, again obvious from dimension
counting. However, its soft order starts at $\tau^{3}$, while naive
power counting for both type A and type B expansion for $M_{\text{pole},s^{3}}$
gives $\tau^{2}$. Somehow the two schemes at this order are identical.
Terms higher than $s^{4}$ will not contribute until $\tau^{3}$.
All these terms can then be added into the type A expansion, leading
to type B soft theorems, 
\begin{align}
\left.M_{n+2}\right|_{\text{type B}}= & \left.M_{n+2}\right|_{\text{type A}}\nonumber \\
 & +\left(\left.M_{\text{pole},s^{2}}\right|_{\text{type B}}-\left.M_{\text{pole},s^{2}}\right|_{\text{type A}}\right)\nonumber \\
 & +\left(\left.M_{\text{pole},s^{3}}\right|_{\text{type B}}-\left.M_{\text{pole},s^{3}}\right|_{\text{type A}}\right)\nonumber \\
 & +\cdots\cdots\,.
 \label{eq:AtoB}
\end{align}
We demonstrate these results using examples from the last subsection.

\subsubsection{Type B double soft theorems with vanishing single soft limits}

Here we start with theories with vanishing single soft limits since
they are simpler. As their type A double soft limits vanish, the type
B limits are simply the differences coming from pole diagrams. The
expansion can be determined from (\ref{eq:AtoB}) to be 
\begin{align}
\left.M_{n+2}\left(\tau p,\tau q,\cdots\right)\right|_{\text{type B}} & =\sum_{i=0}^{\lambda_{d}}\tau^{i}S^{(i)}M_{n}\left(\cdots\right)+O\left(\tau^{\lambda_{d}+1}\right), &  & \lambda_{d}=2\lambda+1
\end{align}
since $\left.M_{n+2}\right|_{\text{type A}}$ expansion is valid up
to order $2\lambda+1$, 
\begin{align}
\left.M_{n+2}\right|_{\text{type A}} & =0+O\left(\tau^{\lambda_{d}+1}\right). &  & \lambda_{d}=2\lambda+1
\end{align}
For DBI, the tree-level four-point amplitude only contains $s^{2}$
terms. Therefore, at tree level, 
\begin{align}
\left.M_{n+2}\right|_{\text{type B}}= & \left.M_{n+2}\right|_{\text{type A}}\nonumber \\
 & +\left(\left.M_{\text{pole},s^{2}}\right|_{\text{type B}}-\left.M_{\text{pole},s^{2}}\right|_{\text{type A}}\right) \nonumber \\
= & \left(\left.M_{\text{pole},s^{2}}\right|_{\text{type B}}-\left.M_{\text{pole},s^{2}}\right|_{\text{type A}}\right)+O\left(\tau^{4}\right) \label{eq:AtoB_DBI_tree}
\end{align}
because of vanishing type A double soft limit, 
\begin{align}
\left.M_{n+2}\right|_{\text{type A}} & =0+O\left(\tau^{\lambda_{d}+1}\right), &  & \lambda_{d}=2\lambda+1=3
\end{align}
This leads to soft operators up to $\tau^{3}$, 
\begin{align}
S_{\text{DBI}}^{(0)}= & 0\nonumber \\
S_{\text{DBI}}^{(1)}= & c_{4}^{(2)}\sum_{i}\left[\frac{\left[k_{i}\cdot\left(p-q\right)\right]^{2}}{k_{i}\cdot\left(p+q\right)}+k_{i}\cdot\left(p+q\right)\right],\nonumber \\
S_{\text{DBI}}^{(2)}= & 2c_{4}^{(2)}\sum_{i}\left\{ \frac{2\left(k_{i}\cdot p\right)\left(k_{i}\cdot q\right)}{\left[k_{i}\cdot\left(p+q\right)\right]^{2}}\left(p\cdot q\right)\right.+\left.\frac{k_{i}\cdot\left(p-q\right)}{\left[k_{i}\cdot\left(p+q\right)\right]}\left(p_{\mu}^{\text{ }}q_{\nu}J_{i}^{\mu\nu}\right)+k_{i}\cdot\left(p+q\right)\right\} \nonumber \\
S_{\text{DBI}}^{(3)}= & 2c_{4}^{(2)}\sum_{i}\left\{ \frac{\left(p_{\mu}q_{\nu}J^{\mu\nu}\right)^{2}}{k_{i}\cdot\left(p+q\right)}\right.\left.+\frac{(p\cdot q)^{2}}{\left[k_{i}\cdot\left(p+q\right)\right]}\left(\frac{3}{2}+\frac{\left[k_{i}\cdot\left(p-q\right)\right]{}^{2}}{\left[k_{i}\cdot\left(p+q\right)\right]^{2}}\right)\right\}. \label{eq:DBI_B_tree}
\end{align}

After some rearrangements, these operators reduce to the ones derived
in \cite{soft2He} except for factors of $c_{4}^{(2)}$. Since the
soft theorems here relate $s^{m}$ terms in $M_{n+2}$ to $s^{m-1}$ terms
in $M_{n}$, the coupling constant is essential for consistency of
mass dimension. This issue would be more prominent for type B theorems
with nonvanishing single soft limits.

For tree-level amplitudes of special Galileon, the four point vertex
only contains $s^{3}$ terms. Thus, only $M_{\text{pole},s^{3}}$
needs to be considered, giving 
\begin{align}
S_{\text{sGal}}^{(0)}= & S_{\text{sGal}}^{(1)}=S_{\text{sGal}}^{(2)}=0\nonumber \\
S_{\text{sGal}}^{(3)}= & 3c_{4}^{(3)}\left(p\cdot q\right)\sum_{i}\left[\frac{\left[k_{i}\cdot\left(p-q\right)\right]^{2}}{k_{i}\cdot\left(p+q\right)}+k_{i}\cdot\left(p+q\right)\right],\nonumber\\
S_{\text{sGal}}^{(4)}= & 6c_{4}^{(3)}\left(p\cdot q\right)\sum_{i}\left\{ \frac{2\left(k_{i}\cdot p\right)\left(k_{i}\cdot q\right)}{\left[k_{i}\cdot\left(p+q\right)\right]^{2}}\left(p\cdot q\right)\right.+\left.\frac{k_{i}\cdot\left(p-q\right)}{\left[k_{i}\cdot\left(p+q\right)\right]}\left(p_{\mu}^{\text{ }}q_{\nu}J_{i}^{\mu\nu}\right)+k_{i}\cdot\left(p+q\right)\right\} \nonumber \\
S_{\text{sGal}}^{(5)}= & 6c_{4}^{(3)}\left(p\cdot q\right)\sum_{i}\left\{ \frac{\left(p_{\mu}q_{\nu}J^{\mu\nu}\right)^{2}}{k_{i}\cdot\left(p+q\right)}\right.\left.+\frac{(p\cdot q)^{2}}{\left[k_{i}\cdot\left(p+q\right)\right]}\left(\frac{3}{2}+\frac{\left[k_{i}\cdot\left(p-q\right)\right]{}^{2}}{\left[k_{i}\cdot\left(p+q\right)\right]^{2}}\right)\right\} .
\end{align}
These soft operators match the results in \cite{soft2He} as well,
again up to the factors of $c_{4}^{(3)}$.

\subsubsection{Type B double soft theorems for nonvanishing soft limits}

For theories with nonvanishing soft limits, 
\begin{align}
\left.M_{n+2}\left(\tau p,\tau q,\cdots\right)\right|_{\text{type B}} & =\sum_{i=0}^{\lambda_{d}}\tau^{i}S^{(i)}M_{n}\left(\cdots\right)+O\left(\tau^{\lambda_{d}}\right), &  & \lambda_{d}=\lambda
\end{align}
where the difference of pole diagrams and the original type A double
soft limit both contribute. For dilaton, with $\lambda=1$, the $s^{2}$
vertex adds an additional term to the original type A operator, giving
\begin{align}
S_{\text{cDBI}}^{(1)}= & \left(p+q\right)^{\mu}\sum_{i}\left[\left(\frac{D-2}{2}+k_{i}\cdot\partial_{i}\right)\partial_{i,\mu}-\frac{1}{2}k_{i,\mu}\partial_{i}^{2}\right]\left[\left(n+1\right)\frac{D-2}{2}-D+\sum_{i}k_{i}\cdot\partial_{i}\right]\nonumber \\
 & +c_{4}^{(2)}\sum_{i}\left[\frac{\left[k_{i}\cdot\left(p-q\right)\right]^{2}}{k_{i}\cdot\left(p+q\right)}+k_{i}\cdot\left(p+q\right)\right]. \label{eq:cDBI_B_tree}
\end{align}
Note that this operator mixes terms with different mass dimensions.
Therefore, the coupling constant $c_{4}^{(2)}$, omitted in results
in \cite{soft2He,soft2Huang}, is crucial here to produce sensible
result. To our knowledge this soft theorem has not been derived elsewhere.
We have performed numerical tests on 4, 6, 8 point amplitudes of $s^{2}$and
$s^{3}$orders to verify the result.

\subsection{Loop Correction of Double Soft Theorems}

Attempts has been made elsewhere to determine whether loop correction
will modify double soft theorems by using current algebra (e.g. \cite{soft2Huang}).
However, as the different expansion schemes were not distinguished,
it was difficult to identify which of the theorems will be modified
by loop correction. Our formulation provides a clear way to address
this issue. Type A theorems would be protected from loop correction
by symmetry, while the fate of type B depends on the new vertices
generated by loop corrections, as we discuss below.

\subsubsection{Type A}

Our derivation of type A double soft theorems only relies on single
soft theorems, which are directly implied by symmetry. It does not
involve the explicit form of any vertex. Therefore, as long as the
symmetry is not anomalous, type A theorems should be valid even if
loop corrections are considered, regardless of soft order.

As an example, consider the amplitude of bosonic string disucssed
in \cite{soft2Huang}, which follows the same broken symmetry and
soft theorems (\ref{eq:type_A_0}) of DBI. At tree level, its four
point vertex contains $s^{2}$, but one loop correction gives an additional
$s^{3}$ term, $M_{\text{4},s^{3}}=c_{4}^{(3)}\left(s^{3}+t^{3}+u^{3}\right)$.
If we follow traditional approach of Weinberg soft theorems, the pole
diagrams, $\sum_{i}\tilde{M}_{4,s^{3}}\left(k_{i},p,q\right)\frac{1}{\ell_{i}}\tilde{M}_{n}\left(\cdots,\ell_{i},\cdots\right)$,
seems to modify order $\tau^{2}$ and $\tau^{3}$ double soft theorems,
by dimention counting. However, those theorems are symmetry-protected
and will not be modified. This means that symmetry forces the loop
correction from pole and gut diagrams cancel each other. This would
be the case for all other type A theorems, e.g. that of dilaton and
special Galileon amplitudes.

To investigate the theorems discussed in \cite{soft2He,soft2Huang},
we must consider type B scheme instead.

\subsubsection{Type B}

As opposed to type A theorems, type B theorems depend on the explicit
form of four point vertices. Therefore, loop corrections that introduce
new terms to four point vertex may modify type B soft theorems. (\ref{eq:AtoB_s2}),
(\ref{eq:AtoB_s3}) and similar expressions for $M_{\text{pole},s^{m}}$
indicates on which order the modification would occur. Also, note
that loop correction only modifies double soft theorems instead of
destroying it. That is, $M_{n+2}$ can still be reduced to lower point
amplitude $M_{n}$, only with modified soft operators.

We again consider the amplitude of bosonic string discussed in \cite{soft2Huang},
and the type B theorems in (\ref{eq:DBI_B_tree}). Since loop correction
generates an $s^{3}$ four-point vertex among higher order of $s$,
the type B double soft theorem is no longer given by (\ref{eq:AtoB_DBI_tree})
, but by 
\begin{align}
\left.M_{n+2}\right|_{\text{type B}}= & \left.M_{n+2}\right|_{\text{type A}}\nonumber \\
 & +\left(\left.M_{\text{pole},s^{2}}\right|_{\text{type B}}-\left.M_{\text{pole},s^{2}}\right|_{\text{type A}}\right)\nonumber \\
 & +\left(\left.M_{\text{pole},s^{3}}\right|_{\text{type B}}-\left.M_{\text{pole},s^{3}}\right|_{\text{type A}}\right)\nonumber \\
 & +\cdots\cdots\label{eq:AtoB_DBI_loop}
\end{align}
where $\left(\left.M_{\text{pole},s^{3}}\right|_{\text{type B}}-\left.M_{\text{pole},s^{3}}\right|_{\text{type A}}\right)$
and higher order terms are loop corrections of type B theorems. Naive
power counting also indicates that the term $\left(\left.M_{\text{pole},s^{3}}\right|_{\text{type B}}-\left.M_{\text{pole},s^{3}}\right|_{\text{type A}}\right)$
will modify $\tau^{2}$ theorem, as stated in \cite{soft2Huang}.
However, (\ref{eq:AtoB_s3}) states that modification starts at $\tau^{3}$,
since $\left(\left.M_{\text{pole},s^{3}}\right|_{\text{type B}}-\left.M_{\text{pole},s^{3}}\right|_{\text{type A}}\right)=\mathcal{O}\left(\tau^{3}\right)$.
More explicitly, contribution of $M_{\text{pole},s^{3}}$ at order
$\mathcal{O}\left(\tau^{2}\right)$ is identical for type B and type
A, thus protected by symmetry. This explains why $\tau^{2}$ theorem
is intact. Clearly, this result comes from the structure of $s^{3}$
vertex itself, that is, $\left(\left.M_{\text{pole},s^{3}}\right|_{\text{type B}}-\left.M_{\text{pole},s^{3}}\right|_{\text{type A}}\right)$.
It is not simply a result of symmetry as conjectured in \cite{soft2Huang}.
Our consideration also predicts that $\tau^{3}$ theorem will be modified,
\begin{align}
S_{\text{DBI}}^{(3)}= & 2c_{4}^{(2)}\sum_{i}\left\{ \frac{\left(p_{\mu}q_{\nu}J^{\mu\nu}\right)^{2}}{k_{i}\cdot\left(p+q\right)}\right.\left.+\frac{(p\cdot q)^{2}}{\left[k_{i}\cdot\left(p+q\right)\right]}\left(\frac{3}{2}+\frac{\left[k_{i}\cdot\left(p-q\right)\right]{}^{2}}{\left[k_{i}\cdot\left(p+q\right)\right]^{2}}\right)\right\} \nonumber \\
 & +3c_{4}^{(3)}(p\cdot q)\sum_{i}\left(\frac{[k_{i}\cdot(p-q)]^{2}}{\left[k_{i}\cdot\left(p+q\right)\right]}\right)M_{n}(\cdots k_{i},\cdots) \nonumber \\
 & +\cdots\cdots
\end{align}
as has been confirmed by explicit computation of loop correction in
\cite{soft2Huang}.

The above procedure provides an effective way to predict the behavior
of type B theorems against loop correction. In general, loop correction
modifies type B theorems by adding further terms of $\left(\left.M_{\text{pole},s^{m}}\right|_{\text{type B}}-\left.M_{\text{pole},s^{m}}\right|_{\text{type A}}\right)$
in (\ref{eq:AtoB}). These terms can be computed, as in (\ref{eq:AtoB_s2})
and (\ref{eq:AtoB_s3}), which may not contribute until a certain
order. Thus, type B theorems prior to this order are still free from
modication of loop corrections.

As a further example, consider the type B theorem for dilaton in (\ref{eq:cDBI_B_tree}).
The soft operators are only of $\tau^{0}$ and $\tau^{1}$ order,
while the lowest possible order of loop correction, coming from $s^{3}$
vertices, starts at $\tau^{3}$. Therefore, these type B double soft
theorems would be protected from any loop correction, as has been
verified by explicit loop calculations \cite{soft2Huang}. As for
special Galileon amplitudes, we would need to compute $\left.M_{\text{pole},s^{4}}\right|_{\text{type B}}-\left.M_{\text{pole},s^{4}}\right|_{\text{type A}}$
and beyond.

\section{Extension to Multiple Soft Theorems}

Our formulation can be readily extended to multiple soft theorems, that is, the momenta of more than two legs are taken to vanish. 
In the following, we denote multiple soft theorems with $m$ soft legs as $m$-soft theorem. 
Type A $m$-soft theorems can be derived directly from single soft theorems as before, and type B theorems can be derived from type A $m$-soft theorems, double to $(m-1)$-soft type B and type A theorems, and vertices up to ($m+2$)-point. 
As in the case for double soft theorems, type A are symmetry protected, whereas type B would be modified by loop corrections. 

\subsection{Type A Multiple Soft Theorems}

For an ($n+m$) point amplitude $M_{n+m}$ with $m$ external momenta taken soft, the type A $m$-soft expansion is simply $m$-variate Taylor expansion,
\begin{align}
&M_{n+m}\left.\left(\tau p_{1},\cdots,\tau p_{m},k_{1},\cdots,k_{n}\right)\right|_{\text{type A}}=\left.M_{n+m}\left(\tau_{1}p_{1},\cdots,\tau_{m}p_{m},k_{1},\cdots,k_{n}\right)\right|_{\tau_{1}=\cdots=\tau_{m}=\tau=0} \nonumber \\
=&M_{n+2}\left(0,0,\cdots\right)\nonumber \\
&+\left[\left(\tau_{1}\frac{\partial}{\partial\tau_{1}}\right)+\cdots+\left(\tau_{m}\frac{\partial}{\partial\tau_{m}}\right)\right]\left.M_{n+m}\left(\tau_{1}p_{1},\cdots,\tau_{m}p_{m},k_{1},\cdots,k_{n}\right)\right|_{\tau_{1}=\cdots=\tau_{m}=\tau=0} \nonumber \\
&+\cdots\,,
\end{align}
where $p_1$ through $p_m$ are the soft momenta, and $k_1$ through $k_n$ are the remaining hard momenta. 
The order $\tau^{i}$ expansion would be
\begin{align}
\sum_{r_{1}\cdots r_{m}}^{r_{1}+\cdots+r_{m}=i}\tau^{i}\frac{\partial^{r_{1}}}{\partial\tau_{1}^{r_{1}}}\frac{\partial^{r_{2}}}{\partial\tau_{2}^{r_{2}}}\cdots\frac{\partial^{r_{m}}}{\partial\tau_{m}^{r_{m}}}\left.M_{n+m}\left(\tau_{1}p_{1},\cdots,\tau_{m}p_{m},k_{1},\cdots,k_{n}\right)\right|_{\tau_{1}=\cdots=\tau_{m}=\tau=0}\,.
\label{eq:multi_ith}
\end{align}

For non-vanishing single soft theorems up to order $\lambda$, type A $m$-soft theorems can be derived by applying single soft theorems $m$ times, 
\begin{align}
M_{n+m} & \left.\left(p_{1},\cdots,p_{m},k_{1},\cdots,k_{n}\right)\right|_{\text{type A}} \nonumber \\
= & \left(S_{n+m-1}^{(0)}+p_{1}\cdot S_{n+m-1}^{(1)}+\cdots+p_{1}^{\lambda}\cdot S_{n+m-1}^{(\lambda)}\right)M_{n+m-1}\left(p_{2},\cdots\right)+O\left(p_{1}^{\lambda+1}\right)\nonumber \\
= & \left(S_{n+m-1}^{(0)}+p_{1}\cdot S_{n+m-1}^{(1)}+\cdots+p_{1}^{\lambda}\cdot S_{n+m-1}^{(\lambda)}\right)\cdots\left(S_{n}^{(0)}+p_{m}\cdot S_{n}^{(1)}+\cdots p_{m}^{\lambda}\cdot S_{n}^{(\lambda)}\right)M_{n}\left(\cdots\right)\nonumber \\
 & +\cdots\nonumber \\
= & \left[S_{n+m-1}^{(0)}S_{n+m-2}^{(0)}\cdots S_{n}^{(0)}+p_{1}^{\mu}S_{n+m-1}^{(1)}S_{n+m-2}^{(0)}\cdots S_{n}^{(0)}+p_{2}^{\mu}S_{n+m-1}^{(1)}S_{n+m-2}^{(1)}\cdots S_{n}^{(0)}+\cdots\right]M_{n}\left(\cdots\right)\nonumber \\
 & +\cdots\nonumber \\
= & \left[S_{\text{d}}^{(0)}+S_{\text{d}}^{(1)}+\cdots+S_{\text{d}}^{(\lambda_{d})}\right]M_{n}\left(\cdots\right)+\cdots,
\end{align}
as in \eqref{eq:type_A_expand} for double soft theorems. 
For similar reasons, multiple soft theorems exist up to order $\lambda_d = \lambda$. 
This can be seen from \eqref{eq:multi_ith}, 
where in each individual terms, none of the expansion orders $r_1$ through $r_m$ can exceed $\lambda$. 
Thus, terms $S^{(r_1)} S^{r_2} \cdots S^{(r_m)}$ obtainable from single soft but with $\left(r_1+\cdots+r_m\right)>\lambda$ will also be left out from multiple soft theorems, 
so that multiple soft theorems will contain less information than single soft theorems. 

For vanishing single soft theorems, the first expansion in $p_1$ gives zero up to order $\lambda$, 
\begin{align}
M_{n+m}\left(p_{1},\cdots,p_{m},k_{1},\cdots,k_{n}\right)=0+\mathcal{O}\left(p_1^{\lambda+1}\right)\,,
\end{align}
so subsequent expansion in $p_2$ through $p_m$ is trivial and can be done up to arbitrary order. 
This will give us expansion terms up to order $\left( p_1^{\lambda} p_2^{\infty} \cdots p_m^{\infty} \right)$. 
Similarly, expanding first in $p_j$ gives expansion terms up to order $\left( p_1^{\infty} \cdots p_j^{\lambda} \cdots  p_m^{\infty} \right)$. 
That is, a term $\frac{\partial^{r_{1}}}{\partial\tau_{1}^{r_{1}}}\frac{\partial^{r_{2}}}{\partial\tau_{2}^{r_{2}}}\cdots\frac{\partial^{r_{m}}}{\partial\tau_{m}^{r_{m}}}M_{n+m}$ in \eqref{eq:multi_ith} can be obtained if at least one of the exponents $r_j$ does not exceed $\lambda$, or that $\min\{ r_j \}\leq \lambda$. 
Therefore, type A $m$-soft theorem can be obtained up to order $\lambda_d =m \left( \lambda+1 \right) -1$, or
\begin{align}
&M_{n+m}\left.\left(\tau p_{1},\cdots,\tau p_{m},k_{1},\cdots,k_{n}\right)\right|_{\text{type A}} \nonumber \\
=&0+\mathcal{O}\left(\tau^{m \left( \lambda+1 \right)}\right)
\end{align}
At this order, all possible terms that can be obtained from $m$ consecutive single soft theorems would be included, 
so the multiple soft theorems contain equivalent information as vanishing single soft theorems. 

As is the case for double soft theorems, type A multiple soft theorems are directly derived from single soft theorems, 
so that they are protected from loop corrections by underlying symmetry. 

\subsection{Type B Multiple Soft Theorems}

Type B multiple soft expansion is obtained from single variable Taylor expansion using identical expansion parameter $\tau$, 
\begin{align}
&M_{n+m}\left.\left(\tau p_{1},\cdots,\tau p_{m},k_{1},\cdots,k_{n}\right)\right|_{\text{type B}} \nonumber \\
=&M_{n+m}\left(0,\cdots,0,k_{1},\cdots,k_{n}\right)\nonumber\\
&+\left.\tau\frac{\partial}{\partial\tau}M_{n+m}\left(\tau p_{1},\cdots,\tau p_{m},k_{1},\cdots,k_{n}\right)\right|_{\tau=0}\nonumber\\
&+\left.\frac{1}{2}\tau^{2}\frac{\partial^{2}}{\partial\tau^{2}}M_{n+m}\left(\tau p_{1},\cdots,\tau p_{m},k_{1},\cdots,k_{n}\right)\right|_{\tau=0}\nonumber\\
&+\cdots\,.
\label{eq:multi_B}
\end{align}
This would be the natural scheme if, for example, CHY representation is used for derivation. 
For multiple soft theorems there also exist other schemes, where only some of the expansion parameters is taken identical. 
For example, for quadruple soft theorems we may use two expansion parameters, $\tau$ and $\epsilon$,
\begin{align}
&\left.M_{n+m}\left(\tau p_{1},\tau p_{2},\epsilon p_{3},\epsilon p_{4}, k_{1},\cdots,k_{n}\right)\right|_{\tau =\epsilon =0} \nonumber \\
=&M_{n+m}\left( 0,0, 0,0,\cdots \right) \nonumber \\
&+ \tau \left\{\left[ \frac{\partial}{\partial \tau} M_{n+m}\left( \tau p_1,\tau p_2, 0,0,\cdots \right) \right] \bigg|_{\tau = 0}
+  \left[ \frac{\partial}{\partial \epsilon} M_{n+m}\left(  0,0,\epsilon p_3,\epsilon p_4,\cdots \right) \bigg|_{\epsilon = 0} \right] \right\} \nonumber \\
&+\cdots 
\end{align}
Here we will not consider such cases. 
Instead we only focus on the type B expansion scheme \eqref{eq:multi_B}, 
which would naturally arise from methods such as CHY representation. 

To derive type B $m$-soft soft theorems, we again need its difference from type A theorems, 
\begin{align}
M_{n+m}\big|_{\text{type B}}-M_{n+m}\big|_{\text{type A}}\,.
\end{align}
Again, only the diagrams that become singular under $m$-soft limit behave differently under the two schemes. 
They are denoted by pole diagrams as usual, which include diagrams with $(r-2)$ of the $m$  soft momenta attached to an $r$-point vertex, 
where $4 \leq r \leq m+2$. 
These are the only cases where an inner propagator goes on-shell. 
\begin{figure}
\centering
\includegraphics[height=3.8cm]{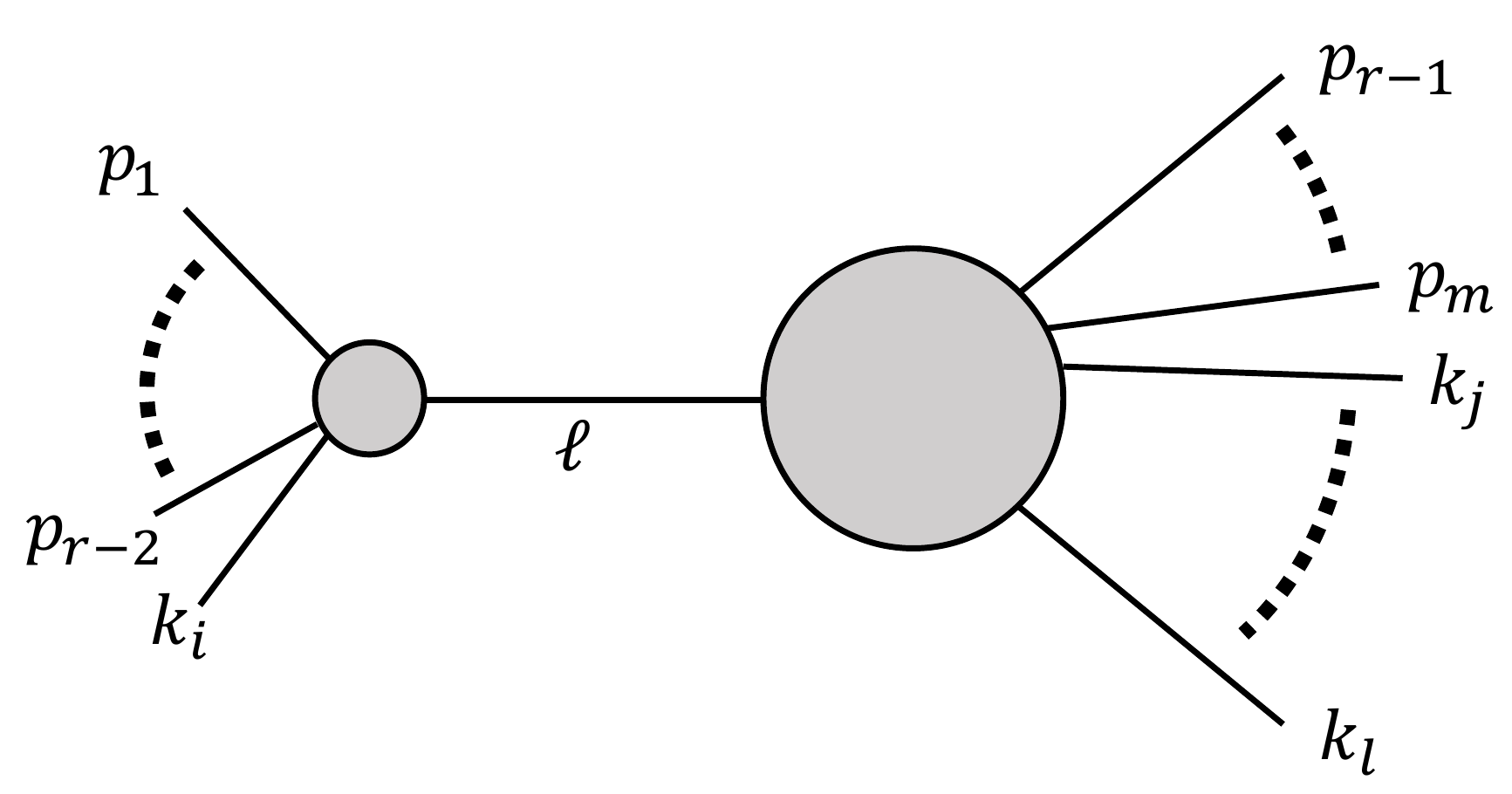}
\caption{An example of pole diagram, with a $r$-point vertex.}
\label{fig:pole_multi}
\end{figure}
An example of pole diagram is shown in Fig.\ref{fig:pole_multi}, which can be expressed as
\begin{align}
M_{r}\left(p_{1},\cdots,p_{r-2},k_{i}\right)\frac{1}{\ell^{2}}M_{n+m-r+2}\left(p_{r-1},\cdots,p_{m},\ell,\cdots\right)
\label{eq:pole_multi}
\end{align}
where $\ell = k_i + p_1 + \cdots + p_{r-2}$. 
Thus we again have
\begin{align}
\left.M_{n+m}\right|_{\text{type B}}=\left.M_{n+m}\right|_{\text{type A}}+\left(\left.M_{\text{pole}}\right|_{\text{type B}}-\left.M_{\text{pole}}\right|_{\text{type A}}\right)
\label{eq:multi_AtoB}
\end{align}
so we need to expand the pole diagrams in the two different schemes. 
The situation is more complicated than double soft theorems, since the lower point amplitude $M_{n+m-r+2}$ also contains soft momenta. 
However, once double to ($m-1$)-soft theorems are derived, the expansion of $M_{n+m-r+2}$ is straightforward. 
For example, consider type A expansion of \eqref{eq:pole_multi}. The expansion on $p_1$ through $p_{r-2}$ can be carried out directly, while that on $p_{r-1}$ through $p_m$ can be performed using type A ($m-r+1$)-soft theorems. 
Expanding the two subsets of momenta separately is legitimate, since we use distinct expansion parameters in type A for each momentum. 
On the other hand, using the same strategy on type B $m$-soft theorems seems illegal, as this introduces two expansion parameters, $\tau_1$ for $p_1 \sim p_{r-2}$, and $\tau_2$ for $p_{r-1} \sim p_m$. 
To address this, let us consider possible singularities from the pole diagrams. 
The first singularity comes from the propagator, which develops when $\tau_1=0$. 
The other ones all come from the lower-point amplitude $M_{n+m-r+2}$, which develop when $\tau_2=0$. 
Since no other singularities would be introduced if $\tau_1=\tau_2=0$ simultaneously, this expansion is equivalent to taking a single expansion parameter, $\tau_1=\tau_2=\tau$. 
Thus, our procedure is actually valid to derive type B soft theorems. 
Once the two expansions are evaluated, \eqref{eq:multi_AtoB} readily gives type B $m$-soft theorems. 

To sum up, type B $m$-soft theorems can be recursively derived from type B and type A $n$-soft theorems with $n<m$, together with four-point to $m+2$ point vertices. 
The procedure is straightforward in principle, but gets cumbersome for larger $m$. 
Since the information of vertices is used, type B theorems in general will be modified by loop corrections.

\section{Fixing amplitudes by soft theorems\label{sec:fix_ampitude}}

Since we proved that double soft theorems can be derived from single
soft theorems, the former contain no more information other than the
four-point vertex. To test this, we can also examine their constraints
on explicit amplitudes. Soft theorems relate higher point amplitudes
to lower point ones, constraining their forms.
We will start from using single soft theorem and respect unitarity in the beginning to fix DBI and
dilaton amplitudes, which was done in \cite{periodic,dilaton} but
we still present it here. Then we relax our unitary assumption and
fix the amplitude by only imposing locality and single soft theorem.
This has been done for DBI in \cite{DBI_locality,DBI_locality2}. However, we note that the amplitude is in fact not completely fixed; the coefficient of the factorization channel is still undetermined, though other terms are indeed fixed. 
Performing this test on dilaton amplitudes, we find identical result, which we have checked up to eight-point.
Then we move to double soft theorem. 
We will show that for DBI amplitudes, imposing locality and type A double soft theorem give identical result, that is, only the coefficient of factorization channel is left undetermined. 
Imposing type B instead, this ambiguity is removed, fixing the amplitude completely. 
Dilaton amplitudes, however cannot be totally fixed by imposing locality and both kinds of 
double soft theorems. Imposing type B also fixes the coefficient of factorization channel, though some other terms are still undetermined. Even if starting with unitary assumption, we
still cannot totally fix dilaton amplitude by using double soft theorem. 
We therefore see that only type B double soft theorems may contain further information than single soft theorems, while some information from nonvanishing single soft theorem may be lost in double soft theorems. This is consistent with our analytical results.

\subsection{Applying single soft theorem to fix amplitudes}

There are two kinds of single soft theorems, vanishing and non-vanishing.
We follow the approach in \cite{periodic,dilaton} to fix amplitudes. 
\begin{itemize}
\item Vanishing soft limit in certain degree $\lambda$. ( 0 for NLSM, 1
for DBI, 2 for sGal)
\item Non-vanishing soft limit, e.g. Dilaton.
\end{itemize}
To fix the ($n+1$)-point amplitude, we write down the ($n+1$)-point
ansatz, which contains some coefficients to be determined. We take
one leg soft, applying the soft theorem, which forces the ansatz with
one leg taken soft to vanish or relate to lower n-point amplitude.
\begin{itemize}
\item Vanishing soft limit:
\end{itemize}
\begin{align}
\tau^{0}: & \left.M_{n+1\_\text{ansatz}}\right|_{\mathcal{O}(\tau^{0})}=0\nonumber \\
\tau^{1}: & \left.M_{n+1\_\text{ansatz}}\right|_{\mathcal{O}(\tau^{1})}=0\nonumber \\
\vdots\nonumber \\
\tau^{\lambda}: & \left.M_{n+1\_\text{ansatz}}\right|_{\mathcal{O}(\tau^{\lambda})}=0
\end{align}
\begin{itemize}
\item Non-vanishing soft limit:
\begin{align}
\text{Leading :\ensuremath{\left.M_{n+1\_\text{ansatz}}\right|_{\mathcal{O}(\tau^{0})}}=} & \left.\mathcal{S}^{(0)}M_{n}\right|_{\mathcal{O}(\tau^{0})}\nonumber \\
\text{Sub-leading :\ensuremath{\left.M_{n+1\_\text{ansatz}}\right|_{\mathcal{O}(\tau^{1})}}}= & \left.\left(\mathcal{S}^{(0)}+\tau\mathcal{S}^{(1)}\right)M_{n}\right|_{\mathcal{O}(\tau^{1})}\nonumber \\
\vdots
\end{align}
\end{itemize}
After putting numerics into $s_{ij}$ , we have sets of linear equations
of those undetermined coefficients. We solve those linear equations
for those coefficients. Here we will explicitly show how to apply
the above procedure to some cases like DBI, dilaton and conformal
DBI. 

\subsubsection{DBI}

Let's take DBI six-point $s^{3}$ amplitude for example. One part
of ansatz for six-point $s^{3}$ amplitude comes from local operator,
which is represented by two polynomial $c_{6p,1}(s_{12}^{3}+\mathcal{P}_{6})$
and $c_{6p,2}(s_{123}^{3}+\mathcal{P}_{6})$. The other part is from
factorization channel of two four-point $s^{2}$ amplitude, which
means we respect unitarity in the beginning \footnote{$s_{i \cdots k} = (p_i + \cdots + p_k)^2$}
\begin{equation}
\left(c_{4}^{(2)}\right)^{2}\left((s_{12}^{2}+s_{13}^{2}+s_{23}^{2})\frac{1}{s_{123}}(s_{45}^{2}+s_{46}^{2}+s_{56}^{2})+\mathcal{P}_{6}\right)
\end{equation}
where $P_{n}$ denotes summing over permutation of n elements. The
six-point $s^{3}$ ansatz is :

\begin{equation}
M_{6\text{\_ansatz}}^{(3)}=c_{6p,1}(s_{12}^{3}+\mathcal{P}_{6})+c_{6p,2}(s_{123}^{3}+\mathcal{P}_{6})+(c_{4}^{(2)})^{2}\left((s_{12}^{2}+s_{13}^{2}+s_{23}^{2})\frac{1}{s_{123}}(s_{45}^{2}+s_{46}^{2}+s_{56}^{2})+\mathcal{P}_{6}\right)\label{eq:M6s3_U}
\end{equation}
The leading and sub-leading soft theorem \eqref{eq:0_single} force
the $\mathcal{O}(\tau^{0})$ and $\mathcal{O}(\tau^{1})$ amplitude
to vanish 

\begin{align}
\text{Leading :}\left.M_{6\_\text{ansatz}}^{(3)}\right|_{\mathcal{O}(\tau^{0})}= & 0\nonumber \\
\text{Sub-leading :\ensuremath{\left.\text{\ensuremath{M}}_{6\_\text{ansatz}}^{(3)}\right|_{\mathcal{O}(\tau^{1})}}=} & 0
\end{align}
We can solve the $c_{6p,1}$ and $c_{6p,2}$ in terms of $c_{4}^{(2)}$,
which means the DBI six-point amplitude is fixed up to one four-point
coupling constant. That is,

\begin{equation}
M_{6\text{\_DBI}}^{(3)}=(c_{4}^{(2)})^{2}\left[-(s_{123}^{3}+\mathcal{P}_{6})+\left((s_{12}^{2}+s_{13}^{2}+s_{23}^{2})\frac{1}{s_{123}}(s_{45}^{2}+s_{46}^{2}+s_{56}^{2})+\mathcal{P}_{6}\right)\right]\label{eq:M6_DBI_sol_U}
\end{equation}
which reproduces the results in \cite{periodic}. 

\subsubsection{Dilaton}

Dilaton six-point $s^{3}$ amplitude needs the lower point information
as an input. The inputs are five-point $s^{3}$ amplitude $A_{5}^{(3)}=c_{5}^{(3)}(s_{12}^{3}+\mathcal{P\mathrm{_{5}}})$
and four-point $s^{2}$ amplitude $A_{4}^{(2)}=c_{4}^{(2)}(s_{12}^{2}+\mathcal{P}_{4})$.
The six-point $s^{3}$ ansatz is the same as DBI . Then we apply the
leading $(\tau^{0})$ and sub-leading $(\tau^{1})$ soft theorem \eqref{eq:cDBI_single}
to relate our six-point ansatz to lower five-point amplitude, 

\begin{align}
\text{Leading:}\left.M_{6\_\text{ansatz}}^{(3)}\right|_{\mathcal{O}(\tau^{0})}= & \left.\mathcal{S}^{(0)}M_{5}^{(3)}\right|_{\mathcal{O}(\tau^{0})}\nonumber \\
\text{Sub-leading:}\left.M_{6\_\text{ansatz}}^{(3)}\right|_{\mathcal{O}(\tau^{1})}= & \text{\ensuremath{\left.\left(\mathcal{S}^{(0)}+\tau\mathcal{S}^{(1)}\right)M_{5}^{(3)}\right|_{\mathcal{O}(\tau^{1})}}}
\end{align}
Notice that leading soft theorem also contains sub-leading piece $(\tau^{1})$
, which needs to be considered when we apply sub-leading soft theorem.
The two unknown $c_{6p,1}$ and $c_{6p,2}$ can be solved in terms
of $c_{5}^{(3)}$ and $c_{4}^{(2)}$. In 4D, for example we get

\begin{align}
M_{6}^{(3)} & =-c_{5}^{(3)}(s_{12}^{3}+\mathcal{P}_{6})-\left(\frac{c_{5}^{(3)}}{2}+(c_{4}^{(2)})^{2}\right)(s_{123}^{3}+\mathcal{P}_{6})\nonumber \\
 & +(c_{4}^{(2)})^{2}\left((s_{12}^{2}+s_{13}^{2}+s_{23}^{2})\frac{1}{s_{123}}(s_{45}^{2}+s_{46}^{2}+s_{56}^{2})+\mathcal{P}_{6}\right)\label{eq:M6_dilaton_sol_U}
\end{align}
We can use the above six-point $s^{3}$ amplitude as an input to fix
seven-point $s^{3}$ amplitude.

\begin{equation}
M_{7}^{(3)}=c_{5}^{(3)}(s_{12}^{3}+\mathcal{P}_{7})+\left(c_{5}^{(3)}+3(c_{4}^{(2)})^{2}\right)(s_{123}^{3}+\mathcal{P}_{6})-(c_{4}^{(2)})^{2}M_{7}^{\text{fac}}
\end{equation}

\begin{align}
M_{7}^{\text{fac}}= & (s_{12}^{2}+s_{13}^{2}+s_{23}^{2})\frac{1}{s_{123}}(s_{45}^{2}+s_{46}^{2}+s_{47}^{2}+s_{56}^{2}+s_{57}^{2}+s_{67}^{2}+(s_{45}+s_{46}+s_{47})^{2}\nonumber \\
 & +(s_{45}+s_{56}+s_{57})^{2}++(s_{46}+s_{56}+s_{67})^{2}+(s_{47}+s_{57}+s_{67})^{2})+\mathcal{P}_{7}
\end{align}
which is identical to the result in \cite{dilaton}. In 6D, the explicit dilaton amplitudes are different, but still totally fixed in terms of lower point amplitude. 

\subsubsection{Conformal DBI}

Conformal DBI (cDBI) is a sub-class of dilaton. We can apply dilaton
single soft theorem \eqref{eq:cDBI_single} to fix cDBI amplitude. For
cDBI six-point $s^{3}$ amplitude, its soft limit is dilaton soft
operator \eqref{eq:cDBI_single} acting on five-point $s^{3}$ amplitude.
But cDBI five-point $s^{3}$ amplitude is zero, so its soft limit
is the same as DBI, i.e. vanishing. For cDBI six-point $s^{3}$, we
get the same amplitude as six-point DBI since the soft limit happens
to be the same. To get seven-point $s^{3}$ amplitude, we use the
cDBI six-point $s^{3}$ amplitude. After applying single soft theorem,
we can fix the amplitude. 

\subsection{Soft theorem + Locality $\protect\longrightarrow$ Unitarity}

The above six-point ansatz has assumed factorized form, which means
we respect unitarity in the very beginning. What if we do not assume
unitarity first, can we fix the amplitude by soft theorem and locality(graph structure)? The answer is yes. This has been checked on DBI
amplitudes \cite{DBI_locality,DBI_locality2}, but not yet on dilaton, which we
show below. 

Again we take the six-point $s^{3}$ amplitude for dilaton as an example,
but the ansatz should be modified. Local operator remains unchanged.
Now the numerator sitting on the poles are not necessarily two lower
four-point amplitudes, but any $s^{4}$ independent polynomials which
respect locality. For example, the $s^{4}$ polynomial sitting on
$s_{123}$ pole must respect $(1,2,3)$ permutation invariance and
$(4,5,6)$ permutation invariance. We find out there are 11 independent
polynomials satisfying such condition, which means there are 11 unknown
coefficients to be solved. Now the six-point ansatz looks like:

\begin{equation}
\widetilde{M}_{6\_\text{ansatz}}^{(3)}=c_{6p,1}(s_{12}^{3}+\mathcal{P}_{6})+c_{6p,2}(s_{123}^{3}+\mathcal{P}_{6})+\sum_{i=1}^{11}c_{6f,i}\left(\frac{1}{s_{123}}F_{(1,2,3)(4,5,6)}+\mathcal{P}_{6}\right)\label{eq:M6s3}
\end{equation}
where $\widetilde{M}_{\text{n\_ansatz}}$ denotes ansatz without assuming
factorization, and $F_{(1,2,3)(4,5,6)}$ denote $s^{4}$ polynomial
with $(1,2,3)$ permutation invariance and $(4,5,6)$ permutation
invariance. We apply soft leading($\tau^{0}$) and sub-leading($\tau^{1})$
soft theorem and solve for the coefficients. The result is
\begin{align}
M_{6}^{(3)} & =-c_{5}^{(3)}(s_{12}^{3}+\mathcal{P}_{6})-\left(\frac{c_{5}^{(3)}}{2}+c_{6f}\right)(s_{123}^{3}+\mathcal{P}_{6})\nonumber \\
 & +c_{6f}\left((s_{12}^{2}+s_{13}^{2}+s_{23}^{2})\frac{1}{s_{123}}(s_{45}^{2}+s_{46}^{2}+s_{56}^{2})+\mathcal{P}_{6}\right)\label{eq:M6_dilaton_sol_noU}
\end{align}
We find that this solution indeed factorize to lower four-point amplitudes
on poles, that is, the functional form of $F$ reduces to that given
by Feynman diagrams, 
\begin{equation}
F_{(1,2,3)(4,5,6)}=(s_{12}^{2}+s_{13}^{2}+s_{23}^{2})(s_{45}^{2}+s_{46}^{2}+s_{56}^{2})\label{eq:factorizes}
\end{equation}
 which means that unitarity is an emergent property. However, the
coefficient $c_{6f}$, which we knew to be related to $\left(c_{4}^{(2)}\right)^{2}$
when unitarity was assumed, is still undetermined. Therefore, while
the functional form of six-point amplitude is fixed, some coefficients
cannot be related to lower point coefficients. Therefore, there is
still some information left out, compared to the case where unitarity
is imposed together with single soft theorems. 

Once the six-point $s^{3}$ amplitude is fixed (not completely but
to a certain extent), we feed it as an input to determine the seven-point
$s^{3}$ amplitude. For seven-point $s^{3}$ amplitude, there are
41 independent polynomials respect $(1,2,3)$ and $(4,5,6,7)$ permutation
invariance sitting on the $s_{123}$ pole, then we sum over 35 channels
and add 2 contact terms, which respect $\mathcal{P}7$ permutation
invariance. 

\begin{equation}
\widetilde{M}_{7\_\text{ansatz}}^{(3)}=c_{7p,1}(s_{12}^{3}+\mathcal{P}_{7})+c_{7p,2}(s_{123}^{3}+\mathcal{P}_{7})+\sum_{i=1}^{41}c_{7f,i}\left(\frac{1}{s_{123}}F_{(1,2,3)(4,5,6,7)}+\mathcal{P}_{7}\right)\label{eq:M7s3}
\end{equation}
where $F_{(1,2,3)(4,5,6,7)}$ denote $s^{4}$ polynomial with $(1,2,3)$
permutation invariance and $(4,5,6,7)$ permutation invariance. The
solution can factorize on poles as well, but still some coefficients
cannot be expressed in terms of lower point coefficients. 

Now we can determine the eight-point $s^{3}$ amplitude with seven-point
$s^{3}$ amplitude as an input. We find the eight-point $s^{3}$ amplitude
can also has a factorized solution. There are 51 polynomials on the
$s_{123}$ pole, which respect $(1,2,3)$ and $(4,5,6,7,8)$ permutation
invariance. There are 36 polynomials on the $s_{1234}$ pole, which
respect $(1,2,3,4)$ and$(5,6,7,8)$ permutation invariance. There
are 2 contact terms respecting $\mathcal{P}8$ permutation invariance.
\begin{align}
\widetilde{M}_{8\_\text{ansatz}}^{(3)}= & c_{8p,1}(s_{12}^{3}+\mathcal{P}_{8})+c_{8p,2}(s_{123}^{3}+\mathcal{P}_{8})+\sum_{i=1}^{51}c_{8f,i}\left(\frac{1}{s_{123}}F_{(1,2,3)(4,5,6,7,8)}+\mathcal{P}_{8}\right)\nonumber \\
 & +\sum_{i=1}^{36}\bar{c}_{8f,i}\left(\frac{1}{s_{1234}}\bar{F}_{(1,2,3,4)(5,6,7,8)}+\mathcal{P}_{8}\right)
\end{align}
where $F_{(1,2,3)(4,5,6,7,8)}$ denote $s^{4}$ polynomial with $(1,2,3)$
permutation invariance and $(4,5,6,7,8)$ permutation invariance,
$\bar{F}_{(1,2,3,4)(5,6,7,8)}$ denote $s^{4}$ polynomial with $(1,2,3,4)$
permutation invariance and $(5,6,7,8)$ permutation invariance. We
have checked the solution can factorize when $s_{123}$ or $s_{1234}$
go on shell.We expect higher point dilaton $s^{3}$ amplitude can
have factorized solution after imposing leading and sub-leading soft
theorem. But for higher s power ($s\geq4$), we cannot have factorized
solution.

This procedure can also apply to cDBI $s^{3}$ amplitude. For six-point,
we use the same $\widetilde{M}_{6\_\text{ansatz}}^{(3)}$ in \eqref{eq:M6s3}
but apply vanishing soft theorem since there is no five- point $s^{3}$
amplitude for cDBI. Again we find the unique answer which can serve
as an input for determine seven-point $s^{3}$ amplitude. For seven-point,
we use $\widetilde{M}_{7\_\text{ansatz}}^{(3)}$ in \eqref{eq:M7s3}
and apply soft theorem relating it to the cDBI six-point $s^{3}$
amplitude that we just obtained. We do this procedure recursively
up to eight-point. 

To sum up, since the factorization form of amplitude can be enforced
by single soft theorems and locality, we can see that unitarity can
be an emergent property. However, in terms of fixing amplitudes, the
coefficients of factorization channels are left undetermined, as opposed
to the case when unitarity is assumed beforehand. Therefore, single
soft theorems and locality do not provide complete information on
amplitudes. 

We have also used the same amplitude fixing approach on DBI amplitude.
The situation is identical to dilaton amplitudes. For the $s^{3}$
terms for six point amplitude, the solution is 
\begin{equation}
M_{6\text{\_DBI}}^{(3)}=c_{6f}\left[-(s_{123}^{3}+\mathcal{P}_{6})+\left((s_{12}^{2}+s_{13}^{2}+s_{23}^{2})\frac{1}{s_{123}}(s_{45}^{2}+s_{46}^{2}+s_{56}^{2})+\mathcal{P}_{6}\right)\right]\label{eq:M6_DBI_sol_noU}
\end{equation}
where the factorization form is enforced, but the coefficient $c_{6f}$
remains undetermined. That factorization form is enforced reproduces
the result in \cite{DBI_locality,DBI_locality2}. 

\subsection{Applying double soft theorems to fix amplitudes}

We have shown single soft theorem can be applied to fix amplitudes.
Actually we can use double soft theorem to do the same thing. One
interesting question is whether double soft theorem gives more constraints
than single soft theorem. Once the four-point vertex is specified
beforehand, the answer is no. This is reasonable since we have shown
that double soft theorem can be derived from single soft theorem and
the four-point vertex. We test both type A and type B double soft
theorems. When applying numerics explicitly for type A, the two soft
momenta are rescales by the same factor ($p\rightarrow\tau p$, $q\rightarrow\tau q$);
for type B, they are rescaled by different factors ($p\rightarrow\tau_{p}p$,
$q\rightarrow\tau_{q}q$). If we need the first order ($\tau^{1}$)
order coefficient, for example, we extract the coefficient of $\tau^{1}$
for type B or coefficient of ($\tau_{p}^{1}$$,\tau_{q}^{0}$) and
($\tau_{p}^{0}$$,\tau_{q}^{1}$) for type A. We also use the two
kinds of ansatz, either unitarity is assumed or not, e.g. \eqref{eq:M6s3_U}
and \eqref{eq:M6s3} for six-point amplitudes, and compare the results. 

\subsubsection{DBI}

Here we show how to apply double soft theorem to fix DBI six-point
$s^{3}$ amplitude. Here we use both kinds of 6-point $s^{3}$ ansatz,
\ref{eq:M6s3_U} and \ref{eq:M6s3}. The input is DBI four-point amplitude
$A_{4}=c_{4}(s_{12}^{2}+s_{13}^{2}+s_{23}^{2})$ . Now we apply both
kinds of double soft theorems, \eqref{eq:type_A_0} and \eqref{eq:DBI_B_tree}
directly, 

\begin{align}
\tau^{0}:\left.M_{6\_\text{ansatz}}^{(3)}\right|_{\mathcal{O}(\tau^{0})}= & 0\nonumber \\
\tau^{1}:\left.M_{6\_\text{ansatz}}^{(3)}\right|_{\mathcal{O}(\tau^{1})}= & \left.\tau S^{(1)}M_{4}\right|_{\mathcal{O}(\tau^{1})}\nonumber \\
\tau^{2}:\left.M_{6\_\text{ansatz}}^{(3)}\right|_{\mathcal{O}(\tau^{2})}= & \left.\left(\tau S^{(1)}+\tau^{2}S^{(2)}\right)M_{4}\right|_{\mathcal{O}(\tau^{2})}\nonumber \\
\tau^{3}:\left.M_{6\_\text{ansatz}}^{(3)}\right|_{\mathcal{O}(\tau^{3})}= & \left.\left(\tau\mathcal{S}^{(1)}+\tau^{2}\mathcal{S}^{(2)}+\tau^{3}\mathcal{S}^{(3)}\right)M_{4}\right|_{\mathcal{O}(\tau^{3})}
\end{align}
and compare the results. 

Using the ansatz \eqref{eq:M6s3_U} where unitarity is assumed (factorization
form), both type A and type B give a unique solution for all coefficients
after imposing soft constraint up to $\tau^{2}$ order, which is identical
to that fixed by single soft theorems. When using \eqref{eq:M6s3}
where only locality is assumed, the situation is different. Type A
theorems gives the result \eqref{eq:M6_DBI_sol_noU}, where factorization
form is enforced but the coefficient $c_{6f}$ is undetermined, identical
to the case of single soft theorems. This indicates that type A theorems
are equivalent to single soft theorems, consistent with our derivation.
Type B theorems, however, gives the result \eqref{eq:M6_DBI_sol_U},
where the coefficient $c_{6f}$ is also solved, so that the amplitude
is completely fixed in terms of lower point ones. Therefore, type
B double soft theorems indeed contain more information, as expected. 

\subsubsection{Dilaton}

For dilaton we discuss general ($n+2$)-point amplitude. We write
down the ansatz $M_{n+2\_\text{ansatz}}$ assuming factorization first,
then apply type A and type B double soft theorems, \eqref{eq:dilaton_A}
and \eqref{eq:cDBI_B_tree}, to relate (n+2)-point and n-point amplitudes,

\begin{align}
\tau^{0}:\left.M_{n+2\_\text{ansatz}}\right|_{\mathcal{O}(\tau^{0})}= & \left.\mathcal{S}^{(0)}M_{n}\right|_{\mathcal{O}(\tau^{0})}\nonumber \\
\tau^{1}:\left.M_{n+2\_\text{ansatz}}\right|_{\mathcal{O}(\tau^{1})}= & \left.\left(\mathcal{S}^{(0)}+\tau\left(p+q\right)^{\mu}\mathcal{S_{\mu}}^{(1)}\right)M_{n}\right|_{\mathcal{O}(\tau^{1})}
\end{align}

Whether unitarity is assumed or not, neither type A nor type B theorems
can fix the amplitudes completely. More specifically, the coefficients
$c_{6,p1}$ and $c_{6,p2}$ \eqref{eq:M6s3_U} and \eqref{eq:M6s3}
would be related to each other, but still not completely fixed in
terms of lower point amplitudes. From our analytical derivation in
subsection \ref{subsec:type_A_cDBI}, it is a reasonable result. There,
we found that single soft theorems gives a piece $\left(p\cdot S_{n+1}^{(1)}\right)\left(q\cdot S_{n}^{(1)}\right)$,
giving partial information on order $\left(p^{1},q^{1}\right)$. However
this piece is discarded when forming type A double soft theorems,
since other contribution to order $\tau^{2}$ can not be obtained.
Thus, type A theorems contains less information than single soft theorems.
Since the derivation of type B comes from type A, this information
also left out in type B theorems. 

When unitarity is not assumed, unitarity (factorization form) can be enforced by both type A and type B. 
However, the coefficient $c_{6f}$ can only be fixed by type B double soft theorems. 
Therefore, type B double soft theorems indeed contain some additional information, while type A theorems simply contain less information than single soft theorems. This is also consistent
with our analytical derivation. 

\section{Conclusion and Outlook\label{sec:conclusion}}

In this article, we provide a theory-independent procedure to derive double soft theorems and identify their connection to symmetry. 
The procedure stems from combining some known relations among single and double soft theorems. 
First, there are two kinds of double soft theorems, type A and B,
with different expansion schemes;
second, type A can be derived from single soft theorems; 
third, type B can be obtained from type A theorems by adding the information of four-point vertex. 
The first fact explains the obstacle to derive double soft theorem from current algebra. 
To address this problem, we combine the latter two relations into a theory-independent way to work out double soft theorems from single soft theorems. 
This helps us clarify how all these kinds of theorems are related
to symmetry. Type A theorems are directly implied by single soft theorems,
whereas type B theorems can be obtained from type A using information
from four-point vertex. Explicit formulas have been presented for arbitrary single scalar theory, but the procedure is also applicable to more general theories as well. 

Using our procedure, we have reproduced all double soft theorems
of Dirac-Born-Infield, dilaton and special Galileon theory, some missing
in previous works using current algebra, as well as deriving a new
subleading theorem for dilaton. Since this approach is based on single
soft theorems, which have been derived using current algebra, the
symmetry basis of those double soft theorems is established. We have
also provided a technical explanation of why DBI possess double soft
theorems up to higher order than dilaton, despite having similar structure
of current algebra.

Our formulation also provides a clear-cut way to determine whether
a double soft theorem withstands loop correction. Since type A double
soft theorems can be derived solely from single soft theorems, they
are protected as long as the symmetry is not anomalous. On the other
hand, type B are related to type A by the explicit form of four-point
vertex, so they may be modified, depending on the characteristics
of loop correction terms of four-point vertex. Using this procedure
we have clarified the behavior of double soft theorems of DBI and
dilaton under loop corrections, where discussions in literatures are
limited to conjectures.

Apart from solving issues of double soft theorems, our relation clarifies
whether double soft theorems contain more information than single
soft theorems. In principle, type A theorems contain no more information
than single soft theorems, while type B theorems contain additional
information from four-point vertex, which includes the field-dependent
part of the nonlinear symmetry transformation. However, for theories
with non-vanishing single soft theorems, some technicality of grouping
expansion orders prevent double soft theorems from fully incorporating
the information from single soft theorems. Therefore, double soft
theorems may contain less information in this case. To test this,
we separately apply single and double soft theorems to constrain the
form of amplitudes numerically. 
We investigate whether higher-point amplitude can be completely fixed by lower-point ones, 
and whether unitarity (factorization) can emerge from locality (the presence of Feynman propagators), 
by imposing soft theorems. 
For DBI and special Galileon amplitudes, with vanishing single soft theorems, 
both single and double soft theorems can enforce unitarity and fix all other coefficients in terms of the coefficient of factorization channels. 
This last coefficient, however, can only be fixed by type B double soft theorems. 
For dilaton, with non-vanishing single soft limits, single soft theorem and locality can force
unitarity and fix all coefficients except for that of factorization channels for $s^{3}$ amplitudes. 
Both type A and type B double soft theorem can enforce unitarity, but the coefficient of factorization channel can only be fixed by type B, while both give less constraint on the remaining coefficients. 
These numerical tests further corroborate our analytical result. 

We also show how to extend our result to multiple soft theorems, which is straightforward. The distinction of expansion schemes, the derivation procedure, and the modification by loop correction can all be directly applied on multiple soft theorems. 

It would be interesting to apply our method to a generalized kind of soft theorems, where
the soft limit of an amplitude reduces to an extended theory with additional fields, instead of the original theory. 
Some theories with vanishing single soft limit possess this kind of extended soft theorems at higher soft order, as shown in \cite{soft_extended} by CHY representation. 
For example, for nonlinear sigma model (NLSM) the usual single soft theorem is the Adler's zero at leading order, but at subleading order there is actually an extended soft theorem. 
The situation is similar for Dirac-Born-Infield and special Galileon theories. 
Recently,  it was shown that for NLSM, such extended theorem comes from the field-dependent part of  the symmetry transformation, and is valid at all loop level \cite{NLSM_phi3}. 
Our formulation can be used in these cases to obtain double soft theorems. 
However, the second soft expansion in the derivation of type A theorems in \eqref{eq:type_A_expand} requires the single soft theorem for the extended theory, which has not been derived yet. 
Also, deriving type B theorems requires vertices involving the new fields in the extended theory, in addition to the original ones. 
Therefore, if double soft theorems can be obtained, they may incorporate further information about the extended theory. 
We leave this task for the future. 

\section*{Acknowledgements}
We would like to thank Yu-tin Huang, and Congkao Wen for suggesting the problem and subsequent discussions. We also like to thank Chuan-Tsung Chan for useful comments. The work of Z. Z. Li,  H. H. Lin, and S. Q. Zhang are supported by Taiwan MoST grant 106-2628-M-002 -012 -MY3.

\appendix

\section{Type A and Type B Expansions for Pole Diagrams\label{sec:type_a_calc}}

\subsection{$s^{2}$ terms}

Type B expansion for pole diagrams is straight forward. For a four
point vertex, 
\begin{equation}
M_{4,s^{2}}(k_{1},k_{2},k_{3},k_{4})=c_{4}^{(2)}\left(s^{2}+t^{2}+u^{2}\right)
\end{equation}
The pole diagrams together have the form 
\begin{equation}
\sum_{i}\tilde{M}_{4,s^{2}}\left(k_{i},p,q\right)\frac{1}{\ell_{i}}\tilde{M}_{n}\left(\cdots,\ell_{i},\cdots\right)=\sum_{i}2\frac{c_{4}^{(2)}\left((k_{i}\cdot p)^{2}+(k_{i}\cdot q)^{2}+(p,q)^{2}\right)}{\left[k_{i}\cdot\left(p+q\right)+p\cdot q\right]}\tilde{M}_{n}\left(\cdots k_{i}+p+q,\cdots\right)
\end{equation}
We first separate out the four point vertex and the propagator, 
\begin{equation}
2\frac{c_{4}^{(2)}\left((k_{i}\cdot p)^{2}+(k_{i}\cdot q)^{2}+(p,q)^{2}\right)}{\left[k_{i}\cdot\left(p+q\right)+p\cdot q\right]}
\end{equation}
Its expansion in $\tau$ can be expressed as
\begin{align}
 & \frac{\left((k_{i}\cdot p)^{2}+(k_{i}\cdot q)^{2}+(p\cdot q)^{2}\right)}{\left[k_{i}\cdot\left(p+q\right)+p\cdot q\right]}=0 &  & \tau^{0} \nonumber \\
 & \hspace{0.4in}+\frac{\left((k_{i}\cdot p)^{2}+(k_{i}\cdot q)^{2}\right)}{\left[k_{i}\cdot\left(p+q\right)\right]} &  & \tau^{1} \nonumber \\
 & \hspace{0.4in}-\frac{\left((k_{i}\cdot p)^{2}+(k_{i}\cdot q)^{2}\right)}{\left[k_{i}\cdot\left(p+q\right)\right]^{2}}\left(p\cdot q\right) &  & \tau^{2} \nonumber \\
 & \hspace{0.4in}+\frac{(p\cdot q)^{2}}{\left[k_{i}\cdot\left(p+q\right)\right]}+\frac{\left((k_{i}\cdot p)^{2}+(k_{i}\cdot q)^{2}\right)}{\left[k_{i}\cdot\left(p+q\right)\right]^{3}}\left(p\cdot q\right)^{2} &  & \tau^{3}
\end{align}
On the other hand, expansion of $M_{n}$ is

\begin{equation}
M_{n}\left(\cdots,k_{i},\cdots\right)+\left(p+q\right)\cdot\frac{\partial}{\partial k_{i}}M_{n}\left(\cdots,k_{i},\cdots\right)+\frac{1}{2}\left[\left(p+q\right)\cdot\frac{\partial}{\partial k_{i}}\right]^{2}M_{n}\left(\cdots,k_{i},\cdots\right)\label{eq:Mn_soft-1}
\end{equation}

Therefore the type B expansion of pole diagrams can be expanded as
follows,

\begin{align}
 & \left.M_{\text{pole},s^{2}}\right|_{\text{type B}}=0 &  & \tau^{0}\nonumber \\
 & \hspace{0.3in}+2c_{4}^{(2)}\sum_{i}\frac{\left((k_{i}\cdot p)^{2}+(k_{i}\cdot q)^{2}\right)}{\left[k_{i}\cdot\left(p+q\right)\right]}M_{n}\left(\cdots k_{i},\cdots\right) &  & \tau^{1}\nonumber \\
 & \begin{aligned}\hspace{0.3in}+2c_{4}^{(2)}\sum_{i} & \left\{ \frac{\left((k_{i}\cdot p)^{2}+(k_{i}\cdot q)^{2}\right)}{\left[k_{i}\cdot\left(p+q\right)\right]}\left(p+q\right)\cdot\frac{\partial}{\partial k_{i}}\right.\\
 & \left.\frac{\left((k_{i}\cdot p)^{2}+(k_{i}\cdot q)^{2}\right)}{\left[k_{i}\cdot\left(p+q\right)\right]^{2}}\left(p\cdot q\right)\right\} M_{n}\left(\cdots k_{i},\cdots\right)
\end{aligned}
 &  & \tau^{2}\nonumber \\
 & \begin{aligned}\hspace{0.3in}+2c_{4}^{(2)}\sum_{i} & \left\{ \frac{(k_{i}\cdot p)^{2}+(k_{i}\cdot q)^{2}}{2k_{i}\cdot\left(p+q\right)}\left[\left(p+q\right)\cdot\frac{\partial}{\partial k_{i}}\right]^{2}\right.\\
 & -\frac{\left((k_{i}\cdot p)^{2}+(k_{i}\cdot q)^{2}\right)}{\left[k_{i}\cdot\left(p+q\right)\right]^{2}}\left(p\cdot q\right)\left(p+q\right)\cdot\frac{\partial}{\partial k_{i}}+\frac{(p\cdot q)^{2}}{\left[k_{i}\cdot\left(p+q\right)\right]}\\
 & \left.+\frac{\left((k_{i}\cdot p)^{2}+(k_{i}\cdot q)^{2}\right)}{\left[k_{i}\cdot\left(p+q\right)\right]^{3}}\left(p\cdot q\right)^{2}\right\} M_{n}\left(\cdots,k_{i},\cdots\right)
\end{aligned}
 &  & \tau^{3}\label{eq:pole_expand-1}
\end{align}

Type A expansion requires double-variable Taylor expansion. We first
try to obtain as much terms as possible: 
\begin{align}
M_{\text{pole},s^{2}}\left(\cdots,0,0\right)= & 0\nonumber \\
q\cdot\frac{\partial}{\partial q}M_{\text{pole},s^{2}}\left(\cdots,0,q\right)|_{q=0}= & \sum_{i}2c_{4}^{(2)}k_{i}\cdot q\left[M_{n}\left(\cdots,k_{i},\cdots\right)\right]\label{eq:gut_p0q1}\\
\frac{1}{2}q^{2}\frac{\partial^{2}}{\partial q^{2}}M_{\text{pole},s^{2}}\left(\cdots,0,q\right)|_{q=0}= & 2\sum_{i}c_{4}^{(2)}(k_{i}\cdot q)\left[q\cdot\frac{\partial}{\partial k_{i}}\tilde{M}_{n}\left(\cdots,k_{i},\cdots\right)\right]\label{eq:gut_p0q2}\\
\frac{1}{3!}q^{3}\frac{\partial^{3}}{\partial q^{3}}M_{\text{pole},s^{2}}\left(\cdots,0,q\right)|_{q=0}= & 2\sum_{i}c_{4}^{(2)}(k_{i}\cdot q)\frac{1}{2}\left[\left(q\cdot\frac{\partial}{\partial k_{i}}\right)^{2}\tilde{M}_{n}\left(\cdots,k_{i},\cdots\right)\right]\label{eq:gut_p0q3}
\end{align}
\begin{align}
 & \left.\left(\left.p\frac{\partial}{\partial p}M_{\text{pole},s^{2}}\left(\cdots,p,q\right)\right|_{p=0}\right)\right|_{q=0}=-2c_{4}^{(2)}\sum_{i}\left(k_{i}\cdot p\right)M_{n}\left(\cdots\ell_{i},\cdots\right)\label{eq:gut_p1q0}\\
 & \left.\left(q\cdot\frac{\partial}{\partial q}\right)\left(p\cdot\frac{\partial}{\partial p}\right)M_{\text{pole},s^{2}}\left(\cdots,p,q\right)\right|_{p=q=0}\nonumber \\
 & \hspace{0.2in}=-2c_{4}^{(2)}\sum_{i}\left(p\cdot q+(k_{i}\cdot p)q\frac{\partial}{\partial k_{i}}-(k_{i}\cdot q)p\frac{\partial}{\partial k_{i}}\right)M_{n}\left(\cdots,k_{i},\cdots\right)\\
 & \frac{1}{2}\left.\left(q\cdot\frac{\partial}{\partial q}\right)\left(q\cdot\frac{\partial}{\partial q}\right)\left(p\cdot\frac{\partial}{\partial p}\right)M_{\text{pole},s^{2}}\left(\cdots,p,q\right)\right|_{p=q=0}\nonumber \\
 & \hspace{0.2in}\begin{aligned}=-2c_{4}^{(2)}\sum_{i} & \left[\left(p\cdot q\right)\left(q\cdot\frac{\partial}{\partial k_{i}}\right)+\frac{1}{2}(k_{i}\cdot p)\left(q\cdot\frac{\partial}{\partial k_{i}}\right)^{2}-(k_{i}\cdot q)\left(p\cdot\frac{\partial}{\partial k_{i}}\right)\left(q\cdot\frac{\partial}{\partial k_{i}}\right)\right]M_{n}\left(\cdots,k_{i},\cdots\right)\end{aligned}
\end{align}
\begin{align}
p\cdot\frac{\partial}{\partial p}M_{\text{pole},s^{2}}\left(\cdots,p,0\right)|_{p=0}= & \sum_{i}2c_{4}^{(2)}\left(k_{i}\cdot p\right)M_{n}\left(\cdots,k_{i},\cdots\right)\\
\left.\left(\left.q\frac{\partial}{\partial q}M_{\text{pole},s^{2}}\left(\cdots,p,q\right)\right|_{q=0}\right)\right|_{p=0}= & -2c_{4}^{(2)}\sum_{i}\left(k_{i}\cdot q\right)M_{n}\left(\cdots\ell_{i},\cdots\right)\\
\frac{1}{2}p^{2}\frac{\partial^{2}}{\partial p^{2}}M_{\text{pole},s^{2}}\left(\cdots,0,p\right)|_{p=0}= & -2\sum_{i}c_{4}^{(2)}(k_{i}\cdot p)\left[p\cdot\frac{\partial}{\partial k_{i}}\tilde{M}_{n}\left(\cdots,p_{i},\cdots\right)\right]\\
\frac{1}{3!}p^{3}\frac{\partial^{3}}{\partial p^{3}}M_{\text{pole},s^{2}}\left(\cdots,p,0\right)|_{p=0}= & -2\sum_{i}\frac{1}{2}c_{4}^{(2)}(k_{i}\cdot p)\left[\left(p\cdot\frac{\partial}{\partial k_{i}}\right)^{2}\tilde{M}_{n}\left(\cdots,k_{i},\cdots\right)\right]
\end{align}
\begin{align}
 & \left.\left(p\cdot\frac{\partial}{\partial p}\right)\left(q\cdot\frac{\partial}{\partial q}\right)M_{\text{pole},s^{2}}\left(\cdots,p,q\right)\right|_{p=q=0}\nonumber \\
 & \hspace{0.3in}=2c_{4}^{(2)}\sum_{i}\left(p\cdot q+(k_{i}\cdot q)p\frac{\partial}{\partial k_{i}}-(k_{i}\cdot p)q\frac{\partial}{\partial k_{i}}\right)M_{n}\left(\cdots,k_{i},\cdots\right)\\
 & \frac{1}{2}\left.\left(p\cdot\frac{\partial}{\partial p}\right)\left(p\cdot\frac{\partial}{\partial p}\right)\left(q\cdot\frac{\partial}{\partial q}\right)M_{\text{pole},s^{2}}\left(\cdots,p,q\right)\right|_{p=q=0}\nonumber \\
 & \hspace{0.3in}=2c_{4}^{(2)}\sum_{i}\left[\left(p\cdot q\right)\left(p\cdot\frac{\partial}{\partial k_{i}}\right)+\frac{1}{2}(k_{i}\cdot q)\left(p\frac{\partial}{\partial k_{i}}\right)^{2}-(k_{i}\cdot p)\left(p\frac{\partial}{\partial k_{i}}\right)\left(q\frac{\partial}{\partial k_{i}}\right)\right]M_{n}\left(\cdots,k_{i},\cdots\right)
\end{align}
We have obtained inequivalent terms involving mixed derivatives of
different orderings. Apparently we would have trouble trying combine
them. However, the difference between those terms actually contribute
to the next order. We take the terms $\left(q\cdot\frac{\partial}{\partial q}\right)\left(p\cdot\frac{\partial}{\partial p}\right)M_{\text{pole},s^{2}}$
and $\left(p\cdot\frac{\partial}{\partial p}\right)\left(q\cdot\frac{\partial}{\partial q}\right)M_{\text{pole},s^{2}}$
as an example. When we performed our expansion, the momenta $k_{i}$
are treated as objects of order $\left(p^{0},q^{0}\right)$. However,
since momentum conservation must be maintained, some of them must
carry dependence on $p$, $q$. Thus, our expression for $\left(q\cdot\frac{\partial}{\partial q}\right)\left(p\cdot\frac{\partial}{\partial p}\right)M_{\text{pole},s^{2}}$
actually contributes to order $\left(p^{1},q^{2}\right)$ as well
as the obvious $\left(p^{1},q^{1}\right)$. Similarly, $\left(p\cdot\frac{\partial}{\partial p}\right)\left(q\cdot\frac{\partial}{\partial q}\right)M_{\text{pole},s^{2}}$
contributes to next order, but to $\left(p^{2},q^{1}\right)$ instead
of $\left(p^{1},q^{2}\right)$, leading to inequivalence. Therefore,
the two mixed derivative terms should be combined according to the
coefficient of $\left(p^{2},q^{1}\right)$ and $\left(p^{1},q^{2}\right)$
dictated by double-variate Taylor expansions. Other ordering ambiguities
can be resolved in a similar fashion.

Collecting all the results, we can evaluate the difference between
type A and type B expansion of pole diagrams, 
\begin{align}
 & \left.\left[\left.M_{\text{pole},s^{2}}\left(\cdots,\tau p,\tau q\right)\right|_{\text{type B}}-\left.M_{\text{pole},s^{2}}\left(\cdots,\tau p,\tau q\right)\right|_{\text{type A}}\right]\right|_{\tau^{0}}=0-0=0 \nonumber \\
 & \left.\left[\left.M_{\text{pole},s^{2}}\left(\cdots,\tau p,\tau q\right)\right|_{\text{type B}}-\left.M_{\text{pole},s^{2}}\left(\cdots,\tau p,\tau q\right)\right|_{\text{type A}}\right]\right|_{\tau^{1}} \nonumber \\
= & \sum_{i}\left[\frac{-2c_{4}^{(2)}k_{i}\cdot\left(p+q\right)+2c_{4}k_{i}\cdot\left(p+q\right)}{2}+\frac{2c_{4}^{(2)}\left[(k_{i}\cdot p)^{2}+(k_{i}\cdot q)^{2}\right]}{k_{i}\cdot\left(p+q\right)}\right]M_{n}\left(\cdots,p_{i},\cdots\right) \nonumber \\
= & 2c_{4}^{(2)}\sum_{i}\left[\frac{1}{2}\frac{\left[k_{i}\cdot\left(p-q\right)\right]^{2}}{k_{i}\cdot\left(p+q\right)}+\frac{1}{2}k_{i}\cdot\left(p+q\right)+\frac{k_{i}\cdot\left(p-q\right)}{k_{i}\cdot\left(p+q\right)}\left(p_{\mu}^{\text{ }}q_{\nu}J_{i}^{\mu\nu}\right)\right]M_{n}\left(\cdots,k_{i},\cdots\right)
\end{align}
Second order can be obtained, 
\begin{align}
 & \left.\left[\left.M_{\text{pole},s^{2}}\left(\cdots,\tau p,\tau q\right)\right|_{\text{type B}}-\left.M_{\text{pole},s^{2}}\left(\cdots,\tau p,\tau q\right)\right|_{\text{type A}}\right]\right|_{\tau^{2}} \nonumber \\
= & 2c_{4}^{(2)}\sum_{i}\left[-\frac{\left((k_{i}\cdot p)^{2}+(k_{i}\cdot q)^{2}\right)}{\left[k_{i}\cdot\left(p+q\right)\right]^{2}}+1\right]\left(p\cdot q\right)M_{n}\left(\cdots k_{i},\cdots\right) \nonumber \\
 & -2c_{4}^{(2)}\sum_{i}(k_{i}\cdot q)\left[q\cdot\frac{\partial}{\partial k_{i}}\tilde{M}_{n}\left(\cdots,k_{i},\cdots\right)\right]-2c_{4}^{(2)}\sum_{i}(k_{i}\cdot p)\left[p\cdot\frac{\partial}{\partial k_{i}}\tilde{M}_{n}\left(\cdots,k_{i},\cdots\right)\right] \nonumber \\
 & +\sum_{i}2\frac{c_{4}^{(2)}\left((k_{i}\cdot p)^{2}+(k_{i}\cdot q)^{2}\right)}{\left[k_{i}\cdot\left(p+q\right)\right]}\left(p+q\right)\cdot\frac{\partial}{\partial k_{i}}M_{n}\left(\cdots k_{i},\cdots\right) \nonumber \\
= & 2c_{4}^{(2)}\sum_{i}\frac{2\left(k_{i}\cdot p\right)\left(k_{i}\cdot q\right)}{\left[k_{i}\cdot\left(p+q\right)\right]^{2}}\left(p\cdot q\right)M_{n}\left(\cdots k_{i},\cdots\right) \nonumber \\
 & +2c_{4}^{(2)}\sum_{i}\left[\frac{(k_{i}\cdot q)^{2}-\left(k_{i}\cdot p\right)\left(k_{i}\cdot q\right)}{\left[k_{i}\cdot\left(p+q\right)\right]}\left(p\cdot\frac{\partial}{\partial k_{i}}\right)+\frac{(k_{i}\cdot p)^{2}-\left(k_{i}\cdot q\right)\left(k_{i}\cdot p\right)}{\left[k_{i}\cdot\left(p+q\right)\right]}\left(q\cdot\frac{\partial}{\partial k_{i}}\right)\right]M_{n}\left(\cdots k_{i},\cdots\right) \nonumber \\
= & 2c_{4}^{(2)}\sum_{i}\left\{ \frac{2\left(k_{i}\cdot p\right)\left(k_{i}\cdot q\right)}{\left[k_{i}\cdot\left(p+q\right)\right]^{2}}\left(p\cdot q\right)+\frac{k_{i}\cdot\left(p-q\right)}{k_{i}\cdot\left(p+q\right)}\left(p_{\mu}^{\text{ }}q_{\nu}J_{i}^{\mu\nu}\right)\right\} M_{n}\left(\cdots k_{i},\cdots\right)
\end{align}
For third order, the terms without derivatives 
\begin{align}
 & 2c_{4}^{(2)}\sum_{i}\left[\frac{(p\cdot q)^{2}}{\left[k_{i}\cdot\left(p+q\right)\right]}+\frac{\left((k_{i}\cdot p)^{2}+(k_{i}\cdot q)^{2}\right)}{\left[k_{i}\cdot\left(p+q\right)\right]^{3}}\left(p\cdot q\right)^{2}\right]M_{n}\left(\cdots,k_{i},\cdots\right)\nonumber \\
= & 2c_{4}^{(2)}\sum_{i}\frac{(p\cdot q)^{2}}{\left[k_{i}\cdot\left(p+q\right)\right]}\left(1+\frac{\left((k_{i}\cdot p)^{2}+(k_{i}\cdot q)^{2}\right)}{\left[k_{i}\cdot\left(p+q\right)\right]}\right)M_{n}\left(\cdots,k_{i},\cdots\right)\nonumber \\
= & 2c_{4}^{(2)}\sum_{i}\frac{(p\cdot q)^{2}}{\left[k_{i}\cdot\left(p+q\right)\right]}\left(\frac{3}{2}+\frac{\left[k_{i}\cdot\left(p-q\right)\right]{}^{2}}{\left[k_{i}\cdot\left(p+q\right)\right]^{2}}\right)M_{n}\left(\cdots,k_{i},\cdots\right)
\end{align}
terms with single derivative, 
\begin{align}
 & 2c_{4}^{(2)}\sum\left(1-\frac{\left((k_{i}\cdot p)^{2}+(k_{i}\cdot q)^{2}\right)}{\left[k_{i}\cdot\left(p+q\right)\right]^{2}}\right)_{i}\left(p\cdot q\right)\left[\left(p+q\right)\cdot\frac{\partial}{\partial k_{i}}\right]\nonumber \\
= & 2c_{4}^{(2)}\sum_{i}\frac{2\left(k_{i}\cdot p\right)\left(k_{i}\cdot q\right)}{\left[k_{i}\cdot\left(p+q\right)\right]^{2}}\left(p\cdot q\right)\left[\left(p+q\right)\cdot\frac{\partial}{\partial k_{i}}\right]M_{n}\left(\cdots,k_{i},\cdots\right)
\end{align}
terms with double derivative, 
\begin{align}
\hspace{0.2in}2c_{4}^{(2)}\sum_{i} & \left\{ \frac{1}{2}\left[k_{i}\cdot\left(p-q\right)\right]\left[\left(q\cdot\frac{\partial}{\partial k_{i}}\right)^{2}-\left(p\cdot\frac{\partial}{\partial k_{i}}\right)^{2}-\left[k_{i}\cdot\left(p+q\right)\right]\left(q\cdot\frac{\partial}{\partial k_{i}}\right)\left(p\cdot\frac{\partial}{\partial k_{i}}\right)\right]\right. \nonumber \\
 & \left.+\frac{1}{2}\frac{\left((k_{i}\cdot p)^{2}+(k_{i}\cdot q)^{2}\right)}{\left[k_{i}\cdot\left(p+q\right)\right]}\left[\left(p+q\right)\cdot\frac{\partial}{\partial k_{i}}\right]^{2}\right\} M_{n}\left(\cdots,k_{i},\cdots\right) \nonumber \\
=2c_{4}^{(2)}\sum_{i} & \frac{1}{2}\left\{ \frac{2\left(k_{i}\cdot p\right)^{2}\left(q\cdot\frac{\partial}{\partial k_{i}}\right)^{2}+2\left(k_{i}\cdot q\right)^{2}\left(p\cdot\frac{\partial}{\partial k_{i}}\right)^{2}}{k_{i}\cdot\left(p+q\right)}\right. \nonumber \\
 & \left.-\frac{4\left(k_{i}\cdot p\right)\left(k_{i}\cdot q\right)\left(q\cdot\frac{\partial}{\partial k_{i}}\right)\left(p\cdot\frac{\partial}{\partial k_{i}}\right)}{k_{i}\cdot\left(p+q\right)}\right\} M_{n}\left(\cdots,k_{i},\cdots\right) \nonumber \\
=2c_{4}^{(2)}\sum_{i} & \left\{ \frac{\left[\left(k_{i}\cdot p\right)\left(q\cdot\frac{\partial}{\partial k_{i}}\right)-\left(k_{i}\cdot q\right)\left(p\cdot\frac{\partial}{\partial k_{i}}\right)\right]^{2}}{k_{i}\cdot\left(p+q\right)}\right. \nonumber \\
 & \left.-\frac{2\left(k_{i}\cdot p\right)\left(k_{i}\cdot q\right)}{\left[k_{i}\cdot\left(p+q\right)\right]^{2}}\left(p\cdot q\right)\left[\left(p+q\right)\cdot\frac{\partial}{\partial k_{i}}\right]\right\} M_{n}\left(\cdots,k_{i},\cdots\right) \nonumber \\
=2c_{4}^{(2)}\sum_{i} & \left\{ \frac{\left(p_{\mu}q_{\nu}J^{\mu\nu}\right)^{2}}{k_{i}\cdot\left(p+q\right)}-\frac{2\left(k_{i}\cdot p\right)\left(k_{i}\cdot q\right)}{\left[k_{i}\cdot\left(p+q\right)\right]^{2}}\left(p\cdot q\right)\left[\left(p+q\right)\cdot\frac{\partial}{\partial k_{i}}\right]\right\} M_{n}\left(\cdots,k_{i},\cdots\right)
\end{align}
which gives 
\begin{align}
 & \left.\left[\left.M_{\text{pole},s^{2}}\left(\cdots,\tau p,\tau q\right)\right|_{\text{type B}}-\left.M_{\text{pole},s^{2}}\left(\cdots,\tau p,\tau q\right)\right|_{\text{type A}}\right]\right|_{\tau^{3}}\nonumber \\
= & 2c_{4}^{(2)}\sum_{i}\left\{ \frac{\left(p_{\mu}q_{\nu}J^{\mu\nu}\right)^{2}}{k_{i}\cdot\left(p+q\right)}\right.\left.+\frac{(p\cdot q)^{2}}{\left[k_{i}\cdot\left(p+q\right)\right]}\left(\frac{3}{2}+\frac{\left[k_{i}\cdot\left(p-q\right)\right]{}^{2}}{\left[k_{i}\cdot\left(p+q\right)\right]^{2}}\right)\right\} M_{n}\left(\cdots,k_{i},\cdots\right)
\end{align}

\subsection{$s^{3}$ terms}

The expansion for $M_{\text{pole},s^{3}}$ can be obtained in a similar
fashion. For a four point vertex, 
\begin{equation}
M_{4,s^{2}}(k_{1},k_{2},k_{3},k_{4})=c_{4}^{(3)}\left(s^{3}+t^{3}+u^{3}\right)
\end{equation}
The pole diagrams together have the form 
\begin{equation}
\sum_{i}\tilde{M}_{4,s^{3}}\left(k_{i},p,q\right)\frac{1}{\ell_{i}}\tilde{M}_{n}\left(\cdots,\ell_{i},\cdots\right)=4c_{4}^{(3)}\sum_{i}\frac{(k_{i}\cdot p)^{3}+(k_{i}\cdot q)^{3}+(p,q)^{3}}{\left[k_{i}\cdot\left(p+q\right)+p\cdot q\right]}\tilde{M}_{n}\left(\cdots k_{i}+p+q,\cdots\right)
\end{equation}
Type B expansion can be expressed as follows,
\begin{align}
 & \left.M_{\text{pole},s^{3}}\right|_{\text{type B}}=0 &  & \tau^{0},\tau^{1}\nonumber \\
 & \hspace{0.3in}+4c_{4}^{(3)}\sum_{i}\left[(k_{i}\cdot p)^{2}-(k_{i}\cdot p)(k_{i}\cdot q)+(k_{i}\cdot q)^{2}\right]M_{n}\left(\cdots k_{i},\cdots\right) &  & \tau^{2}\nonumber \\
 & \begin{aligned}\hspace{0.3in}+4c_{4}^{(3)}\sum_{i} & \left\{ \left[(k_{i}\cdot p)^{2}-(k_{i}\cdot p)(k_{i}\cdot q)+(k_{i}\cdot q)^{2}\right]\left(p+q\right)\cdot\frac{\partial}{\partial k_{i}}M_{n}\left(\cdots k_{i},\cdots\right)\right.\\
 & \left.-\frac{\left((k_{i}\cdot p)^{3}+(k_{i}\cdot q)^{3}\right)}{\left[k_{i}\cdot\left(p+q\right)\right]^{2}}\left(p\cdot q\right)\right\} M_{n}\left(\cdots k_{i},\cdots\right)
\end{aligned}
 &  & \tau^{3}\nonumber \\
 & \begin{aligned}\hspace{0.3in}+4c_{4}^{(3)}\sum_{i} & \left\{ \left[(k_{i}\cdot p)^{2}-(k_{i}\cdot p)(k_{i}\cdot q)+(k_{i}\cdot q)^{2}\right]\left[\left(p+q\right)\cdot\frac{\partial}{\partial k_{i}}\right]^{2}\right.\\
 & -\frac{\left((k_{i}\cdot p)^{3}+(k_{i}\cdot q)^{3}\right)}{\left[k_{i}\cdot\left(p+q\right)\right]^{2}}\left(p\cdot q\right)\left(p+q\right)\cdot\frac{\partial}{\partial k_{i}}+\frac{(p\cdot q)^{2}}{\left[k_{i}\cdot\left(p+q\right)\right]}\\
 & \left.+\frac{(k_{i}\cdot p)^{3}+(k_{i}\cdot q)^{3}}{\left[k_{i}\cdot\left(p+q\right)\right]^{3}}\left(p\cdot q\right)^{2}\right\} M_{n}\left(\cdots,k_{i},\cdots\right)
\end{aligned}
 &  & \tau^{4} \nonumber \\
 & \begin{aligned}\hspace{0.3in}+4c_{4}^{(3)}\sum_{i} & \left\{ \left[(k_{i}\cdot p)^{2}-(k_{i}\cdot p)(k_{i}\cdot q)+(k_{i}\cdot q)^{2}\right]\left[\left(p+q\right)\cdot\frac{\partial}{\partial k_{i}}\right]^{3}\right.\\
 & -\frac{\left((k_{i}\cdot p)^{3}+(k_{i}\cdot q)^{3}\right)}{\left[k_{i}\cdot\left(p+q\right)\right]^{2}}\left(p\cdot q\right)\left[\left(p+q\right)\cdot\frac{\partial}{\partial k_{i}}\right]^{2}\\
 & +\frac{(k_{i}\cdot p)^{3}+(k_{i}\cdot q)^{3}}{\left[k_{i}\cdot\left(p+q\right)\right]^{3}}\left(p\cdot q\right)^{2}\left(p+q\right)\cdot\frac{\partial}{\partial k_{i}}\\
 & \left.-\frac{(k_{i}\cdot p)^{3}+(k_{i}\cdot q)^{3}}{\left[k_{i}\cdot\left(p+q\right)\right]^{4}}\left(p\cdot q\right)^{3}\right\} M_{n}\left(\cdots,k_{i},\cdots\right)
\end{aligned}
 &  & \tau^{5} \label{eq:pole_expand-1-1}
\end{align}

Type A derivation is more complicated. We list all the derivable terms.
If $p$ is expanded first, we get 
\begin{align}
 & M_{\text{pole},s^{3}}\left(\cdots,0,0\right)=q\cdot\frac{\partial}{\partial q}M_{\text{pole},s^{3}}\left(\cdots,0,q\right){}_{q=0}=0\nonumber \\
 & \left.\frac{1}{m!}q^{m}\frac{\partial^{m}}{\partial q^{m}}M_{\text{pole},s^{3}}\left(\cdots,0,q\right)\right|{}_{q=0}=4c_{4}^{(3)}\sum_{i}(k_{i}\cdot q)^{2}\left[\frac{1}{\left(m-2\right)!}\left(q\cdot\frac{\partial}{\partial k_{i}}\right)^{m-2}\tilde{M}_{n}\left(\cdots,k_{i},\cdots\right)\right], &  & m\geq2\label{eq:gut_p0q2-1}
\end{align}
\begin{align}
 & \left.\left(\left.p\frac{\partial}{\partial p}M_{\text{pole},s^{3}}\left(\cdots,p,q\right)\right|_{p=0}\right)\right|_{q=0}=0\nonumber \\
 & \left.\left(q\cdot\frac{\partial}{\partial q}\right)\left(p\cdot\frac{\partial}{\partial p}\right)M_{\text{pole},s^{3}}\left(\cdots,p,q\right)\right|_{p=q=0}=-4c_{4}^{(3)}\sum_{i}\left(k_{i}\cdot p\right)\left(k_{i}\cdot q\right)\left(\cdots,k_{i},\cdots\right)\nonumber \\
 & \frac{1}{m!}\left.\left(q^{m}\cdot\frac{\partial}{\partial q^{m}}\right)\left(p\cdot\frac{\partial}{\partial p}\right)M_{\text{pole},s^{3}}\left(\cdots,p,q\right)\right|_{p=q=0}\nonumber \\
 & \hspace{0.3in}\begin{aligned}=-4c_{4}^{(3)}\sum_{i} & \left[\frac{1}{\left(m-1\right)!}\left(k_{i}\cdot p\right)\left(k_{i}\cdot q\right)\left(q\frac{\partial}{\partial k_{i}}\right)^{m-1}\right.\\
 & \left.+\frac{1}{\left(m-2\right)!}\left(p\cdot q\right)\left(k_{i}\cdot q\right)\left(q\frac{\partial}{\partial k_{i}}\right)^{m-2}\right]M_{n}\left(\cdots,k_{i},\cdots\right), &  & m\geq2
\end{aligned}
\end{align}
\begin{align}
 & \left.\left(\left.p^{2}\frac{\partial}{\partial p^{2}}M_{\text{pole},s^{3}}\left(\cdots,p,q\right)\right|_{p=0}\right)\right|_{q=0}=4c_{4}^{(3)}\sum_{i}(k_{i}\cdot p)^{2}M_{n}\left(\cdots,k_{i},\cdots\right)\label{eq:gut_p1q0-1-1}\\
 & \left.\frac{1}{2}\left(q\cdot\frac{\partial}{\partial q}\right)\left(p^{2}\cdot\frac{\partial}{\partial p^{2}}\right)M_{\text{pole},s^{3}}\left(\cdots,p,q\right)\right|_{p=q=0}\nonumber \\
 & \hspace{0.3in}=4c_{4}^{(3)}\sum_{i}\left[(k_{i}\cdot p)^{2}\left(q\cdot\frac{\partial}{\partial k_{i}}\right)+\left(k_{i}\cdot p\right)\left(p\cdot q\right)\right]M_{n}\left(\cdots,k_{i},\cdots\right)\\
 & \left.\frac{1}{2!2!}\left(q^{2}\cdot\frac{\partial}{\partial q^{2}}\right)\left(p^{2}\cdot\frac{\partial}{\partial p^{2}}\right)M_{\text{pole},s^{3}}\left(\cdots,p,q\right)\right|_{p=q=0}\nonumber \\
 & \hspace{0.3in}=4c_{4}^{(3)}\sum_{i}\left[\frac{1}{2}(k_{i}\cdot p)^{2}\left(q\cdot\frac{\partial}{\partial k_{i}}\right)^{2}+\left(k_{i}\cdot p\right)\left(p\cdot q\right)\left(q\cdot\frac{\partial}{\partial k_{i}}\right)+\left(p\cdot q\right)^{2}\right]M_{n}\left(\cdots,k_{i},\cdots\right)\\
 & \left.\frac{1}{3!2!}\left(q^{3}\cdot\frac{\partial}{\partial q^{3}}\right)\left(p^{2}\cdot\frac{\partial}{\partial p^{2}}\right)M_{\text{pole},s^{3}}\left(\cdots,p,q\right)\right|_{p=q=0}\nonumber \\
 & \hspace{0.3in}\begin{aligned}=4c_{4}^{(3)}\sum_{i} & \left[\frac{1}{3!}(k_{i}\cdot p)^{2}\left(q\cdot\frac{\partial}{\partial k_{i}}\right)^{3}+\frac{1}{2}\left(k_{i}\cdot p\right)\left(p\cdot q\right)\left(q\cdot\frac{\partial}{\partial k_{i}}\right)^{2}\right.\\
 & \left.+\left(p\cdot q\right)^{2}\left(q\cdot\frac{\partial}{\partial k_{i}}\right)\right]M_{n}\left(\cdots,k_{i},\cdots\right)
\end{aligned}
\end{align}
Terms where $q$ is expanded first can be obtained similarly. The
terms are then combined, where ordering ambiguity of mixed derivatives
resolved by the same prescription as for $M_{\text{pole},s^{2}}$.
The result is 
\begin{align}
 & \left[\left.M_{\text{pole},s^{3}}\left(\cdots,\tau p,\tau q\right)\right|_{\text{type B}}-\left.M_{\text{pole},s^{3}}\left(\cdots,\tau p,\tau q\right)\right|_{\text{type A}}\right]\nonumber \\
= & 0 &  & \tau^{0},\tau^{1},\tau^{2}\nonumber \\
 & +3c_{4}^{(3)}(p\cdot q)\sum_{i}\frac{[k_{i}\cdot(p-q)]^{2}}{k_{i}\cdot\left(p+q\right)}M_{n}(\cdots k_{i},\cdots) &  & \tau^{3}\nonumber \\
 & \begin{aligned}+6c_{4}^{(3)}(p\cdot q)\sum_{i} & \left\{ \frac{k_{i}\cdot\left(p-q\right)}{k_{i}\cdot\left(p+q\right)}\left(p_{\mu}^{\text{ }}q_{\nu}J_{i}^{\mu\nu}\right)\right.\\
 & \left.+2\frac{(k_{i}\cdot p)(k_{i}\cdot q)(p\cdot q)}{\left[k_{i}\cdot\left(p+q\right)\right]^{2}}+k_{i}\cdot(p+q)\right\} M_{n}(\cdots k_{i},\cdots)
\end{aligned}
 &  & \tau^{4}\nonumber \\
 & \begin{aligned}+6c_{4}^{(3)}\left(p\cdot q\right)\sum_{i} & \left\{ \frac{\left(p_{\mu}q_{\nu}J^{\mu\nu}\right)^{2}}{k_{i}\cdot\left(p+q\right)}+2\frac{(k_{i}\cdot p)(k_{i}\cdot q)(p\cdot q)}{\left[k_{i}\cdot\left(p+q\right)\right]^{2}}\left[\left(p+q\right)\cdot\frac{\partial}{\partial k_{i}}\right]\right.\\
 & \left.+\frac{\left(p\cdot q\right)^{2}}{k_{i}\cdot\left(p+q\right)}\left(-\frac{1}{2}+\frac{\left[k_{i}\cdot\left(p-q\right)\right]^{2}}{2\left[k_{i}\cdot\left(p+q\right)\right]^{2}}\right)\right\} M_{n}\left(\cdots k_{i},\cdots\right)
\end{aligned}
 &  & \tau^{5}
\end{align}
Note that the result starts at $\tau^{3}$despite the fact that both
$\left.M_{\text{pole},s^{3}}\right|_{\text{type B}}$ and $\left.M_{\text{pole},s^{3}}\right|_{\text{type A}}$
start at $\tau^{2}$. Their expansion at this order is identical,
thus eventually cancel out.


\begin{thebibliography}{1}
\bibitem{Weinberg}
S.~Weinberg, \emph{{Infrared photons and gravitons}},
  \href{http://dx.doi.org/10.1103/PhysRev.140.B516}{\emph{Phys. Rev.} {\bf 140}
  (1965) B516--B524}.
  
\bibitem{Adler}
S.~L. Adler, 
\emph{{Consistency conditions on the strong interactions implied by a partially conserved axial vector current}},
\href{http://dx.doi.org/10.1103/PhysRev.137.B1022}{\emph{Phys. Rev.} {\bf  137} (1965) B1022--B1033}.
  
\bibitem{Low}
F.~E.~Low, 
\emph{{Bremsstrahlung of very low-energy quanta in elementary particle collisions}}, 
\href{http://dx.doi.org/10.1103/PhysRev.110.974}{\emph{Phys. Rev.} {\bf 110}} (1958) 974-977.

\bibitem{photon_Kroll}
T.~H.~Burnett and N.~M.~Kroll, 
\emph{{Extension of the Low soft photon theorem}}, 
\href{http://dx.doi.org/10.1103/PhysRevLett.20.86}{\emph {Phys. Rev. Lett.} {\bf 20}} (1968) 86.

\bibitem{photon_Duca}
V.~Del~Duca, 
\emph{{High-energy Bremsstrahlung theorems for soft photons}}, 
\href{http://dx.doi.org/10.1016/0550-3213(90)90392-Q}{\emph{Nucl. Phys.} {\bf B345}} (1990) 369-388.

\bibitem{grav_Gross}
D.~J.~Gross and R.~Jackiw, 
\emph{{Low-energy theorem for graviton scattering}}, 
\href{https://doi.org/10.1103/PhysRev.166.1287}{\emph{Phys. Rev.} {\bf 166}} (1968) 1287-1292.

\bibitem{grav_Jackiw}
R.~Jackiw, 
\emph{{Low-energy theorems for massless bosons: photons and gravitons}}, \href{https://doi.org/10.1103/PhysRev.168.1623}{\emph{Phys. Rev.} {\bf 168}} (1968) 1623-1633.

\bibitem{grav_White}
C.~D.~White, 
\emph{{Factorization Properties of Soft Graviton Amplitudes}}, 
\href{https://doi.org/10.1007/JHEP05(2011)060}{\emph{JHEP} {\bf 1105}} (2011) 060, 
\href{https://arxiv.org/abs/1103.2981}{[arXiv:1103.2981]}.

\bibitem{recent_Cachazo}
F.~Cachazo and A.~Strominger, 
\emph{{Evidence for a New Soft Graviton Theorem}}, 
\href{https://arxiv.org/abs/1404.4091}{[1404.4091]}.

\bibitem{recent_Casali}
E. Casali, 
\emph{{Soft sub-leading divergences in Yang-Mills amplitudes}}, 
\href{https://doi.org/10.1007/JHEP08(2014)077}{\emph{JHEP} {\bf 08}} (2014) 077, 
\href{https://arxiv.org/abs/1404.5551}{1404.5551}.

\bibitem{soft2He}
F.~Cachazo, S.~He and E.~Y. Yuan, \emph{{New Double Soft Emission Theorems}},
\href{https://journals.aps.org/prd/abstract/10.1103/PhysRevD.92.065030}{\emph{Phys. Rev.} {\bf D93} (2016) 045032}
  \href{https://arxiv.org/abs/1503.04816}{{\tt 1503.04816}}.
  
\bibitem{DiVecchia} 
  P.~Di Vecchia, R.~Marotta and M.~Mojaza,
 \textit{Double-soft behavior of the dilaton of spontaneously broken conformal invariance},
 \href{https://link.springer.com/article/10.1007%2FJHEP09%282017%29001}{\emph{JHEP} {\bf 1709}, 001 (2017)},
  [\href{https://arxiv.org/abs/1705.06175}{{\tt 1705.06175}}].
  
\bibitem{soft2Huang}
A. L. Guerrieri, Y.-t. Huang, Z. Li, C. Wen, 
\textit{On the Exactness of Soft Theorems}, 
[\href{https://arxiv.org/abs/1705.10078}{{\tt 1705.10078}}].


\bibitem{gluon_AB}
T.~Klose, T.~McLoughlin, D.~Nandan, J.~Plefka and G.~Travaglini
\emph{{Double-Soft Limits of Gluons and Gravitons}}
\href{http://dx.doi.org/10.1007/JHEP07(2015)135}{\emph{JHEP} {\bf 07}} (2015)135
\href{http://arxiv.org/abs/arXiv:1504.05558}{[1504.05558]}

\bibitem{grav_AB}
A.~Volovich, C.~Wen, M.~Zlotnikov,
\emph{{Double Soft Theorems in Gauge and String Theories}},
\href{https://doi.org/10.1007/JHEP07(2015)095}{\emph{JHEP} {\bf 07}} (2015) 095,
\href{https://arxiv.org/abs/1504.05559}{[1504.05559]}.

\bibitem{current_is_A}
T.~McLoughlin and D.~Nandan,
\emph{{Multi-Soft gluon limits and extended current algebras at null-infinity}},
\href{http://dx.doi.org/10.1007/JHEP08(2017)124}{\emph{JHEP} {\bf 08}}  (2017) 124,
\href{http://arxiv.org/abs/arXiv:1610.03841}{1610.03841}.

\bibitem{SCET}
A.~J.~Larkoski, D~Neill, I~W.~Stewart
\emph{{Soft Theorems from Effective Field Theory}}
\href{https://doi.org/10.1007/JHEP06(2015)077}{\emph{JHEP},{\bf 06}} (2009) 077, 
\href{https://arxiv.org/abs/1412.3108}{1412.3108}


\bibitem{nonlinear}
I.~Low, \emph{{Double Soft Theorems and Shift Symmetry in Nonlinear Sigma Models}}, 
\href{http://dx.doi.org/10.1103/PhysRevD.93.045032}{\emph{Phys. Rev.} {\bf D93} (2016) 045032}, [\href{https://arxiv.org/abs/1512.01232}{{\tt  1512.01232}}].

\bibitem{periodic}
C.~Cheung, K.~Kampf, J.~Novotny, C.-H. Shen and J.~Trnka, \emph{{A Periodic Table of Effective Field Theories}},
  \href{http://dx.doi.org/10.1007/JHEP02(2017)020}{\emph{JHEP} {\bf 02} (2017)
  020}, [\href{https://arxiv.org/abs/1611.03137}{{\tt 1611.03137}}].

\bibitem{DBI_locality}
N. Arkani-Hamed, L. Rodina and J. Trnka, 
\emph{{Locality and Unitarity from Singularities and Gauge Invariance,}} 
[\href{https://arxiv.org/abs/1612.02797}{\tt 1612.02797]}.

\bibitem{DBI_locality2}
L.~Rodina, \emph{{Uniqueness from gauge invariance and the Adler zero}},
  [\href{https://arxiv.org/abs/1612.06342}{{\tt 1612.06342}}].

\bibitem{cDBI_single}
P.~Di~Vecchia, R.~Marotta, M.~Mojaza and J.~Nohle, \emph{{New soft theorems for
  the gravity dilaton and the Nambu-Goldstone dilaton at subsubleading order}},
  \href{http://dx.doi.org/10.1103/PhysRevD.93.085015}{\emph{Phys. Rev.} {\bf
  D93} (2016) 085015}, [\href{https://arxiv.org/abs/1512.03316}{{\tt
  1512.03316}}].

\bibitem{dilaton}
M.~Bianchi, A.~L. Guerrieri, Y.-t. Huang, C.-J. Lee and C.~Wen,
  \emph{{Exploring soft constraints on effective actions}},
  \href{http://dx.doi.org/10.1007/JHEP10(2016)036}{\emph{JHEP} {\bf 10} (2016)
  036}, [\href{https://arxiv.org/abs/1605.08697}{{\tt 1605.08697}}].
  
\bibitem{soft_extended}
F.~Cachazo, P.~Cha, and S.~Mizera,
   \emph{{Extensions of Theories from Soft Limits}},
   \href{https://link.springer.com/article/10.1007%2FJHEP06%282016%29170}{\emph{JHEP} {\bf 06}, 170 (2016)},
   [\href{https://arxiv.org/abs/1604.03893}{1604.03893}]
  
\bibitem{NLSM_phi3}
I.~Low and Z.~Yin,
\emph{{Ward Identity and Scattering Amplitudes in Nonlinear Sigma Models}},
[\href{https://arxiv.org/abs/1709.08639}{{\tt 1709.08639}}].

\bibitem{AB_ref}
T.~Klose, T.~McLoughlin, D.~Nandan, J.~Plefka and G.~Travaglini, 
\emph{{Double-Soft Limits of Gluons and Gravitons}},
\href{http://dx.doi.org/10.1007/JHEP07(2015)135}{\emph{JHEP} {\bf 07}} (2015) 135,
[\href{https://arxiv.org/abs/1504.05558}{{\tt 1504.05558}}].

  
\end{thebibliography}
\end{document}